\documentclass[aps,prr,twocolumn,amsmath,amssymb,superscriptaddress,longbibliography]{revtex4-1}
\usepackage[english]{babel}
\usepackage{xcolor}
\usepackage{amssymb}
\usepackage{dcolumn}
\usepackage{bm}
\usepackage{graphicx}
\usepackage{amsmath}
\usepackage{braket}

\usepackage{graphicx}        

\graphicspath{{pict/}{}}

\usepackage[normalem]{ulem}

\usepackage{dcolumn}
\usepackage{bm}
\usepackage[pdfstartview=FitH, CJKbookmarks=true, bookmarksnumbered=true, bookmarksopen=true, colorlinks=true, pdfborder=001, citecolor=blue, linkcolor=blue, urlcolor=blue, linktocpage=true] {hyperref}

\begin{document}
\title{Dissipation-selected photon-phonon pair and photon-bundle emission in a tripartite Mollow system} 
	
\author{Jiayuan Zhang}
\affiliation{Guangdong Provincial Key Laboratory of Quantum Metrology and Sensing $\&$ School of Physics and Astronomy, Sun Yat-Sen University (Zhuhai Campus), Zhuhai 519082, China}

\author{Geng Zhao}
\affiliation{Guangdong Provincial Key Laboratory of Quantum Metrology and Sensing $\&$ School of Physics and Astronomy, Sun Yat-Sen University (Zhuhai Campus), Zhuhai 519082, China}

\author{Jing Tang}
\email{jingtang@gdut.edu.cn}
\affiliation{School of Physics and Optoelectronic Engineering, Guangdong University of Technology, Guangzhou 510006, China}
\affiliation{Guangdong Provincial Key Laboratory of Sensing Physics and System Integration Applications, Guangdong University of Technology, Guangzhou, 510006, China}

\author{Yuangang Deng}
\email{dengyg3@mail.sysu.edu.cn}
\affiliation{Guangdong Provincial Key Laboratory of Quantum Metrology and Sensing $\&$ School of Physics and Astronomy, Sun Yat-Sen University (Zhuhai Campus), Zhuhai 519082, China}

\date{\today}

\begin{abstract}
Dissipation is generally regarded as detrimental to nonclassical-state generation, yet whether it can determine which quantum resource is emitted from an open hybrid quantum system remains largely unexplored. Here, we show that the symmetry of bosonic dissipation channels can select and switch quantum-emission pathways in a tripartite Mollow system, where a strongly driven two-level emitter simultaneously couples to cavity photons and motional phonons. In the Mollow regime, higher-order tripartite interactions generate resonant photon-phonon pair excitations, whose emitted quantum states are subsequently selected by dissipation. Under symmetric dissipation, the system produces strongly correlated photon-phonon pairs, enabling both single-pair emission and correlated multipair bundles with pronounced nonclassical correlations. Breaking the dissipation symmetry progressively suppresses the hybrid correlations and dynamically decouples the photonic and phononic degrees of freedom. In the strongly asymmetric regime, long-lived phonons accumulate in highly excited Fock states, while rapidly leaking cavity photons are converted into high-purity single-photon and two-photon bundles. The crossover between these distinct emission regimes is characterized through time-resolved correlation functions, Fock-state distributions, and Wigner quasiprobability distributions. Our results establish dissipation symmetry as a resource for steering quantum emission between hybrid and purely photonic nonclassical states, providing a route toward programmable hybrid quantum emitters and dissipative quantum-state engineering.
\end{abstract}

\maketitle 

\section{Introduction}

Nonclassical quantum states constitute a fundamental resource for quantum science and technology. Their unique properties, arising from quantum superposition, entanglement, and squeezing, enable functionalities beyond classical limits in quantum communication, quantum computation, precision metrology, and quantum sensing \cite{nonclassical_superposition_thermal,nonclassical_superposition_in_any_physical_theory,nonclassical_entanglement,nonclassical_entanglement_of_light,nonclassical_entanglement_of_photon,nonclassical_squeezing_Gaussian_States,nonclassical_squeezing_cooling_of_oscillator,nonclassical_squeezing_spin,quantum_metrology_1,quantum_metrology_2,quantum_metrology_3,quantum_metrology_4,quantum_metrology_5,quantum_metrology_6}. Consequently, the controlled generation and manipulation of nonclassical excitations have become central objectives in modern quantum science \cite{nonclassical_processing_capacity,nonclassical_processing_capacity_b,nonclassical_measurement_sensitivity,quantum_entanglement,quantum_entanglement_1,quantum_entanglement_2,nonclassical_quantum_computing,nonclassical_quantum_computing_1,nonclassical_quantum_computing_2}. In particular, engineered quantum emitters capable of producing tailored nonclassical states provide essential building blocks for scalable quantum technologies.

Among the various forms of nonclassical emission, correlated multiquantum states have attracted considerable interest because of their potential applications in Heisenberg-limited metrology, quantum lithography, biological sensing, and quantum information processing \cite{Multiquanta_physics_in_metrology,Multiquanta_physics_in_lithography,Multiquantum_Biological_measurement,Multiquantum_Biological_measurement_1}. Recent theoretical and experimental advances have demonstrated multiphoton-bundle emission and multiquanta-state generation in a wide range of platforms, including post-selected systems, Rydberg ensembles, and cavity-, circuit-, and waveguide-QED architectures \cite{Multiquantum_post_selection,Multiquantum_Rydberg_atom,Multiquantum_Rydberg_atom_1,Multiquantum_Rydberg_atom_2,Rydberg_PB,Multiquantum_QED_platform,Multiquantum_QED_platform_1,Multiquantum_QED_platform_2,Multiquantum_QED_platform_3,Multiquantum_QED_platform_4,Multiquantum_QED_platform_5,Quantum_Entangled_Pairs_Bin-qian,Optics_14-26,Tau_min_of_Bundle_Emission,Emitters_of_N-photon_bundles,deng2021motional,circuit_PB_1,circuit_PB_2,circuit_PB_3}. Unlike conventional single-photon emitters, bundle emitters release energy in correlated groups of multiple quanta, providing new opportunities for deterministic nonclassical light sources beyond the single-photon paradigm \cite{quantum_metrology,quantum_image,quantum_information_processing_1,quantum_information_processing_2}. Meanwhile, hybrid quantum resources involving multiple bosonic degrees of freedom, such as correlated photon-phonon pairs, have emerged as promising candidates for quantum networking and quantum transduction because they distribute quantum correlations across distinct physical platforms \cite{Quantum_Entangled_Pairs_Bin-qian,Correlations_between_Photons_and_Phonons,photon_pairs_by_free_electrons}.

Hybrid systems combining photons, phonons, and quantum emitters provide a natural platform for generating such complex quantum resources \cite{Hybrid_quantum_resource_1,Hybrid_quantum_resource_2,Hybrid_quantum_resource_4,Hybrid_quantum_resource_5,Hybrid_quantum_resource_6}. In particular, tripartite interactions enable coherent conversion between atomic, photonic, and phononic excitations, giving rise to multipartite entanglement, squeezed-state engineering, quantum transduction, and hybrid quantum networking \cite{tripartite_JC_two_quantized_harmonic_oscillators,tripartite_Topological_Fizeau_Drag_photon,tripartite_cavity_atom_photon,tripartite_genuinely_entangled_state,tripartite_interactions_by_Prof_Deng,tripartite_panjianwei,tripartite_photon-phonon-magnon,tripartite_photons_phonons_qubit_excitation}. Strong tripartite coupling can coherently convert atomic excitations into correlated photon-phonon excitations, thereby extending quantum emission beyond purely photonic states \cite{complex_quantum_resources_1}. Under strong coherent driving, these systems enter the Mollow regime, where the dressed-state structure supports higher-order photon-phonon resonances and resonant multiphoton-multiphonon transitions. Such higher-order processes provide a versatile platform for generating hybrid quantum resources beyond conventional single-photon emission.

 \begin{figure*}[t]
\includegraphics[width=0.9\textwidth]{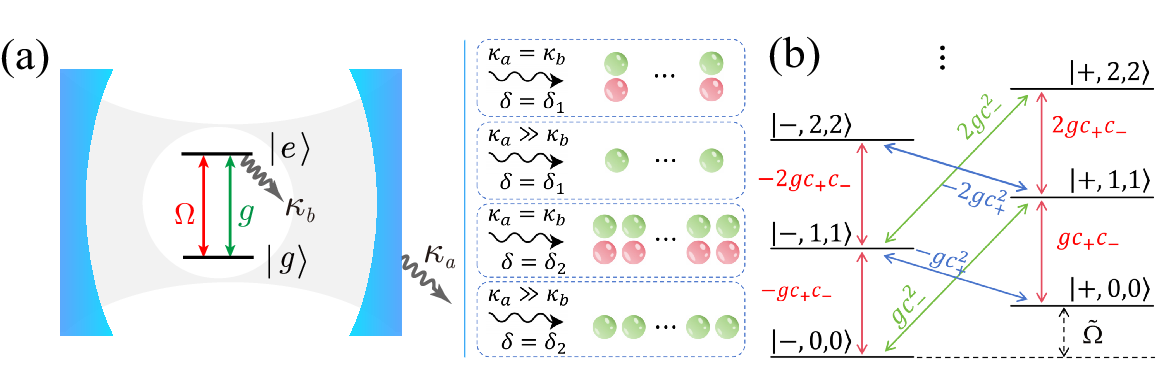}	     
\caption{
(a) Schematic of the dissipative tripartite Mollow system. A driven two-level emitter is coupled to a cavity-photon mode  and a motional phonon mode through the tripartite interaction. By tuning the dissipation ratio $\kappa_a/\kappa_b$, the system can be switched between photon-phonon pair emission and photon-bundle emission at the single-pair ($\delta=\delta_1$) and multipair ($\delta=\delta_2$) resonances. 
(b) Dressed-state structure in the Mollow regime. Strong driving forms the manifolds $|\pm,n,n\rangle$, and the tripartite interaction mediates higher-order resonant transitions between $|+,0,0\rangle$ and $|-,N,N\rangle$, providing the coherent origin of dissipation-selected quantum emission.
}
\label{fig:model}
\end{figure*}

While coherent interactions determine the quantum resources that can be generated, the emitted quantum states of an open system are ultimately governed by the competition between coherent dynamics and dissipation. Most previous studies have regarded dissipation primarily as a source of decoherence or, more recently, as a resource for stabilizing nonclassical steady states through reservoir engineering \cite{dissipation_nonclassical_1,dissipation_nonclassical_2,dissipation_nonclassical_3,dissipation_nonclassical_4,quantum_engineering_1,quantum_engineering_2,quantum_engineering_3,quantum_engineering_4}. However, a fundamentally different question remains largely unexplored: can dissipation determine which quantum resource is emitted from a given coherent excitation process ? More specifically, can the same coherent interaction produce qualitatively different quantum-emission channels solely by changing the symmetry of the dissipation ? Answering this question would establish dissipation as a mechanism for quantum-resource selection, opening new opportunities for steering nonclassical emission in open hybrid quantum systems.

In this work, we address this question using a tripartite Mollow system consisting of a strongly driven two-level emitter coupled simultaneously to cavity photons and motional phonons. In the strong-driving regime, cavity-enhanced Stokes processes generate effective higher-order spin-photon-phonon interactions, enabling resonant multiphoton-multiphonon transitions and correlated photon-phonon excitations. We show that the symmetry of the bosonic dissipation channels determines the nature of the emitted quantum states. In particular, symmetric dissipation favors the emission of correlated photon-phonon pairs, whereas dissipation asymmetry can transform the emission into single-photon and photon-bundle output. To characterize this dissipation-selected quantum emission, we analyze correlation functions, Fock-state distributions, and phase-space quasiprobability distributions. Our results establish a route toward dissipation-controlled quantum emitters and hybrid quantum-state engineering, revealing how the symmetry of loss channels can select between hybrid and purely photonic nonclassical resources.

This paper is organized as follows. In Sec.~\ref{Sec2}, we introduce the tripartite Mollow system and derive the effective spin-photon-phonon Hamiltonian. In Sec.~\ref{Sec3}, we identify the multiphoton-multiphonon resonances and the associated photon-phonon super-Rabi oscillations. Section~\ref{Sec4} introduces the quantum-statistical diagnostics used to characterize the emission properties. In Sec.~\ref{Sec5}, we demonstrate dissipation-selected quantum emission and show how dissipation symmetry switches the system between photon-phonon pair emission and photon-bundle emission. Finally, a summary is given in Sec.~\ref{Sec6}.

\section{Model and hybrid Hamiltonian}\label{Sec2}

\textit{Model}.---We consider a single alkaline-earth-metal atom, for instance $^{88}{\rm Sr}$, confined in a one-dimensional harmonic trap and coupled to a single mode of a high-finesse optical cavity. As illustrated in Fig.~\ref{fig:model}(a), the relevant internal degree of freedom is described by a two-level transition between the ground state $|g\rangle$ $(^1S_0)$ and the excited state $|e\rangle$ $(^3P_1)$. The atomic transition is driven by a transverse classical field with Rabi frequency $\Omega$, while the cavity mode couples to the atom through a position-dependent interaction. In a frame rotating at the pump frequency $\omega_p$, the Hamiltonian reads
\begin{align}
\hat{\cal H}_0
=&~\frac{\hat{p}_x^2}{2M}
+V(\hat{x})
+\Delta_{\sigma}\hat{\sigma}^{\dagger}\hat{\sigma}
+\Delta_a\hat{a}^{\dagger}\hat{a}
\nonumber\\
&+g_0\sin(k_L\hat{x})
\left(\hat{a}^{\dagger}\hat{\sigma}
+\hat{a}\hat{\sigma}^{\dagger}\right)
+\Omega\hat{\sigma}_x .
\label{H0}
\end{align}
Here $\hat{a}$ is the annihilation operator of the cavity mode, $\hat{\sigma}=|g\rangle\langle e|$ is the atomic lowering operator, and $\hat{\sigma}_x=\hat{\sigma}+\hat{\sigma}^{\dagger}$. The parameters $g_0$ and $k_L$ denote the single-atom cavity coupling strength and cavity wave vector, respectively. The detunings are defined as $\Delta_a=\omega_a-\omega_p$ and $\Delta_{\sigma}=\omega_{\sigma}-\omega_p$, where $\omega_a$, $\omega_{\sigma}$, and $\omega_p$ are the cavity, atomic-transition, and pump frequencies. The atomic center-of-mass motion is governed by a spin-independent harmonic trapping potential $V(\hat{x})=\frac{1}{2}M\omega_b^2\hat{x}^2$, where $M$ is the atomic mass and $\omega_b$ is the trap frequency.

The motional degree of freedom is quantized as
\begin{equation}
\hat{x}=x_{\rm zpf}(\hat{b}+\hat{b}^{\dagger}),\qquad x_{\rm zpf}=\sqrt{\frac{\hbar}{2M\omega_b}},
\end{equation}
where $\hat{b}$ is the annihilation operator for the motional mode and $x_{\rm zpf}$ is the zero-point fluctuation amplitude. In the Lamb-Dicke regime, $k_Lx_{\rm zpf}\ll1$, the spatially dependent coupling can be expanded to leading order as $\sin(k_L\hat{x})\simeq k_L\hat{x}$. Substituting the quantized displacement into Eq.~\eqref{H0} and omitting the constant zero-point energy, we obtain
\begin{align}
\hat{\cal H}_1
=&~\omega_b\hat{b}^{\dagger}\hat{b}
+\Delta_a\hat{a}^{\dagger}\hat{a}
+\Delta_{\sigma}\hat{\sigma}^{\dagger}\hat{\sigma}
\nonumber\\
&+g
(\hat{b}+\hat{b}^{\dagger})
\left(\hat{a}^{\dagger}\hat{\sigma}
+\hat{a}\hat{\sigma}^{\dagger}\right)
+\Omega\hat{\sigma}_x ,
\label{H_LD}
\end{align}
where $g=g_0k_Lx_{\rm zpf}$ is the zero-point-motion-assisted spin-photon-phonon coupling strength.

We next transform to a rotating frame in which the cavity and motional modes have residual detunings $\delta_a$ and $\delta_b$, respectively. Near the blue-sideband resonance, the dominant resonant process is the Stokes process in which an atomic excitation is converted into a correlated photon-phonon pair, and vice versa. Retaining this near-resonant contribution and neglecting rapidly oscillating terms under the rotating-wave approximation, the effective tripartite Hamiltonian in a rotating reference frame at the frequency $\omega'$ ($\approx \omega_b$) becomes~\cite{tripartite_interactions_by_Prof_Deng}
\begin{align}
\hat{\cal H}_2
=&~\delta_a \hat{a}^{\dagger}\hat{a}
+\delta_b \hat{b}^{\dagger}\hat{b}
+\Delta_{\sigma}\hat{\sigma}^{\dagger}\hat{\sigma}
+g\left(
\hat{a}^{\dagger}\hat{b}^{\dagger}\hat{\sigma}
+\hat{a}\hat{b}\hat{\sigma}^{\dagger}
\right)
+\Omega\hat{\sigma}_x,
\label{H_tripartite}
\end{align}
where the effective detuning are $\delta_a = \Delta_a+\omega'$, $\delta_b= \omega_b-\omega'$ and $\Delta_{\sigma}=-\Delta+\delta_a+\delta_b$ with photon-phonon-atom detuning $\Delta=\omega_a+\omega_b-\omega_{\sigma}$.

Equation~\eqref{H_tripartite} describes the coherent exchange between an atomic excitation and a correlated photon-phonon pair. This interaction is the microscopic origin of the resonant multiphoton-multiphonon processes considered below. When the coherent drive is sufficiently strong, the atom enters the Mollow regime and the dressed-state structure supports effective higher-order transitions between $|+,0,0\rangle$ and $|-,N,N\rangle$. These coherent transitions generate the hybrid quantum resources, while the dissipation channels introduced in the following sections determine the properties of the emitted quantum states.

Under strong driving, the atom enters the Mollow regime and the bare atomic states are replaced by the dressed states shown in Fig.~\ref{fig:model}(b). In contrast to the conventional Mollow ladder encountered in cavity-QED systems~\cite{Emitters_of_N-photon_bundles,Mollow_Triplet_origin,Optics_14-26,N-Phonon_Bundle_Emission_via_the_Stokes_Process}, each manifold of the present tripartite system is characterized by the hybrid states $|\pm,n_a,n_b\rangle$, where $n_a$ and $n_b$ denote the photon and phonon populations, respectively. The atomic dressed states are
\begin{equation}
|+\rangle=c_+|e\rangle+c_-|g\rangle,
\qquad
|-\rangle=c_-|e\rangle-c_+|g\rangle,
\end{equation}
with eigenenergies $E_{\pm}=({\Delta_{\sigma}\pm\tilde{\Omega}})/{2}$ and the generalized Rabi frequency $\tilde{\Omega}=\sqrt{\Delta_{\sigma}^{2}+4\Omega^{2}}$.

\textit{Nth-order effective Hamiltonian}.---In the strong-driving regime, virtual intermediate states enable higher-order transitions between different manifolds of the Mollow ladder. To describe these processes, we introduce the dressed-state transition operator $\hat{\tilde{\sigma}}
=|-\rangle\langle+|$. Substituting $\hat{\sigma}=c_-^2\hat{\tilde{\sigma}}^{\dagger}-c_+^2\hat{\tilde{\sigma}}+c_+c_-\hat{\tilde{\sigma}}_z$ into Eq.~(\ref{H_tripartite}) and performing perturbation theory in the strong-driving limit $g\ll\tilde{\Omega}$, we obtain an effective Hamiltonian describing resonant transitions between the states $|+,0,0\rangle$ and $|-,N,N\rangle$,
\begin{align}
\hat{\cal H}_{\rm eff}^{(N)}
=&~
\delta_a\hat{a}^{\dagger}\hat{a}
+\delta_b\hat{b}^{\dagger}\hat{b}
+\frac{\tilde{\Omega}}{2}
\hat{\tilde{\sigma}}_z
\nonumber\\
&
+g_{\rm eff}^{(N)}
\left(
\hat{a}^{N}\hat{b}^{N}\hat{\tilde{\sigma}}^{\dagger}
+\hat{a}^{\dagger N}
\hat{b}^{\dagger N}
\hat{\tilde{\sigma}}
\right),
\label{eq:H_eff^n}
\end{align}
where the effective coupling strength of the $N$th-order process is (see Appendix \ref{AppA})
\begin{align}
g_{\rm eff}^{(N)}
=
(N^2)^{N-1}
\frac{
c_-^{N-1}
c_+^{N+1}
}{
[(N-1)!]^2
\tilde{\Omega}^{\,N-1}
}
(-g)^N
\label{eq:g^n}.
\end{align}

Equation~(\ref{eq:H_eff^n}) describes an effective coherent exchange between the states $|+,0,0\rangle$ and $|-,N,N\rangle$. Physically, this process corresponds to the simultaneous creation or annihilation of $N$ correlated photon-phonon pairs. Throughout this work, we refer to the resonance associated with the transition
$|+,0,0\rangle\leftrightarrow|-,N,N\rangle$
as the $N$-pair resonance. In particular, the cases $N=1$, $2$, and $3$ correspond to the one-pair, two-pair, and three-pair resonances, respectively. 
The resonance condition
\begin{equation}
\tilde{\Omega}
=
N(\delta_a+\delta_b),
\end{equation}
selects the desired $N$th-order process while suppressing competing lower-order transitions.

At the $N$-pair resonance, the effective Hilbert space separates into two sectors. For excitation numbers below the resonant threshold, the states remain uncoupled,
\begin{equation}
|\Psi\rangle_{|-,n,n\rangle}
=
|-,n,n\rangle,
\qquad
0\le n < N,
\end{equation}
with eigenenergies
\begin{equation}
E_{|-,n,n\rangle}
=
\left(n-\frac{N}{2}\right)
(\delta_a+\delta_b),
\end{equation}
For $n\ge N$, the resonant coupling hybridizes the states
$|-,n,n\rangle$ and $|+,n-N,n-N\rangle$, yielding the dressed eigenstates
\begin{align}
|\pm,n,n\rangle'
=
\frac{
|-,n,n\rangle
\pm
|+,n-N,n-N\rangle
}{\sqrt{2}},
\end{align}
with eigenenergies
\begin{align}
E_{|\pm,n,n\rangle'}
=
\left(n-\frac{N}{2}\right)
(\delta_a+\delta_b)
\pm
g_{\rm eff}^{(N)}
\frac{n!}{(n-N)!}.
\end{align}
These hybridized manifolds  support the generation of correlated multiphoton-multiphonon excitations through higher-order resonant processes. As we show below, dissipation determines whether these excitations are emitted as correlated photon-phonon pairs or as photon bundles.

\begin{figure}[t]
	\includegraphics[width=0.49\textwidth]{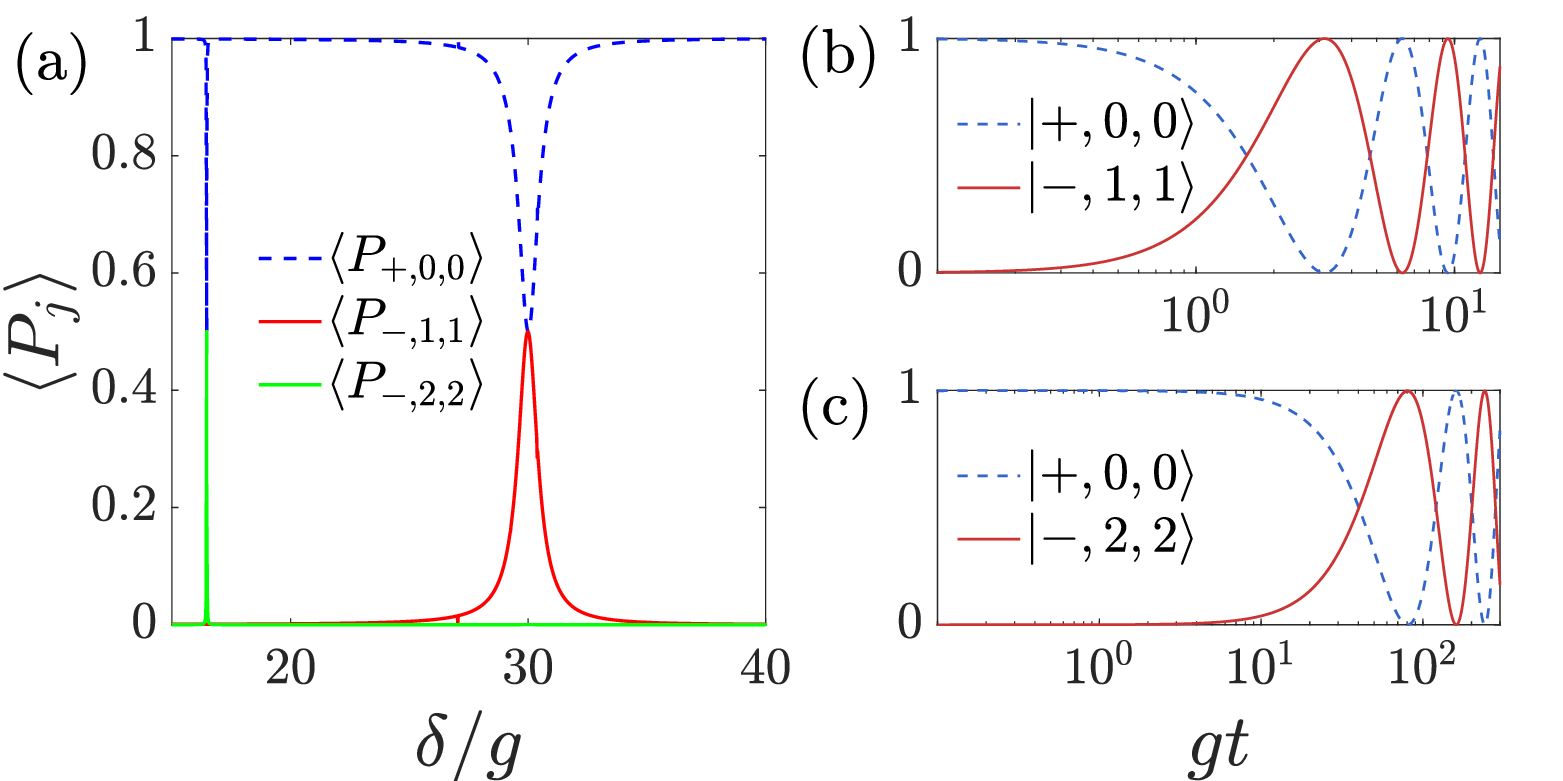}	     
	\caption{(a) Time-averaged population of states $|+,0,0\rangle$,  $|-,1,1\rangle$, $|-,2,2\rangle$ as functions of $\delta$, showing the one-pair, two-pair, and three-pair resonances. 	(b)-(c) Photon-phonon super-Rabi oscillations between $|+,0,0\rangle$ and $|-,1,1\rangle$, $|-,2,2\rangle$, respectively, with $\delta/g=26.25$, $16.47$, corresponding to the one-pair and two-pair resonances. The other parameters are $\Omega/g=30$ and $\Delta/g=60$.}
	\label{fig:resonance}
\end{figure}

\section{Photon-phonon super-Rabi oscillations}\label{Sec3}

The effective Hamiltonian in Eq.~(\ref{eq:H_eff^n}) shows that the tripartite Mollow system can resonantly convert a dressed atomic excitation into correlated photon-phonon bundles. As illustrated in Fig.~\ref{fig:model}(b), a super-Rabi resonance occurs when the dressed-state splitting matches the energy required to create an $N$-pair photon-phonon bundle. For the fundamental process
$|\pm,0,0\rangle\leftrightarrow|\mp,1,1\rangle$,
the resonance condition is $\delta_a+\delta_b=\pm \tilde{\Omega}$. Using $\Delta_{\sigma}=-\Delta+\delta_a+\delta_b$ and
$\Delta=\omega_a+\omega_b-\omega_{\sigma}$, the corresponding resonance detuning for fixing $\delta=\delta_a=\delta_b$ is
\begin{equation}
\delta_1
=
\frac{\Delta^2+4\Omega^2}{4\Delta},
\label{eq:1_resonance_point}
\end{equation}
For an $N$th-order process with $N\geq2$, the relevant transition is $|\pm,0,0\rangle\leftrightarrow|\mp,N,N\rangle$, corresponding to the simultaneous creation or annihilation of $N$ photons and $N$ motional phonons. The resonance condition becomes $2N \delta =\pm\tilde{\Omega}$, which gives
\begin{equation}
\delta_N
=
\frac{
-\Delta
\pm
\sqrt{4(N^2-1)\Omega^2+N^2\Delta^2}
}{
2(N^2-1)
}.
\label{eq:resonance_point}
\end{equation}
Equations~(\ref{eq:1_resonance_point}) and
(\ref{eq:resonance_point}) specify the detunings at which different higher-order sideband processes are selectively activated. In practice, different sideband orders can be addressed by tuning the cavity detuning, the trap frequency, or the driving amplitude. The case $N=1$ corresponds to the fundamental photon-phonon-pair resonance, whereas $N\geq2$ describes multipair resonances involving correlated multiphoton-multiphonon bundles. These resonances provide the coherent pathways underlying the dissipation-selected emission processes discussed below.

To identify these resonances numerically, we first consider the closed-system dynamics in the absence of dissipation. Following the time-averaged population analysis in Ref.~\cite{Two_photon_decay}, a complete population transfer between
$|+,0,0\rangle$ and $|-,N,N\rangle$ is indicated by
\begin{equation}
\left\langle P_{|+,0,0\rangle}\right\rangle
=
\left\langle P_{|-,N,N\rangle}\right\rangle
=
\frac{1}{2},
\end{equation}
where $\left\langle P_{|j\rangle}\right\rangle =\lim_{T\rightarrow\infty}\frac{1}{T}\int_{-T/2}^{T/2}P_{|j\rangle}(t)\,dt$ is the long-time-averaged population of state $|j\rangle$.

Figure~\ref{fig:resonance}(a) shows the time-averaged populations of
$|+,0,0\rangle$, $|-,1,1\rangle$ and $|-,2,2\rangle$ as functions of $\delta$ for
$\Omega/g=30$ and $\Delta/g=60$. The two peaks identify the one-pair and two-pair resonances. The corresponding coherent dynamics are shown in
Figs.~\ref{fig:resonance}(b)--\ref{fig:resonance}(c), where $\delta$ is tuned to the respective resonance conditions.  In each case, the population oscillates predominantly between $|+,0,0\rangle$ and $|-,N,N\rangle$, demonstrating photon-phonon super-Rabi oscillations associated with the coherent generation of $N$ correlated photon-phonon pairs.

The oscillatory dynamics originate from the isolation of an effective two-state subspace. Once the transition
$|+,0,0\rangle\leftrightarrow|-,N,N\rangle$ is resonant, higher manifolds such as
$|+,2N,2N\rangle$ remain far off resonance and are suppressed by the dressed-state energy mismatch. The resulting dynamics therefore correspond to a photon-phonon super-Rabi oscillation, which provides the coherent basis for the dissipation-selected emission processes discussed below.

\section{Quantum-statistical diagnostics}\label{Sec4}

The coherent $N$-pair resonances discussed above generate the hybrid quantum resources available in the tripartite Mollow system. Whether these excitations are emitted as correlated photon-phonon pairs or converted into photon bundles is determined by dissipation. To capture this dissipative selection mechanism, we describe the system dynamics by the Lindblad master equation
\begin{align}
\frac{d\rho}{dt}
=
-i[\hat{\cal H}_2,\rho]
+\gamma{\cal L}[\hat{\sigma}]\rho
+\kappa_a{\cal L}[\hat{a}]\rho
+\kappa_b{\cal L}[\hat{b}]\rho ,
\label{eq:master}
\end{align}
where $\gamma$ is the atomic decay rate and $\kappa_a$ ($\kappa_b$) is the decay rate of the photon (phonon) mode. The Lindblad superoperator is defined as ${\cal L}[\hat{o}]\rho=\frac{1}{2}\left(2\hat{o}\rho\hat{o}^{\dagger}-\hat{o}^{\dagger}\hat{o}\rho-\rho\hat{o}^{\dagger}\hat{o}\right)$. 

In numerical simulation, we set $g/\kappa_a=10$ and $\kappa_b/\gamma=1$. Remarkably, the dissipation ratio $\kappa_a/\kappa_b$ serves as the key parameter controlling the emission properties of the system. When $\kappa_a=\kappa_b$, the two bosonic emission channels are balanced and correlated photon-phonon pairs are favored. In contrast, strong dissipation asymmetry $\kappa_a\gg\kappa_b$ suppresses photon-phonon correlations and converts the system into a photon-bundle emitter.

To identify correlated photon-phonon bundle emission, we introduce the time-delayed second-order correlation function for an $N$-pair bundle~\cite{Quantum_Entangled_Pairs_Bin-qian,phonon_and_photon_phonon_pairs},
\begin{align}
g_{2N,ab}^{(2)}(\tau)
=
\frac{
\left\langle
(\hat{a}\hat{b})^{\dagger N}(0)
(\hat{a}\hat{b})^{\dagger N}(\tau)
(\hat{a}\hat{b})^{N}(\tau)
(\hat{a}\hat{b})^{N}(0)
\right\rangle
}{
\left\langle
(\hat{a}\hat{b})^{\dagger N}
(\hat{a}\hat{b})^{N}
\right\rangle^{2}
},
\label{eq:gNab}
\end{align}
where $\langle\cdots\rangle$ denotes the steady-state expectation value and $\tau$ is the time interval. For an $N$-pair emission process, correlations cannot be evaluated at arbitrarily short times. We therefore use the minimal emission time $\tau_{\min}^{(N)}=\sum_{m=1}^{N}{1}/{m\gamma}$ as the earliest time at which the $N$-bundle correlation is meaningfully defined~\cite{Emitters_of_N-photon_bundles,Tau_min_of_Bundle_Emission}. In practice, $g_{2N,ab}^{(2)}(\tau_{\min}^{(N)})$ can be closely approximated as an effective zero-delay bundle correlation. The reason is that the generalized second-order correlation function  is ill-defined in a short temporal window of width $\tau_{\min}^{(N)}$.

\begin{figure}[t]
\includegraphics[width=0.50\textwidth]{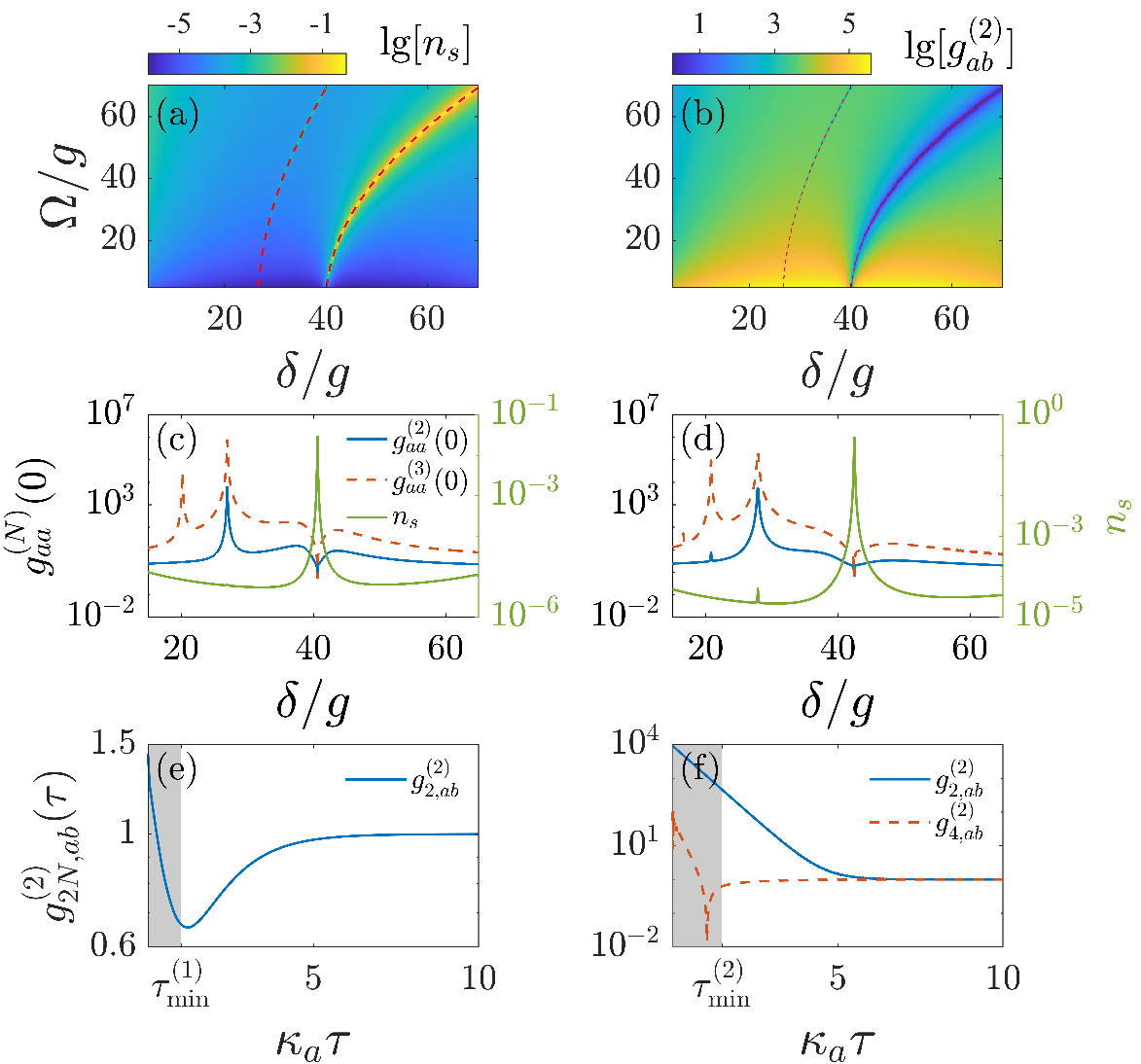}
\caption{
Photon-phonon pair emission under symmetric dissipation, $\kappa_a=\kappa_b$, with $g/\kappa_a=10$ and $\Delta/g=160$. (a) Steady-state population $n_s=n_a=n_b$ and (b) photon-phonon cross-correlation $g_{ab}^{(2)}(0)$ as functions of $\Omega/g$ and $\delta/g$. The dashed curves denote the analytically predicted one-pair and two-pair resonances. 
(c)-(d) Photon autocorrelations $g_{aa}^{(2)}(0)$ and $g_{aa}^{(3)}(0)$ together with the population $n_s$ versus $\delta/g$ for $\Omega/g=10$ and $20$, respectively, revealing the one-pair and two-pair resonances. (e) Time-delayed correlation $g_{2,ab}^{(2)}(\tau)$ at the one-pair resonance, showing antibunching between successive pairs. (f) Time-delayed correlations $g_{2,ab}^{(2)}(\tau)$ and $g_{4,ab}^{(2)}(\tau)$ at the two-pair resonance, confirming correlated two-pair bundle emission.
}
\label{bundle}  
\end{figure}

The statistical signatures of photon-phonon bundle emission are as follows. For a single correlated pair ($N=1$), successive pairs should be antibunched,
\begin{align}
g_{2,ab}^{(2)}(\tau_{\min}^{(1)})
<
g_{2,ab}^{(2)}(\tau),
\qquad
\tau>\tau_{\min}^{(1)},
\end{align}
For an $N$-pair bundle with $N\geq2$, two conditions must be satisfied. First, the single-pair correlation should exhibit bunching,
\begin{align}
g_{2,ab}^{(2)}(\tau_{\min}^{(1)})
>
g_{2,ab}^{(2)}(\tau),
\qquad
\tau>\tau_{\min}^{(1)} ,
\end{align}
which indicates that individual photon-phonon pairs are emitted within the same bundle rather than independently. Second, the $N$-pair bundle correlation should exhibit antibunching,
\begin{align}
g_{2N,ab}^{(2)}(\tau_{\min}^{(N)})
<
g_{2N,ab}^{(2)}(\tau),
\qquad
\tau>\tau_{\min}^{(N)} ,
\end{align}
ensuring temporal separation between successive $N$-pair bundles. Together, these two criteria distinguish genuine $N$-pair bundle emission from uncorrelated multiparticle emission.

To further quantify quantum correlations, we also use equal-time second-order correlation functions
\begin{align}
g^{(2)}_{oo'}(0)
=
\frac{
\left\langle
\hat{o}^{\dagger}\hat{o}'^{\dagger}
\hat{o}'\hat{o}
\right\rangle
}{
\left\langle
\hat{o}^{\dagger}\hat{o}
\right\rangle_{\rm ss}
\left\langle
\hat{o}'^{\dagger}\hat{o}'
\right\rangle
},
\qquad
o,o'\in\{a,b\}.
\label{eq:g2oo}
\end{align}

The quantities $g_{aa}^{(2)}(0)$ and $g_{bb}^{(2)}(0)$ characterize photon and phonon autocorrelations, respectively, while $g_{ab}^{(2)}(0)=g_{ba}^{(2)}(0)$ quantifies photon-phonon cross correlations~\cite{Quantum_Entangled_Pairs_Bin-qian,pengzhong}. 
These observables allow us to identify the transition from correlated photon-phonon pair emission to photon-bundle emission as the dissipation symmetry is varied.

\section{Results}\label{Sec5}

\subsection{Emission of photon-phonon pairs under symmetric dissipation}
We first consider the symmetric dissipation regime, $\kappa_a/\kappa_b=1$. In this case, the photon and phonon loss channels are balanced, so that the excitations generated by the tripartite interaction experience identical dissipation. As a result, the system preferentially emits correlated photon-phonon pairs rather than mode-selective excitations.

To identify the parameter regions supporting pair emission, we map the steady-state response in the $(\Omega,\delta)$ plane for $\Delta/g=160$. Figure~\ref{bundle}(a) shows the mean excitation number
$n_s=n_a=n_b$, where
$n_a=\langle\hat a^\dagger\hat a\rangle$ and
$n_b=\langle\hat b^\dagger\hat b\rangle$. The equality $n_a=n_b$ reflects the pairwise nature of the excitation process. Distinct resonance branches emerge as the driving strength is varied, indicating efficient generation of photon-phonon pairs. The corresponding photon-phonon cross correlation $g_{ab}^{(2)}(0)$ is shown in Fig.~\ref{bundle}(b). The strongest correlation peaks occur along the same resonance branches where the populations are enhanced, demonstrating that these resonances originate from correlated photon-phonon generation.

To determine the order of the emitted bundles, we examine representative detuning cuts at fixed driving strengths. For $\Omega/g=10$, Fig.~\ref{bundle}(c) exhibits a pronounced resonance at $\delta/g\simeq40.63$, corresponding to the one-pair resonance predicted by the dressed-state analysis. The associated time-delayed correlation function shown in Fig.~\ref{bundle}(e) satisfies
$
g_{2,ab}^{(2)}(\tau_{\min}^{(1)})
<
g_{2,ab}^{(2)}(\tau)
$,
demonstrating antibunching between successive emitted pairs and thereby confirming single-pair emission.

Increasing the driving strength to $\Omega/g=20$ activates a higher-order resonance near $\delta/g\simeq27.90$, as shown in Fig.~\ref{bundle}(d). In this regime, the temporal correlations shown in Fig.~\ref{bundle}(f) satisfy
$
g_{2,ab}^{(2)}(\tau_{\min}^{(1)})
>
g_{2,ab}^{(2)}(\tau)
$
and
$
g_{4,ab}^{(2)}(\tau_{\min}^{(2)})
<
g_{4,ab}^{(2)}(\tau)
$.
The first condition indicates that individual photon-phonon pairs are emitted within the same bundle, whereas the second demonstrates antibunching between successive two-pair bundles. Together, these signatures confirm two-pair photon-phonon bundle emission associated with the two-pair resonance.

The above results reveal a direct correspondence between the coherent multiphoton-multiphonon resonances derived in Sec.~III and the emitted quantum states of the open system. Under symmetric dissipation, the hybrid character of the generated excitations is preserved, allowing the coherent $N$-pair resonances to be directly converted into emitted photon-phonon pair bundles. 
By tuning the driving strength and detuning, the system can therefore be switched between different photon-phonon pair-emission channels while maintaining strong photon-phonon correlations. As we show below, this behavior changes qualitatively once the dissipation symmetry is broken.

\begin{figure}[t] 
\centering
\includegraphics[width=0.48\textwidth]{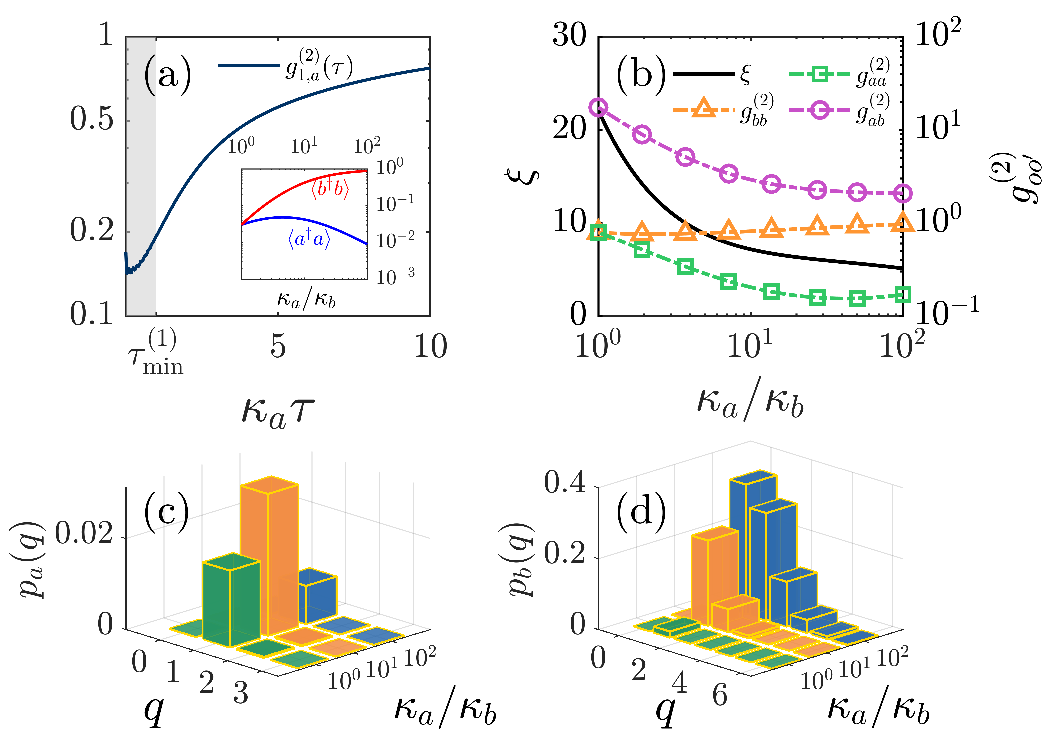}
\caption{Dissipation-induced transition from photon-phonon pair emission to single-photon emission at the first-order resonance for $\delta/g=26$, $\Omega/g=10$, and $\Delta/g=160$. (a) Time-delayed photon correlation function $g_{1,a}^{(2)}(\tau)$ for $\kappa_a/\kappa_b=100$, showing antibunched single-photon emission. Inset: steady-state photon and phonon populations as functions of the dissipation ratio $\kappa_a/\kappa_b$. (b) Cauchy-Schwarz parameter $\xi$, photon autocorrelation $g_{aa}^{(2)}(0)$, phonon autocorrelation $g_{bb}^{(2)}(0)$, and photon-phonon cross-correlation $g_{ab}^{(2)}(0)$ versus $\kappa_a/\kappa_b$. Increasing dissipation asymmetry suppresses the photon-phonon correlations and drives the system toward dynamical decoupling. (c) Steady-state photon Fock-state distributions and (d) phonon Fock-state distributions for $\kappa_a/\kappa_b=1$, $10$, and $100$. Strong dissipation asymmetry concentrates the photon population in the single-photon sector while phonons accumulate in higher Fock states. }
\label{fig:steady fock1}  
\end{figure}

\subsection{Photon emission induced by dissipation-symmetry breaking}

Having established photon-phonon pair emission under symmetric dissipation, we now turn to the asymmetric regime $\kappa_a\gg\kappa_b$. In this limit, cavity photons decay much faster than phonons, so that the two bosonic modes play markedly different roles in the open-system dynamics. The photon mode serves as the dominant output channel, whereas phonons remain long lived and accumulate within the system. As a result, the correlated photon-phonon excitations generated by the tripartite interaction are no longer emitted symmetrically. Instead, the dissipation asymmetry favors photonic emission and fundamentally alters the nature of the emitted quantum states, because photons leave the cavity much faster than phonons, repeated photon emission can occur before the phonon excitation is depleted, naturally favoring photon-only emission.

To characterize the photon-emission statistics in this regime, we introduce the time-delayed bundle correlation function of the cavity mode,

\begin{align}
g_{N,a}^{(2)}(\tau)
=
\frac{
\left\langle
\hat{a}^{\dagger N}(0)
\hat{a}^{\dagger N}(\tau)
\hat{a}^{N}(\tau)
\hat{a}^{N}(0)
\right\rangle
}{
\left\langle
\hat{a}^{\dagger N}
\hat{a}^{N}
\right\rangle^{2}
},
\label{eq:gNa}
\end{align}
where this quantity is the photon-only counterpart of the photon-phonon bundle correlation defined above and is used to distinguish single-photon emission from photon-bundle emission.

\begin{figure}[t] 
\centering
\includegraphics[width=0.48\textwidth]{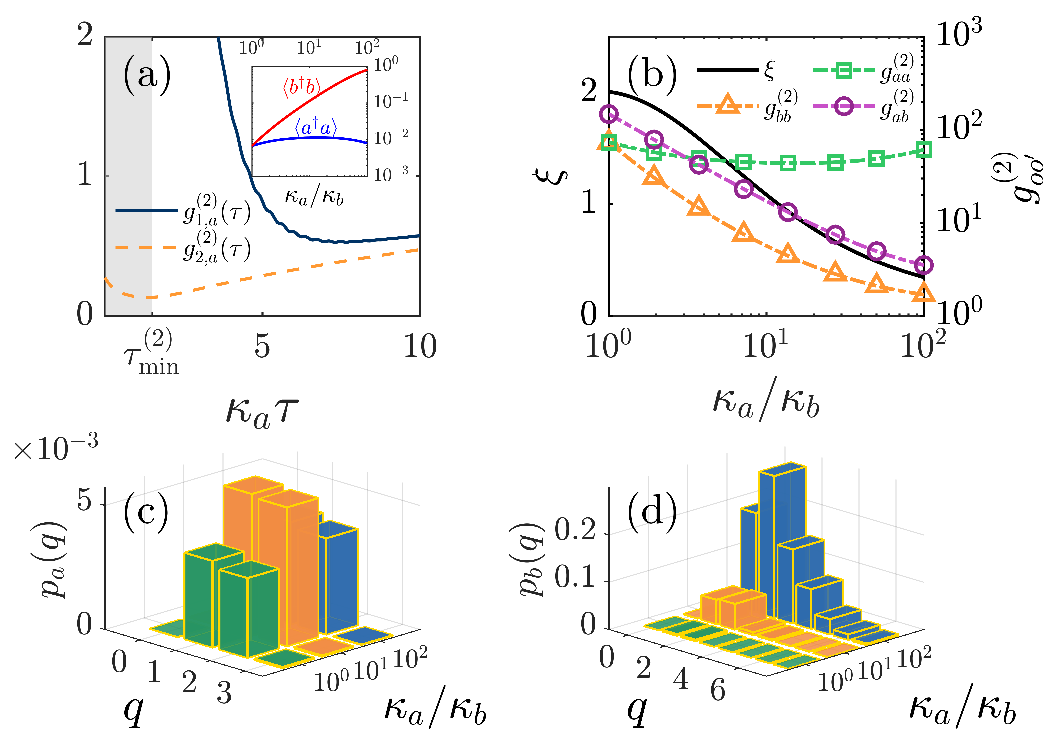}
\caption{
Dissipation-induced transition from photon-phonon pair emission to two-photon bundle emission at the two-pair resonance, with $\delta/g\simeq19.64$, $\Omega/g=30$, and $\Delta/g=90$. (a) Time-delayed photon correlations $g_{1,a}^{(2)}(\tau)$ and $g_{2,a}^{(2)}(\tau)$ for $\kappa_a/\kappa_b=100$, showing bunching of individual photons within a bundle and antibunching between successive two-photon bundles. Inset: steady-state photon and phonon populations as functions of the dissipation ratio $\kappa_a/\kappa_b$. (b) Cauchy-Schwarz parameter $\xi$, photon autocorrelation $g_{aa}^{(2)}(0)$, phonon autocorrelation $g_{bb}^{(2)}(0)$, and photon-phonon cross-correlation $g_{ab}^{(2)}(0)$ versus $\kappa_a/\kappa_b$. Increasing dissipation asymmetry suppresses photon-phonon correlations and drives the system toward dynamical decoupling.
(c) Steady-state photon Fock-state distributions and (d) phonon Fock-state distributions for $\kappa_a/\kappa_b=1$, $10$, and $100$. Strong dissipation asymmetry promotes two-photon bundle emission while driving the phonon population toward higher Fock states.}\label{fig:steady fock2}  
\end{figure}

We first focus on the vicinity of the one-pair resonance, $\delta/g\simeq40.63$, with $\Omega/g=10$ and $\Delta/g=160$. The dissipation ratio is varied as $\kappa_a/\kappa_b=1,10,100$ to track the crossover from photon-phonon pair emission to photonic emission. As shown in the inset of Fig.~\ref{fig:steady fock1}(a), increasing $\kappa_a/\kappa_b$ produces a pronounced population imbalance between the two bosonic modes. The photon population remains low because photons rapidly leave the cavity, whereas the phonon population increases owing to the much longer phonon lifetime. This imbalance is the direct steady-state signature of dissipation-induced filtering.

We next examine the photon statistics in the strongly asymmetric case $\kappa_a/\kappa_b=100$. Figure~\ref{fig:steady fock1}(a) shows that the photon correlation satisfies
$
g_{1,a}^{(2)}(\tau_{\min}^{(1)})
<
g_{1,a}^{(2)}(\tau)
$,
indicating antibunching between successive emitted photons.  Thus, the one-pair resonance is converted into high-purity single-photon emission once the dissipation symmetry is strongly broken. Although the coherent tripartite interaction continues to generate correlated photon-phonon excitations, the asymmetric loss channels strongly favor photonic output.

To quantify the loss of hybrid photon-phonon character, we evaluate the Cauchy-Schwarz parameter
\begin{equation}
\xi
=
\frac{
g_{ab}^{(2)}(0)
}{
\sqrt{
g_{aa}^{(2)}(0)
g_{bb}^{(2)}(0)
}
},
\end{equation}
For classical fields, $\xi\leq1$, whereas $\xi>1$ signals nonclassical photon-phonon correlations~\cite{pengzhong,Cauchy_Schwarz_inequality,Cauchy_Schwarz_inequality_1,Cauchy_Schwarz_inequality_2}. As shown in Fig.~\ref{fig:steady fock1}(b), $\xi$ decreases toward unity as $\kappa_a/\kappa_b$ increases. This behavior demonstrates that the intermode nonclassical correlations characteristic of the symmetric pair-emission regime are gradually suppressed by dissipation-symmetry breaking. The emitted quantum state therefore crosses over from a hybrid photon-phonon resource to a photon-dominated resource.

The steady-state Fock-state distributions provide further insight into this crossover. Figures~\ref{fig:steady fock1}(c) and \ref{fig:steady fock1}(d) show the photon and phonon number distributions,
$p_o(q)=\langle q_o|\rho_o|q_o\rangle$ with $o=a,b$. As $\kappa_a/\kappa_b$ increases from 1 to 100, the photon distribution becomes concentrated in the single-photon sector, whereas the phonon distribution shifts toward higher Fock states. This behavior reflects the strong separation of timescales between the two bosonic modes: photons are rapidly emitted, while phonons remain trapped in the system. Consequently, dissipation-symmetry breaking transforms the photon-phonon pair source into a single-photon emitter accompanied by phonon accumulation.

\begin{figure*}[t] 
\centering
\includegraphics[width=0.98\textwidth]{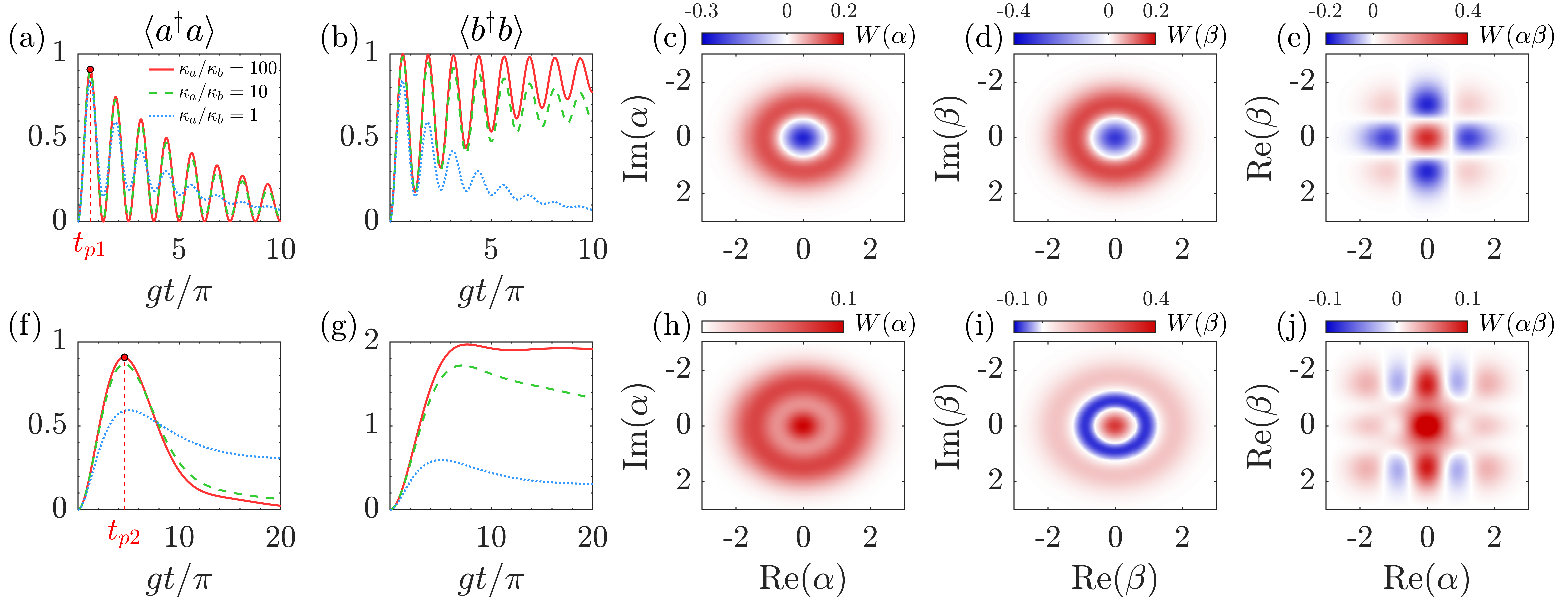}
\caption{
Time-resolved dynamics and phase-space signatures of dissipation-selected emission for $\Omega/g=10$ and $\Delta/g=10$. (a)-(b) Evolution of the photon and phonon populations at the one-pair resonance ($\delta/g=12.5$) for $\kappa_a/\kappa_b=1$, $10$, and $100$. The red dot in (a) marks the first photon-population maximum $t_{p1}$. (c)-(e) Photon Wigner function $W(\alpha)$, phonon Wigner function $W(\beta)$, and two-mode Wigner function $W(\alpha,\beta)$ evaluated at $t_{p1}$ for $\kappa_a/\kappa_b=100$. The negative regions indicate nonclassical single-mode states and photon-phonon correlations. (f)-(g) Evolution of the photon and phonon populations at the two-pair resonance ($\delta=5g$). The red dot in (f) marks the first photon-population maximum $t_{p2}$. (h)-(j) Corresponding Wigner functions $W(\alpha)$, $W(\beta)$, and $W(\alpha,\beta)$ evaluated at $t_{p2}$ for $\kappa_a/\kappa_b=100$. Strong cavity dissipation reduces the negativity of the photon Wigner function, whereas the phonon and two-mode Wigner functions remain nonclassical. The dynamics directly reveal the dissipation-induced separation between a fast photonic output channel and a long-lived phononic storage channel.
} \label{fig:dynamic evolution process} 
\end{figure*}

We next examine the effect of dissipation-symmetry breaking at the two-pair resonance, with $\Omega/g=30$, $\Delta/g=90$, and $\delta/g\simeq19.64$. This setting allows us to investigate whether a coherent two-pair photon-phonon process can be converted into photon-bundle emission by increasing the dissipation asymmetry. The inset of Fig.~\ref{fig:steady fock2}(a) shows the steady-state photon and phonon populations as functions of $\kappa_a/\kappa_b$. As in the one-pair resonance, increasing the dissipation ratio produces a pronounced population imbalance between the two bosonic modes, indicating that the dissipation asymmetry continues to strongly influence the emission dynamics at higher-order resonances.

We then focus on the strongly asymmetric  regime  $\kappa_a/\kappa_b=100$. The time-delayed photon correlations in Fig.~\ref{fig:steady fock2}(a) satisfy
$
g_{1,a}^{(2)}(\tau_{\min}^{(1)})
>
g_{1,a}^{(2)}(\tau)
$
and
$
g_{2,a}^{(2)}(\tau_{\min}^{(2)})
<
g_{2,a}^{(2)}(\tau)
$.
The first inequality indicates that photons are emitted in correlated groups, whereas the second demonstrates antibunching between successive two-photon bundles. Together, these signatures establish two-photon bundle emission. Thus, the two-pair resonance, which under symmetric dissipation produces correlated photon-phonon pairs, is converted into photonic bundle emission when the dissipation symmetry is strongly broken.

The loss of hybrid photon-phonon character is quantified by the Cauchy--Schwarz parameter $\xi$. As shown in Fig.~\ref{fig:steady fock2}(b), $\xi$ decreases rapidly as $\kappa_a/\kappa_b$ increases and approaches the classical bound near $\kappa_a/\kappa_b\simeq10$. Compared with the first-order resonance, this faster reduction of intermode nonclassicality reflects the weaker effective coupling and smaller steady-state population associated with the higher-order process. Hence, higher-order photon-phonon resources are more sensitive to dissipation imbalance.

The steady-state Fock-state distributions in Figs.~\ref{fig:steady fock2}(c) and \ref{fig:steady fock2}(d) provide a complementary picture of this transition.  As $\kappa_a/\kappa_b$ increases, the photon distribution develops dominant weight in the one- and two-photon sectors, consistent with photon-bundle emission, whereas the phonon distribution spreads toward higher Fock states. These distributions confirm the separation of roles induced by dissipation asymmetry: the cavity mode acts as a fast radiative output channel, while the phonon mode stores residual motional excitations. Together, the correlation functions, Cauchy-Schwarz parameter, and Fock-state distributions demonstrate that dissipation-symmetry breaking converts the two-pair photon-phonon resonance into two-photon bundle emission.

\subsection{Time-resolved dynamics and phase-space signatures}

The steady-state analysis identifies the quantum states selected by the dissipation channels, but it does not directly reveal how this selection emerges during the emission process. In particular, the steady state alone cannot show how a coherently generated hybrid photon-phonon excitation evolves into either a correlated photon-phonon pair or a photon bundle. To uncover this microscopic mechanism, we study the real-time evolution from the initial state $|+,0,0\rangle$. This time-resolved analysis  reveals the interplay between coherent tripartite excitation and dissipation.

We first consider the single-pair resonance with $\delta/g=12.5$, $\Omega/g=10$, and $\Delta/g=10$. Figures~\ref{fig:dynamic evolution process}(a) and \ref{fig:dynamic evolution process}(b) show the evolution of the photon and phonon populations for $\kappa_a/\kappa_b=1,10,100$. As the dissipation asymmetry increases, the phonon population builds up to progressively larger values, whereas the photon population remains transient and exhibits damped oscillations. This behavior is the dynamical signature of dissipative filtering: photons generated through the tripartite interaction are continuously emitted into the output channel, while long-lived phonons remain in the system and accumulate over time.

To characterize the quantum state during the emission event, we focus on the time $t_{p1}$ at which the photon population first reaches its maximum, marked by the red dot in Fig.~\ref{fig:dynamic evolution process}(a). For $\kappa_a/\kappa_b=100$, we evaluate the single-mode photon Wigner function $W(\alpha)$, the single-mode phonon Wigner function $W(\beta)$, and the two-mode Wigner function $W(\alpha,\beta)$ of the photon-phonon subsystem. Following Ref.~\cite{entangled_Wigner}, the two-mode Wigner function is defined as
\begin{align}
W(\alpha,\beta)
=
\frac{4}{\pi^2}
{\rm Tr}
\left[
D^{\dagger}(\beta)
D^{\dagger}(\alpha)
\rho^{ab}
D(\alpha)
D(\beta)
\Pi^{ab}
\right],
\end{align}
where $D(o)=\exp(o\hat{o}^{\dagger}-o^*\hat{o})$ is the displacement operator, $\rho^{ab}$ is the reduced density matrix of the photon-phonon subsystem, and
$\Pi^{ab}=\exp(i\pi\hat{a}^{\dagger}\hat{a})\exp(i\pi\hat{b}^{\dagger}\hat{b})$ is the joint parity operator. As shown in Figs.~\ref{fig:dynamic evolution process}(c)--\ref{fig:dynamic evolution process}(e), the photon and phonon Wigner functions both exhibit negative regions, confirming that each mode retains nonclassical character at the emission time. The negativity of the two-mode Wigner function further shows that nonclassical photon-phonon correlations survive despite the strong dissipation asymmetry.

The Fock-state decomposition at $t_{p1}$ clarifies the dominant dynamical pathway. The two-mode state is mainly composed of the correlated state $|1_a1_b\rangle$, whose population exceeds $90\%$, while the state $|0_a1_b\rangle$ has a smaller but visible weight. The latter component represents an intermediate state after the photon has leaked out while the phonon remains stored. All other Fock components contribute less than $1\%$. Thus, the one-pair resonance is dominated by the coherent generation of a photon-phonon pair followed by photon loss from the cavity.

We next analyze the two-pair resonance, $\delta=5g$, for the same dissipation ratios. Starting again from $|+,0,0\rangle$, Figs.~\ref{fig:dynamic evolution process}(f) and \ref{fig:dynamic evolution process}(g) show the evolution of the photon and phonon populations. The dynamics are slower than in the single-pair case because the effective coupling $g_{\rm eff}^{(N=2)}$ is weaker than $g_{\rm eff}^{(N=1)}$. As $\kappa_a/\kappa_b$ increases, the photon population becomes increasingly transient after the initial emission event, while the phonon population approaches a larger long-time value. This again reflects the separation of roles induced by dissipation asymmetry: the photon mode serves as a fast output channel, whereas the phonon mode stores the remaining motional excitation.

At the first photon-number peak $t_{p2}$ for $\kappa_a/\kappa_b=100$, marked by the red dot in Fig.~\ref{fig:dynamic evolution process}(f), the phonon Wigner function $W(\beta)$ exhibits clear negativity, whereas the photon Wigner function $W(\alpha)$ does not show a pronounced negative region [Figs.~\ref{fig:dynamic evolution process}(h) and \ref{fig:dynamic evolution process}(i)]. This difference arises because rapid cavity leakage strongly suppresses the intracavity photon population and washes out its single-mode phase-space negativity. Nevertheless, the two-mode Wigner function $W(\alpha,\beta)$ remains negative, as shown in Fig.~\ref{fig:dynamic evolution process}(j), indicating that two-mode nonclassical correlations survive even when the single-mode photonic signature is strongly weakened.

The Fock-state decomposition at $t_{p2}$ reveals the dominant pathways of the two-pair process. The correlated states $|1_a1_b\rangle$ and $|2_a2_b\rangle$ have the largest populations, with weights $33.9\%$ and $27.9\%$, respectively. The state $|0_a2_b\rangle$ also has a sizable population of $19.1\%$, which signals photon leakage after the generation of a two-pair excitation. In addition, the ground state $|0_a0_b\rangle$ contributes $18.0\%$, while all other states together contribute less than $2\%$. The dynamics are therefore governed by a restricted set of processes: coherent generation of correlated photon-phonon excitations, photon leakage through the cavity output channel, and phonon storage in the long-lived motional mode. This time-dependent picture confirms the microscopic origin of dissipation-selected emission: the coherent interaction creates hybrid photon-phonon resources, while dissipation asymmetry determines how these resources are partitioned into emitted photons and stored phonons.

\section{Conclusion} \label{Sec6}

In summary, we have demonstrated dissipation-selected quantum emission in a tripartite Mollow system, where a strongly driven two-level emitter is simultaneously coupled to cavity photons and motional phonons. The coherent tripartite interaction generates resonant multiphoton-multiphonon processes, while the symmetry of the bosonic dissipation channels determines how these hybrid excitations are emitted. Under symmetric dissipation, the photon and phonon loss channels are balanced, and the system emits strongly correlated photon-phonon pairs, including both single-pair emission and correlated multipair bundles. When the dissipation symmetry is broken, photon-phonon correlations are progressively suppressed and the two bosonic modes acquire distinct dynamical roles: the cavity mode acts as a fast radiative output channel, whereas the motional mode behaves as a long-lived storage channel. In the strongly asymmetric regime, this dissipative filtering converts the same coherent photon-phonon resonances into high-purity single-photon and two-photon bundle emission, accompanied by phonon population buildup in higher Fock states.

This dissipation-induced crossover is characterized by time-delayed correlation functions, Cauchy-Schwarz correlations, Fock-state distributions, Wigner quasiprobability distributions, and time-resolved dynamics. Together, these results show that the same coherent photon-phonon resonance can be converted into qualitatively different quantum-emission channels solely by changing the dissipation symmetry, without modifying the underlying coherent interaction. More broadly, our results reveal a clear division of roles in open hybrid quantum systems: coherent interactions define the available quantum resources, whereas dissipation selects how these resources are emitted. This establishes dissipation asymmetry as a mechanism for quantum-resource selection and provides a route toward dissipation-engineered quantum emitters, hybrid quantum-state conversion, and controllable nonclassical emission in open light-matter systems~\cite{PhysRevLett.126.020402,mi2024stable,li2025programmable}.

\begin{acknowledgments}
{This work was supported by the National Natural Science Foundation of China (Grant No. 12274473, Grant No.12374365, and Grant No. 12135018), Quantum Science and Technology-National Science and Technology Major Project (Grant No.2025ZD0300400), Guangdong Provincial Quantum Science Strategic Initiative (Grant No. GDZX2505001), and Guangdong University of Technology SPOE Seed Foundation (SF2024111504).}
\end{acknowledgments}

\appendix

\section{Derivation of the effective $N$-pair coupling}
\label{AppA}

In the Mollow regime, the generalized Rabi splitting is much larger than the elementary tripartite coupling strength, $\tilde{\Omega}\gg g$. The tripartite interaction can therefore be treated perturbatively in the dressed-state basis. We start from the driven atomic Hamiltonian
\begin{equation}
\hat{\cal H}_{\rm at}
=
\Delta_{\sigma}\hat{\sigma}^{\dagger}\hat{\sigma}
+
\Omega(\hat{\sigma}+\hat{\sigma}^{\dagger}),
\end{equation}
Its eigenstates are
\begin{align}
|+\rangle &= c_+|e\rangle+c_-|g\rangle, \nonumber \\
|-\rangle &= c_-|e\rangle-c_+|g\rangle,
\end{align}
with eigenenergies $E_{\pm}=\frac{\Delta_{\sigma}}{2}\pm\frac{\tilde{\Omega}}{2}$ and the effective Rabi frequency $\tilde{\Omega}=\sqrt{\Delta_{\sigma}^{2}+4\Omega^{2}}$. The coefficients are $c_{\pm}
=
\sqrt{
{2\Omega^{2}}/(
{\tilde{\Omega}^{2}\mp\Delta_{\sigma}\tilde{\Omega}})
}$ and $c_+^2+c_-^2=1$.

Introducing $\hat{\tilde{\sigma}}
=
|-\rangle\langle+|$ and $\hat{\tilde{\sigma}}_z
=
|+\rangle\langle+|
-
|-\rangle\langle-|$, the bare atomic lowering operator becomes
\begin{equation}
\hat{\sigma}
=
c_-^2\hat{\tilde{\sigma}}^{\dagger}
-
c_+^2\hat{\tilde{\sigma}}
+
c_+c_-\hat{\tilde{\sigma}}_z,
\end{equation}
Substitution into the tripartite Hamiltonian gives
\begin{align}
\hat{\cal H}_3
=&~
\delta_a\hat a^\dagger\hat a
+
\delta_b\hat b^\dagger\hat b
+
\frac{\tilde{\Omega}}{2}\hat{\tilde{\sigma}}_z
\nonumber\\
&+
g\hat a\hat b
\left(
c_-^2\hat{\tilde{\sigma}}
-
c_+^2\hat{\tilde{\sigma}}^{\dagger}
+
c_+c_-\hat{\tilde{\sigma}}_z
\right)
\nonumber\\
&+
g\hat a^\dagger\hat b^\dagger
\left(
c_-^2\hat{\tilde{\sigma}}^{\dagger}
-
c_+^2\hat{\tilde{\sigma}}
+
c_+c_-\hat{\tilde{\sigma}}_z
\right).
\label{App_H_dressed}
\end{align}
The interaction terms in Eq.~\eqref{App_H_dressed} couple neighboring manifolds of the Mollow ladder and provide virtual paths connecting distant resonances.

We now consider the $N$th-order resonance
\begin{equation}
\tilde{\Omega}
=
N(\delta_a+\delta_b)=2
N\delta,
\qquad
\delta=\delta_a=\delta_b,
\end{equation}
At this resonance, the states $|+,0,0\rangle$ and $|-,N,N\rangle$ become quasi-degenerate and define the model space. All remaining states constitute the virtual subspace. The Hamiltonian can be partitioned as
\begin{equation}
\hat{\cal H}
=
\begin{pmatrix}
\hat h^{(N)} & \hat V^{(N)}
\\
\hat V^{(N)\dagger} & \hat H^{(N)}
\end{pmatrix},
\end{equation}
where $\hat h^{(N)}$ acts on the model space
$\{|+,0,0\rangle,|-,N,N\rangle\}$, $\hat H^{(N)}$ acts on the virtual subspace, and $\hat V^{(N)}$ couples the two subspaces.

Using quasi-degenerate perturbation theory, the effective Hamiltonian in the model space can be constructed order by order in $g/\tilde{\Omega}$. The off-diagonal matrix element connecting $|+,0,0\rangle$ and $|-,N,N\rangle$ first appears at order $N$. The dominant virtual path can be written schematically as
\begin{equation}
|+,0,0\rangle
\rightarrow
|-,1,1\rangle
\rightarrow
|+,2,2\rangle
\rightarrow
\cdots
\rightarrow
|-,N,N\rangle ,
\end{equation}
where the intermediate states are off resonant by energy scales set by $\tilde{\Omega}$. Evaluating these virtual transition paths yields
\begin{equation}
\langle -,N,N|
\hat{\cal H}_{\rm eff}^{(N)}
|+,0,0\rangle
=
N! \, g_{\rm eff}^{(N)} ,
\end{equation}
with
\begin{equation}
g_{\rm eff}^{(N)}
=
(-g)^N
\frac{
(N^2)^{N-1}
c_-^{N-1}
c_+^{N+1}
}{
[(N-1)!]^2
\tilde{\Omega}^{N-1}
}.
\label{App_geff}
\end{equation}
The factor $N!$ arises from the bosonic matrix element
$\langle N,N|\hat a^{\dagger N}\hat b^{\dagger N}|0,0\rangle=N!$. Equation~\eqref{App_geff} shows the characteristic scaling
$g_{\rm eff}^{(N)}\propto g^N/\tilde{\Omega}^{N-1}$, reflecting the virtual nature of the $(N-1)$ intermediate transitions.

The resulting effective Hamiltonian for the resonant $N$-pair process is
\begin{align}
\hat{\cal H}_{\rm eff}^{(N)}
=&~
\delta_a\hat a^\dagger\hat a
+
\delta_b\hat b^\dagger\hat b
+
\frac{\tilde{\Omega}}{2}\hat{\tilde{\sigma}}_z
\nonumber\\
&+
g_{\rm eff}^{(N)}
\left(
\hat a^N\hat b^N\hat{\tilde{\sigma}}^{\dagger}
+
\hat a^{\dagger N}\hat b^{\dagger N}\hat{\tilde{\sigma}}
\right).
\label{App_HeffN}
\end{align}
Equation~\eqref{App_HeffN} describes the coherent conversion between a dressed atomic excitation and an $N$-pair photon-phonon bundle. This effective interaction is responsible for the higher-order resonances discussed in the main text and forms the microscopic origin of the multipair emission processes.


\begin{thebibliography}{86}%
\makeatletter
\providecommand \@ifxundefined [1]{%
 \@ifx{#1\undefined}
}%
\providecommand \@ifnum [1]{%
 \ifnum #1\expandafter \@firstoftwo
 \else \expandafter \@secondoftwo
 \fi
}%
\providecommand \@ifx [1]{%
 \ifx #1\expandafter \@firstoftwo
 \else \expandafter \@secondoftwo
 \fi
}%
\providecommand \natexlab [1]{#1}%
\providecommand \enquote  [1]{``#1''}%
\providecommand \bibnamefont  [1]{#1}%
\providecommand \bibfnamefont [1]{#1}%
\providecommand \citenamefont [1]{#1}%
\providecommand \href@noop [0]{\@secondoftwo}%
\providecommand \href [0]{\begingroup \@sanitize@url \@href}%
\providecommand \@href[1]{\@@startlink{#1}\@@href}%
\providecommand \@@href[1]{\endgroup#1\@@endlink}%
\providecommand \@sanitize@url [0]{\catcode `\\12\catcode `\$12\catcode
  `\&12\catcode `\#12\catcode `\^12\catcode `\_12\catcode `\%12\relax}%
\providecommand \@@startlink[1]{}%
\providecommand \@@endlink[0]{}%
\providecommand \url  [0]{\begingroup\@sanitize@url \@url }%
\providecommand \@url [1]{\endgroup\@href {#1}{\urlprefix }}%
\providecommand \urlprefix  [0]{URL }%
\providecommand \Eprint [0]{\href }%
\providecommand \doibase [0]{http://dx.doi.org/}%
\providecommand \selectlanguage [0]{\@gobble}%
\providecommand \bibinfo  [0]{\@secondoftwo}%
\providecommand \bibfield  [0]{\@secondoftwo}%
\providecommand \translation [1]{[#1]}%
\providecommand \BibitemOpen [0]{}%
\providecommand \bibitemStop [0]{}%
\providecommand \bibitemNoStop [0]{.\EOS\space}%
\providecommand \EOS [0]{\spacefactor3000\relax}%
\providecommand \BibitemShut  [1]{\csname bibitem#1\endcsname}%
\let\auto@bib@innerbib\@empty
\bibitem [{\citenamefont {Weber}\ \emph {et~al.}(2022)\citenamefont {Weber},
  \citenamefont {Lula-Rocha}, \citenamefont {Pereira}, \citenamefont
  {Trindade}, \citenamefont {Filho}, \citenamefont {Martins}, \citenamefont
  {Santana},\ and\ \citenamefont
  {Vianna}}]{nonclassical_superposition_thermal}%
  \BibitemOpen
  \bibfield  {author} {\bibinfo {author} {\bibfnamefont {P.E.R.}\ \bibnamefont
  {Weber}}, \bibinfo {author} {\bibfnamefont {V.N.A.}\ \bibnamefont
  {Lula-Rocha}}, \bibinfo {author} {\bibfnamefont {J.C.C.}\ \bibnamefont
  {Pereira}}, \bibinfo {author} {\bibfnamefont {M.A.S.}\ \bibnamefont
  {Trindade}}, \bibinfo {author} {\bibfnamefont {L.M.~Silva}\ \bibnamefont
  {Filho}}, \bibinfo {author} {\bibfnamefont {M.G.R.}\ \bibnamefont {Martins}},
  \bibinfo {author} {\bibfnamefont {A.E.}\ \bibnamefont {Santana}}, \ and\
  \bibinfo {author} {\bibfnamefont {J.D.M.}\ \bibnamefont {Vianna}},\
  }\bibfield  {title} {\enquote {\bibinfo {title} {Non-classical properties of
  superposition thermal quantum states},}\ }\href {\doibase
  https://doi.org/10.1016/j.aop.2022.168986} {\bibfield  {journal} {\bibinfo
  {journal} {Annals of Physics}\ }\textbf {\bibinfo {volume} {443}},\ \bibinfo
  {pages} {168986} (\bibinfo {year} {2022})}\BibitemShut {NoStop}%
\bibitem [{\citenamefont {Aubrun}\ \emph {et~al.}(2022)\citenamefont {Aubrun},
  \citenamefont {Lami}, \citenamefont {Palazuelos},\ and\ \citenamefont
  {Pl\'avala}}]{nonclassical_superposition_in_any_physical_theory}%
  \BibitemOpen
  \bibfield  {author} {\bibinfo {author} {\bibfnamefont {Guillaume}\
  \bibnamefont {Aubrun}}, \bibinfo {author} {\bibfnamefont {Ludovico}\
  \bibnamefont {Lami}}, \bibinfo {author} {\bibfnamefont {Carlos}\ \bibnamefont
  {Palazuelos}}, \ and\ \bibinfo {author} {\bibfnamefont {Martin}\ \bibnamefont
  {Pl\'avala}},\ }\bibfield  {title} {\enquote {\bibinfo {title} {Entanglement
  and superposition are equivalent concepts in any physical theory},}\ }\href
  {\doibase 10.1103/PhysRevLett.128.160402} {\bibfield  {journal} {\bibinfo
  {journal} {Phys. Rev. Lett.}\ }\textbf {\bibinfo {volume} {128}},\ \bibinfo
  {pages} {160402} (\bibinfo {year} {2022})}\BibitemShut {NoStop}%
\bibitem [{\citenamefont {Satoor}\ \emph {et~al.}(2021)\citenamefont {Satoor},
  \citenamefont {Fabre}, \citenamefont {Bouhiron}, \citenamefont {Evrard},
  \citenamefont {Lopes},\ and\ \citenamefont
  {Nascimbene}}]{nonclassical_entanglement}%
  \BibitemOpen
  \bibfield  {author} {\bibinfo {author} {\bibfnamefont {Tanish}\ \bibnamefont
  {Satoor}}, \bibinfo {author} {\bibfnamefont {Aur\'elien}\ \bibnamefont
  {Fabre}}, \bibinfo {author} {\bibfnamefont {Jean-Baptiste}\ \bibnamefont
  {Bouhiron}}, \bibinfo {author} {\bibfnamefont {Alexandre}\ \bibnamefont
  {Evrard}}, \bibinfo {author} {\bibfnamefont {Raphael}\ \bibnamefont {Lopes}},
  \ and\ \bibinfo {author} {\bibfnamefont {Sylvain}\ \bibnamefont
  {Nascimbene}},\ }\bibfield  {title} {\enquote {\bibinfo {title} {Partitioning
  dysprosium's electronic spin to reveal entanglement in nonclassical
  states},}\ }\href {\doibase 10.1103/PhysRevResearch.3.043001} {\bibfield
  {journal} {\bibinfo  {journal} {Phys. Rev. Res.}\ }\textbf {\bibinfo {volume}
  {3}},\ \bibinfo {pages} {043001} (\bibinfo {year} {2021})}\BibitemShut
  {NoStop}%
\bibitem [{\citenamefont {Welakuh}\ \emph {et~al.}(2024)\citenamefont
  {Welakuh}, \citenamefont {Tserkis}, \citenamefont {Smart},\ and\
  \citenamefont {Narang}}]{nonclassical_entanglement_of_light}%
  \BibitemOpen
  \bibfield  {author} {\bibinfo {author} {\bibfnamefont {Davis~M.}\
  \bibnamefont {Welakuh}}, \bibinfo {author} {\bibfnamefont {Spyros}\
  \bibnamefont {Tserkis}}, \bibinfo {author} {\bibfnamefont {Scott~E.}\
  \bibnamefont {Smart}}, \ and\ \bibinfo {author} {\bibfnamefont {Prineha}\
  \bibnamefont {Narang}},\ }\bibfield  {title} {\enquote {\bibinfo {title}
  {Cavity-mediated molecular entanglement and generation of non-classical
  states of light},}\ }\href {\doibase 10.1021/acs.jpca.3c05779} {\bibfield
  {journal} {\bibinfo  {journal} {The Journal of Physical Chemistry A}\
  }\textbf {\bibinfo {volume} {128}},\ \bibinfo {pages} {799--806} (\bibinfo
  {year} {2024})}\BibitemShut {NoStop}%
\bibitem [{\citenamefont {Kam}\ \emph {et~al.}(2025)\citenamefont {Kam},
  \citenamefont {Tsesses}, \citenamefont {Ilin}, \citenamefont {Cohen},
  \citenamefont {Lumer}, \citenamefont {Fridman}, \citenamefont {Lotan},
  \citenamefont {Patsyk}, \citenamefont {Nemirovsky-Levy}, \citenamefont
  {Orenstein}, \citenamefont {Segev},\ and\ \citenamefont
  {Bartal}}]{nonclassical_entanglement_of_photon}%
  \BibitemOpen
  \bibfield  {author} {\bibinfo {author} {\bibfnamefont {Amit}\ \bibnamefont
  {Kam}}, \bibinfo {author} {\bibfnamefont {Shai}\ \bibnamefont {Tsesses}},
  \bibinfo {author} {\bibfnamefont {Yigal}\ \bibnamefont {Ilin}}, \bibinfo
  {author} {\bibfnamefont {Kobi}\ \bibnamefont {Cohen}}, \bibinfo {author}
  {\bibfnamefont {Yaakov}\ \bibnamefont {Lumer}}, \bibinfo {author}
  {\bibfnamefont {Lior}\ \bibnamefont {Fridman}}, \bibinfo {author}
  {\bibfnamefont {Stav}\ \bibnamefont {Lotan}}, \bibinfo {author}
  {\bibfnamefont {Anatoly}\ \bibnamefont {Patsyk}}, \bibinfo {author}
  {\bibfnamefont {Liat}\ \bibnamefont {Nemirovsky-Levy}}, \bibinfo {author}
  {\bibfnamefont {Meir}\ \bibnamefont {Orenstein}}, \bibinfo {author}
  {\bibfnamefont {Mordechai}\ \bibnamefont {Segev}}, \ and\ \bibinfo {author}
  {\bibfnamefont {Guy}\ \bibnamefont {Bartal}},\ }\bibfield  {title} {\enquote
  {\bibinfo {title} {Near-field photon entanglement in total angular
  momentum},}\ }\href {\doibase 10.1038/s41586-025-08761-1} {\bibfield
  {journal} {\bibinfo  {journal} {Nature}\ }\textbf {\bibinfo {volume} {640}},\
  \bibinfo {pages} {634--640} (\bibinfo {year} {2025})}\BibitemShut {NoStop}%
\bibitem [{\citenamefont {Liu}\ \emph {et~al.}(2024)\citenamefont {Liu},
  \citenamefont {Ge},\ and\ \citenamefont
  {Zubairy}}]{nonclassical_squeezing_Gaussian_States}%
  \BibitemOpen
  \bibfield  {author} {\bibinfo {author} {\bibfnamefont {Jiru}\ \bibnamefont
  {Liu}}, \bibinfo {author} {\bibfnamefont {Wenchao}\ \bibnamefont {Ge}}, \
  and\ \bibinfo {author} {\bibfnamefont {M.~Suhail}\ \bibnamefont {Zubairy}},\
  }\bibfield  {title} {\enquote {\bibinfo {title} {Classical-nonclassical
  polarity of gaussian states},}\ }\href {\doibase
  10.1103/PhysRevLett.132.240201} {\bibfield  {journal} {\bibinfo  {journal}
  {Phys. Rev. Lett.}\ }\textbf {\bibinfo {volume} {132}},\ \bibinfo {pages}
  {240201} (\bibinfo {year} {2024})}\BibitemShut {NoStop}%
\bibitem [{\citenamefont {Sch{\"a}fermeier}\ \emph {et~al.}(2016)\citenamefont
  {Sch{\"a}fermeier}, \citenamefont {Kerdoncuff}, \citenamefont {Hoff},
  \citenamefont {Fu}, \citenamefont {Huck}, \citenamefont {Bilek},
  \citenamefont {Harris}, \citenamefont {Bowen}, \citenamefont {Gehring},\ and\
  \citenamefont {Andersen}}]{nonclassical_squeezing_cooling_of_oscillator}%
  \BibitemOpen
  \bibfield  {author} {\bibinfo {author} {\bibfnamefont {Clemens}\ \bibnamefont
  {Sch{\"a}fermeier}}, \bibinfo {author} {\bibfnamefont {Hugo}\ \bibnamefont
  {Kerdoncuff}}, \bibinfo {author} {\bibfnamefont {Ulrich~B.}\ \bibnamefont
  {Hoff}}, \bibinfo {author} {\bibfnamefont {Hao}\ \bibnamefont {Fu}}, \bibinfo
  {author} {\bibfnamefont {Alexander}\ \bibnamefont {Huck}}, \bibinfo {author}
  {\bibfnamefont {Jan}\ \bibnamefont {Bilek}}, \bibinfo {author} {\bibfnamefont
  {Glen~I.}\ \bibnamefont {Harris}}, \bibinfo {author} {\bibfnamefont
  {Warwick~P.}\ \bibnamefont {Bowen}}, \bibinfo {author} {\bibfnamefont
  {Tobias}\ \bibnamefont {Gehring}}, \ and\ \bibinfo {author} {\bibfnamefont
  {Ulrik~L.}\ \bibnamefont {Andersen}},\ }\bibfield  {title} {\enquote
  {\bibinfo {title} {Quantum enhanced feedback cooling of a mechanical
  oscillator using nonclassical light},}\ }\href {\doibase 10.1038/ncomms13628}
  {\bibfield  {journal} {\bibinfo  {journal} {Nature Communications}\ }\textbf
  {\bibinfo {volume} {7}},\ \bibinfo {pages} {13628} (\bibinfo {year}
  {2016})}\BibitemShut {NoStop}%
\bibitem [{\citenamefont {Mamaev}\ \emph {et~al.}(2025)\citenamefont {Mamaev},
  \citenamefont {Koppenh\"ofer}, \citenamefont {Pocklington},\ and\
  \citenamefont {Clerk}}]{nonclassical_squeezing_spin}%
  \BibitemOpen
  \bibfield  {author} {\bibinfo {author} {\bibfnamefont {Mikhail}\ \bibnamefont
  {Mamaev}}, \bibinfo {author} {\bibfnamefont {Martin}\ \bibnamefont
  {Koppenh\"ofer}}, \bibinfo {author} {\bibfnamefont {Andrew}\ \bibnamefont
  {Pocklington}}, \ and\ \bibinfo {author} {\bibfnamefont {Aashish~A.}\
  \bibnamefont {Clerk}},\ }\bibfield  {title} {\enquote {\bibinfo {title}
  {Non-gaussian generalized two-mode squeezing: Applications to two-ensemble
  spin squeezing and beyond},}\ }\href {\doibase
  10.1103/PhysRevLett.134.073603} {\bibfield  {journal} {\bibinfo  {journal}
  {Phys. Rev. Lett.}\ }\textbf {\bibinfo {volume} {134}},\ \bibinfo {pages}
  {073603} (\bibinfo {year} {2025})}\BibitemShut {NoStop}%
\bibitem [{\citenamefont {Meyer-Scott}\ \emph {et~al.}(2022)\citenamefont
  {Meyer-Scott}, \citenamefont {Prasannan}, \citenamefont {Dhand},
  \citenamefont {Eigner}, \citenamefont {Quiring}, \citenamefont {Barkhofen},
  \citenamefont {Brecht}, \citenamefont {Plenio},\ and\ \citenamefont
  {Silberhorn}}]{quantum_metrology_1}%
  \BibitemOpen
  \bibfield  {author} {\bibinfo {author} {\bibfnamefont {Evan}\ \bibnamefont
  {Meyer-Scott}}, \bibinfo {author} {\bibfnamefont {Nidhin}\ \bibnamefont
  {Prasannan}}, \bibinfo {author} {\bibfnamefont {Ish}\ \bibnamefont {Dhand}},
  \bibinfo {author} {\bibfnamefont {Christof}\ \bibnamefont {Eigner}}, \bibinfo
  {author} {\bibfnamefont {Viktor}\ \bibnamefont {Quiring}}, \bibinfo {author}
  {\bibfnamefont {Sonja}\ \bibnamefont {Barkhofen}}, \bibinfo {author}
  {\bibfnamefont {Benjamin}\ \bibnamefont {Brecht}}, \bibinfo {author}
  {\bibfnamefont {Martin~B.}\ \bibnamefont {Plenio}}, \ and\ \bibinfo {author}
  {\bibfnamefont {Christine}\ \bibnamefont {Silberhorn}},\ }\bibfield  {title}
  {\enquote {\bibinfo {title} {Scalable generation of multiphoton entangled
  states by active feed-forward and multiplexing},}\ }\href {\doibase
  10.1103/PhysRevLett.129.150501} {\bibfield  {journal} {\bibinfo  {journal}
  {Phys. Rev. Lett.}\ }\textbf {\bibinfo {volume} {129}},\ \bibinfo {pages}
  {150501} (\bibinfo {year} {2022})}\BibitemShut {NoStop}%
\bibitem [{\citenamefont {Qin}\ \emph {et~al.}(2023)\citenamefont {Qin},
  \citenamefont {Deng}, \citenamefont {Zhong}, \citenamefont {Peng},
  \citenamefont {Su}, \citenamefont {Luo}, \citenamefont {Xu}, \citenamefont
  {Wu}, \citenamefont {Gong}, \citenamefont {Liu}, \citenamefont {Wang},
  \citenamefont {Chen}, \citenamefont {Li}, \citenamefont {Liu}, \citenamefont
  {Lu},\ and\ \citenamefont {Pan}}]{quantum_metrology_2}%
  \BibitemOpen
  \bibfield  {author} {\bibinfo {author} {\bibfnamefont {Jian}\ \bibnamefont
  {Qin}}, \bibinfo {author} {\bibfnamefont {Yu-Hao}\ \bibnamefont {Deng}},
  \bibinfo {author} {\bibfnamefont {Han-Sen}\ \bibnamefont {Zhong}}, \bibinfo
  {author} {\bibfnamefont {Li-Chao}\ \bibnamefont {Peng}}, \bibinfo {author}
  {\bibfnamefont {Hao}\ \bibnamefont {Su}}, \bibinfo {author} {\bibfnamefont
  {Yi-Han}\ \bibnamefont {Luo}}, \bibinfo {author} {\bibfnamefont {Jia-Min}\
  \bibnamefont {Xu}}, \bibinfo {author} {\bibfnamefont {Dian}\ \bibnamefont
  {Wu}}, \bibinfo {author} {\bibfnamefont {Si-Qiu}\ \bibnamefont {Gong}},
  \bibinfo {author} {\bibfnamefont {Hua-Liang}\ \bibnamefont {Liu}}, \bibinfo
  {author} {\bibfnamefont {Hui}\ \bibnamefont {Wang}}, \bibinfo {author}
  {\bibfnamefont {Ming-Cheng}\ \bibnamefont {Chen}}, \bibinfo {author}
  {\bibfnamefont {Li}~\bibnamefont {Li}}, \bibinfo {author} {\bibfnamefont
  {Nai-Le}\ \bibnamefont {Liu}}, \bibinfo {author} {\bibfnamefont {Chao-Yang}\
  \bibnamefont {Lu}}, \ and\ \bibinfo {author} {\bibfnamefont {Jian-Wei}\
  \bibnamefont {Pan}},\ }\bibfield  {title} {\enquote {\bibinfo {title}
  {Unconditional and robust quantum metrological advantage beyond n00n
  states},}\ }\href {\doibase 10.1103/PhysRevLett.130.070801} {\bibfield
  {journal} {\bibinfo  {journal} {Phys. Rev. Lett.}\ }\textbf {\bibinfo
  {volume} {130}},\ \bibinfo {pages} {070801} (\bibinfo {year}
  {2023})}\BibitemShut {NoStop}%
\bibitem [{\citenamefont {Nielsen}\ \emph {et~al.}(2023)\citenamefont
  {Nielsen}, \citenamefont {Neergaard-Nielsen}, \citenamefont {Gehring},\ and\
  \citenamefont {Andersen}}]{quantum_metrology_3}%
  \BibitemOpen
  \bibfield  {author} {\bibinfo {author} {\bibfnamefont {Jens A.~H.}\
  \bibnamefont {Nielsen}}, \bibinfo {author} {\bibfnamefont {Jonas~S.}\
  \bibnamefont {Neergaard-Nielsen}}, \bibinfo {author} {\bibfnamefont {Tobias}\
  \bibnamefont {Gehring}}, \ and\ \bibinfo {author} {\bibfnamefont {Ulrik~L.}\
  \bibnamefont {Andersen}},\ }\bibfield  {title} {\enquote {\bibinfo {title}
  {Deterministic quantum phase estimation beyond n00n states},}\ }\href
  {\doibase 10.1103/PhysRevLett.130.123603} {\bibfield  {journal} {\bibinfo
  {journal} {Phys. Rev. Lett.}\ }\textbf {\bibinfo {volume} {130}},\ \bibinfo
  {pages} {123603} (\bibinfo {year} {2023})}\BibitemShut {NoStop}%
\bibitem [{\citenamefont {Afek}\ \emph {et~al.}(2010)\citenamefont {Afek},
  \citenamefont {Ambar},\ and\ \citenamefont
  {Silberberg}}]{quantum_metrology_4}%
  \BibitemOpen
  \bibfield  {author} {\bibinfo {author} {\bibfnamefont {Itai}\ \bibnamefont
  {Afek}}, \bibinfo {author} {\bibfnamefont {Oron}\ \bibnamefont {Ambar}}, \
  and\ \bibinfo {author} {\bibfnamefont {Yaron}\ \bibnamefont {Silberberg}},\
  }\bibfield  {title} {\enquote {\bibinfo {title} {High-noon states by mixing
  quantum and classical light},}\ }\href {\doibase 10.1126/science.1188172}
  {\bibfield  {journal} {\bibinfo  {journal} {Science}\ }\textbf {\bibinfo
  {volume} {328}},\ \bibinfo {pages} {879--881} (\bibinfo {year}
  {2010})}\BibitemShut {NoStop}%
\bibitem [{\citenamefont {Bell}\ \emph {et~al.}(2013)\citenamefont {Bell},
  \citenamefont {Kannan}, \citenamefont {McMillan}, \citenamefont {Clark},
  \citenamefont {Wadsworth},\ and\ \citenamefont
  {Rarity}}]{quantum_metrology_5}%
  \BibitemOpen
  \bibfield  {author} {\bibinfo {author} {\bibfnamefont {Bryn}\ \bibnamefont
  {Bell}}, \bibinfo {author} {\bibfnamefont {Srikanth}\ \bibnamefont {Kannan}},
  \bibinfo {author} {\bibfnamefont {Alex}\ \bibnamefont {McMillan}}, \bibinfo
  {author} {\bibfnamefont {Alex~S.}\ \bibnamefont {Clark}}, \bibinfo {author}
  {\bibfnamefont {William~J.}\ \bibnamefont {Wadsworth}}, \ and\ \bibinfo
  {author} {\bibfnamefont {John~G.}\ \bibnamefont {Rarity}},\ }\bibfield
  {title} {\enquote {\bibinfo {title} {Multicolor quantum metrology with
  entangled photons},}\ }\href {\doibase 10.1103/PhysRevLett.111.093603}
  {\bibfield  {journal} {\bibinfo  {journal} {Phys. Rev. Lett.}\ }\textbf
  {\bibinfo {volume} {111}},\ \bibinfo {pages} {093603} (\bibinfo {year}
  {2013})}\BibitemShut {NoStop}%
\bibitem [{\citenamefont {Dorfman}\ \emph {et~al.}(2016)\citenamefont
  {Dorfman}, \citenamefont {Schlawin},\ and\ \citenamefont
  {Mukamel}}]{quantum_metrology_6}%
  \BibitemOpen
  \bibfield  {author} {\bibinfo {author} {\bibfnamefont {Konstantin~E.}\
  \bibnamefont {Dorfman}}, \bibinfo {author} {\bibfnamefont {Frank}\
  \bibnamefont {Schlawin}}, \ and\ \bibinfo {author} {\bibfnamefont {Shaul}\
  \bibnamefont {Mukamel}},\ }\bibfield  {title} {\enquote {\bibinfo {title}
  {Nonlinear optical signals and spectroscopy with quantum light},}\ }\href
  {\doibase 10.1103/RevModPhys.88.045008} {\bibfield  {journal} {\bibinfo
  {journal} {Rev. Mod. Phys.}\ }\textbf {\bibinfo {volume} {88}},\ \bibinfo
  {pages} {045008} (\bibinfo {year} {2016})}\BibitemShut {NoStop}%
\bibitem [{\citenamefont {Palacios}\ \emph {et~al.}(2024)\citenamefont
  {Palacios}, \citenamefont {Mart{\'\i}nez-Pe{\~n}a}, \citenamefont {Soriano},
  \citenamefont {Giorgi},\ and\ \citenamefont
  {Zambrini}}]{nonclassical_processing_capacity}%
  \BibitemOpen
  \bibfield  {author} {\bibinfo {author} {\bibfnamefont {Ana}\ \bibnamefont
  {Palacios}}, \bibinfo {author} {\bibfnamefont {Rodrigo}\ \bibnamefont
  {Mart{\'\i}nez-Pe{\~n}a}}, \bibinfo {author} {\bibfnamefont {Miguel~C.}\
  \bibnamefont {Soriano}}, \bibinfo {author} {\bibfnamefont {Gian~Luca}\
  \bibnamefont {Giorgi}}, \ and\ \bibinfo {author} {\bibfnamefont {Roberta}\
  \bibnamefont {Zambrini}},\ }\bibfield  {title} {\enquote {\bibinfo {title}
  {Role of coherence in many-body {Quantum} {Reservoir} {Computing}},}\ }\href
  {\doibase 10.1038/s42005-024-01859-4} {\bibfield  {journal} {\bibinfo
  {journal} {Communications Physics}\ }\textbf {\bibinfo {volume} {7}},\
  \bibinfo {pages} {369} (\bibinfo {year} {2024})}\BibitemShut {NoStop}%
\bibitem [{\citenamefont {Yan}\ \emph {et~al.}(2021)\citenamefont {Yan},
  \citenamefont {Qin}, \citenamefont {Qin}, \citenamefont {Su}, \citenamefont
  {Jia}, \citenamefont {Xie},\ and\ \citenamefont
  {Peng}}]{nonclassical_processing_capacity_b}%
  \BibitemOpen
  \bibfield  {author} {\bibinfo {author} {\bibfnamefont {Zhi-Hui}\ \bibnamefont
  {Yan}}, \bibinfo {author} {\bibfnamefont {Ji-Liang}\ \bibnamefont {Qin}},
  \bibinfo {author} {\bibfnamefont {Zhong-Zhong}\ \bibnamefont {Qin}}, \bibinfo
  {author} {\bibfnamefont {Xiao-Long}\ \bibnamefont {Su}}, \bibinfo {author}
  {\bibfnamefont {Xiao-Jun}\ \bibnamefont {Jia}}, \bibinfo {author}
  {\bibfnamefont {Chang-De}\ \bibnamefont {Xie}}, \ and\ \bibinfo {author}
  {\bibfnamefont {Kun-Chi}\ \bibnamefont {Peng}},\ }\bibfield  {title}
  {\enquote {\bibinfo {title} {Generation of non-classical states of light and
  their application in deterministic quantum teleportation},}\ }\href {\doibase
  https://doi.org/10.1016/j.fmre.2020.11.005} {\bibfield  {journal} {\bibinfo
  {journal} {Fundamental Research}\ }\textbf {\bibinfo {volume} {1}},\ \bibinfo
  {pages} {43--49} (\bibinfo {year} {2021})}\BibitemShut {NoStop}%
\bibitem [{\citenamefont {Yang}\ \emph {et~al.}(2025)\citenamefont {Yang},
  \citenamefont {Luo}, \citenamefont {Zhang}, \citenamefont {Wang},
  \citenamefont {Zou}, \citenamefont {Xia},\ and\ \citenamefont
  {Lu}}]{nonclassical_measurement_sensitivity}%
  \BibitemOpen
  \bibfield  {author} {\bibinfo {author} {\bibfnamefont {Y.~A.}\ \bibnamefont
  {Yang}}, \bibinfo {author} {\bibfnamefont {W.-T.}\ \bibnamefont {Luo}},
  \bibinfo {author} {\bibfnamefont {J.-L.}\ \bibnamefont {Zhang}}, \bibinfo
  {author} {\bibfnamefont {S.-Z.}\ \bibnamefont {Wang}}, \bibinfo {author}
  {\bibfnamefont {Chang-Ling}\ \bibnamefont {Zou}}, \bibinfo {author}
  {\bibfnamefont {T.}~\bibnamefont {Xia}}, \ and\ \bibinfo {author}
  {\bibfnamefont {Z.-T.}\ \bibnamefont {Lu}},\ }\bibfield  {title} {\enquote
  {\bibinfo {title} {Minute-scale {Schr{\"o}dinger}-cat state of spin-5/2
  atoms},}\ }\href {\doibase 10.1038/s41566-024-01555-3} {\bibfield  {journal}
  {\bibinfo  {journal} {Nature Photonics}\ }\textbf {\bibinfo {volume} {19}},\
  \bibinfo {pages} {89--94} (\bibinfo {year} {2025})}\BibitemShut {NoStop}%
\bibitem [{\citenamefont {Jiao}\ \emph {et~al.}(2020)\citenamefont {Jiao},
  \citenamefont {Zhang}, \citenamefont {Zhang}, \citenamefont {Miranowicz},
  \citenamefont {Kuang},\ and\ \citenamefont {Jing}}]{quantum_entanglement}%
  \BibitemOpen
  \bibfield  {author} {\bibinfo {author} {\bibfnamefont {Ya-Feng}\ \bibnamefont
  {Jiao}}, \bibinfo {author} {\bibfnamefont {Sheng-Dian}\ \bibnamefont
  {Zhang}}, \bibinfo {author} {\bibfnamefont {Yan-Lei}\ \bibnamefont {Zhang}},
  \bibinfo {author} {\bibfnamefont {Adam}\ \bibnamefont {Miranowicz}}, \bibinfo
  {author} {\bibfnamefont {Le-Man}\ \bibnamefont {Kuang}}, \ and\ \bibinfo
  {author} {\bibfnamefont {Hui}\ \bibnamefont {Jing}},\ }\bibfield  {title}
  {\enquote {\bibinfo {title} {Nonreciprocal optomechanical entanglement
  against backscattering losses},}\ }\href {\doibase
  10.1103/PhysRevLett.125.143605} {\bibfield  {journal} {\bibinfo  {journal}
  {Phys. Rev. Lett.}\ }\textbf {\bibinfo {volume} {125}},\ \bibinfo {pages}
  {143605} (\bibinfo {year} {2020})}\BibitemShut {NoStop}%
\bibitem [{\citenamefont {long Ren}(2022)}]{quantum_entanglement_1}%
  \BibitemOpen
  \bibfield  {author} {\bibinfo {author} {\bibfnamefont {Ya}~\bibnamefont {long
  Ren}},\ }\bibfield  {title} {\enquote {\bibinfo {title} {Nonreciprocal
  optical--microwave entanglement in a spinning magnetic resonator},}\ }\href
  {\doibase 10.1364/OL.451050} {\bibfield  {journal} {\bibinfo  {journal} {Opt.
  Lett.}\ }\textbf {\bibinfo {volume} {47}},\ \bibinfo {pages} {1125--1128}
  (\bibinfo {year} {2022})}\BibitemShut {NoStop}%
\bibitem [{\citenamefont {Jiao}\ \emph {et~al.}(2022)\citenamefont {Jiao},
  \citenamefont {Liu}, \citenamefont {Li}, \citenamefont {Yang}, \citenamefont
  {Kuang},\ and\ \citenamefont {Jing}}]{quantum_entanglement_2}%
  \BibitemOpen
  \bibfield  {author} {\bibinfo {author} {\bibfnamefont {Ya-Feng}\ \bibnamefont
  {Jiao}}, \bibinfo {author} {\bibfnamefont {Jing-Xue}\ \bibnamefont {Liu}},
  \bibinfo {author} {\bibfnamefont {Ying}\ \bibnamefont {Li}}, \bibinfo
  {author} {\bibfnamefont {Ronghua}\ \bibnamefont {Yang}}, \bibinfo {author}
  {\bibfnamefont {Le-Man}\ \bibnamefont {Kuang}}, \ and\ \bibinfo {author}
  {\bibfnamefont {Hui}\ \bibnamefont {Jing}},\ }\bibfield  {title} {\enquote
  {\bibinfo {title} {Nonreciprocal enhancement of remote entanglement between
  nonidentical mechanical oscillators},}\ }\href {\doibase
  10.1103/PhysRevApplied.18.064008} {\bibfield  {journal} {\bibinfo  {journal}
  {Phys. Rev. Appl.}\ }\textbf {\bibinfo {volume} {18}},\ \bibinfo {pages}
  {064008} (\bibinfo {year} {2022})}\BibitemShut {NoStop}%
\bibitem [{\citenamefont {Hu}\ \emph {et~al.}(2023)\citenamefont {Hu},
  \citenamefont {Guo}, \citenamefont {Liu}, \citenamefont {Li},\ and\
  \citenamefont {Guo}}]{nonclassical_quantum_computing}%
  \BibitemOpen
  \bibfield  {author} {\bibinfo {author} {\bibfnamefont {Xiao-Min}\
  \bibnamefont {Hu}}, \bibinfo {author} {\bibfnamefont {Yu}~\bibnamefont
  {Guo}}, \bibinfo {author} {\bibfnamefont {Bi-Heng}\ \bibnamefont {Liu}},
  \bibinfo {author} {\bibfnamefont {Chuan-Feng}\ \bibnamefont {Li}}, \ and\
  \bibinfo {author} {\bibfnamefont {Guang-Can}\ \bibnamefont {Guo}},\
  }\bibfield  {title} {\enquote {\bibinfo {title} {Progress in quantum
  teleportation},}\ }\href {\doibase 10.1038/s42254-023-00588-x} {\bibfield
  {journal} {\bibinfo  {journal} {Nature Reviews Physics}\ }\textbf {\bibinfo
  {volume} {5}},\ \bibinfo {pages} {339--353} (\bibinfo {year}
  {2023})}\BibitemShut {NoStop}%
\bibitem [{\citenamefont {Descamps}\ \emph {et~al.}(2024)\citenamefont
  {Descamps}, \citenamefont {Fabre}, \citenamefont {Saharyan}, \citenamefont
  {Keller},\ and\ \citenamefont {Milman}}]{nonclassical_quantum_computing_1}%
  \BibitemOpen
  \bibfield  {author} {\bibinfo {author} {\bibfnamefont {Eloi}\ \bibnamefont
  {Descamps}}, \bibinfo {author} {\bibfnamefont {Nicolas}\ \bibnamefont
  {Fabre}}, \bibinfo {author} {\bibfnamefont {Astghik}\ \bibnamefont
  {Saharyan}}, \bibinfo {author} {\bibfnamefont {Arne}\ \bibnamefont {Keller}},
  \ and\ \bibinfo {author} {\bibfnamefont {P\'erola}\ \bibnamefont {Milman}},\
  }\bibfield  {title} {\enquote {\bibinfo {title} {Superselection rules and
  bosonic quantum computational resources},}\ }\href {\doibase
  10.1103/PhysRevLett.133.260605} {\bibfield  {journal} {\bibinfo  {journal}
  {Phys. Rev. Lett.}\ }\textbf {\bibinfo {volume} {133}},\ \bibinfo {pages}
  {260605} (\bibinfo {year} {2024})}\BibitemShut {NoStop}%
\bibitem [{\citenamefont {Drmota}\ \emph {et~al.}(2024)\citenamefont {Drmota},
  \citenamefont {Nadlinger}, \citenamefont {Main}, \citenamefont {Nichol},
  \citenamefont {Ainley}, \citenamefont {Leichtle}, \citenamefont {Mantri},
  \citenamefont {Kashefi}, \citenamefont {Srinivas}, \citenamefont {Araneda},
  \citenamefont {Ballance},\ and\ \citenamefont
  {Lucas}}]{nonclassical_quantum_computing_2}%
  \BibitemOpen
  \bibfield  {author} {\bibinfo {author} {\bibfnamefont {P.}~\bibnamefont
  {Drmota}}, \bibinfo {author} {\bibfnamefont {D.~P.}\ \bibnamefont
  {Nadlinger}}, \bibinfo {author} {\bibfnamefont {D.}~\bibnamefont {Main}},
  \bibinfo {author} {\bibfnamefont {B.~C.}\ \bibnamefont {Nichol}}, \bibinfo
  {author} {\bibfnamefont {E.~M.}\ \bibnamefont {Ainley}}, \bibinfo {author}
  {\bibfnamefont {D.}~\bibnamefont {Leichtle}}, \bibinfo {author}
  {\bibfnamefont {A.}~\bibnamefont {Mantri}}, \bibinfo {author} {\bibfnamefont
  {E.}~\bibnamefont {Kashefi}}, \bibinfo {author} {\bibfnamefont
  {R.}~\bibnamefont {Srinivas}}, \bibinfo {author} {\bibfnamefont
  {G.}~\bibnamefont {Araneda}}, \bibinfo {author} {\bibfnamefont {C.~J.}\
  \bibnamefont {Ballance}}, \ and\ \bibinfo {author} {\bibfnamefont {D.~M.}\
  \bibnamefont {Lucas}},\ }\bibfield  {title} {\enquote {\bibinfo {title}
  {Verifiable blind quantum computing with trapped ions and single photons},}\
  }\href {\doibase 10.1103/PhysRevLett.132.150604} {\bibfield  {journal}
  {\bibinfo  {journal} {Phys. Rev. Lett.}\ }\textbf {\bibinfo {volume} {132}},\
  \bibinfo {pages} {150604} (\bibinfo {year} {2024})}\BibitemShut {NoStop}%
\bibitem [{\citenamefont {Giovannetti}\ \emph {et~al.}(2006)\citenamefont
  {Giovannetti}, \citenamefont {Lloyd},\ and\ \citenamefont
  {Maccone}}]{Multiquanta_physics_in_metrology}%
  \BibitemOpen
  \bibfield  {author} {\bibinfo {author} {\bibfnamefont {Vittorio}\
  \bibnamefont {Giovannetti}}, \bibinfo {author} {\bibfnamefont {Seth}\
  \bibnamefont {Lloyd}}, \ and\ \bibinfo {author} {\bibfnamefont {Lorenzo}\
  \bibnamefont {Maccone}},\ }\bibfield  {title} {\enquote {\bibinfo {title}
  {Quantum metrology},}\ }\href {\doibase 10.1103/PhysRevLett.96.010401}
  {\bibfield  {journal} {\bibinfo  {journal} {Phys. Rev. Lett.}\ }\textbf
  {\bibinfo {volume} {96}},\ \bibinfo {pages} {010401} (\bibinfo {year}
  {2006})}\BibitemShut {NoStop}%
\bibitem [{\citenamefont {D'Angelo}\ \emph {et~al.}(2001)\citenamefont
  {D'Angelo}, \citenamefont {Chekhova},\ and\ \citenamefont
  {Shih}}]{Multiquanta_physics_in_lithography}%
  \BibitemOpen
  \bibfield  {author} {\bibinfo {author} {\bibfnamefont {Milena}\ \bibnamefont
  {D'Angelo}}, \bibinfo {author} {\bibfnamefont {Maria~V.}\ \bibnamefont
  {Chekhova}}, \ and\ \bibinfo {author} {\bibfnamefont {Yanhua}\ \bibnamefont
  {Shih}},\ }\bibfield  {title} {\enquote {\bibinfo {title} {Two-photon
  diffraction and quantum lithography},}\ }\href {\doibase
  10.1103/PhysRevLett.87.013602} {\bibfield  {journal} {\bibinfo  {journal}
  {Phys. Rev. Lett.}\ }\textbf {\bibinfo {volume} {87}},\ \bibinfo {pages}
  {013602} (\bibinfo {year} {2001})}\BibitemShut {NoStop}%
\bibitem [{\citenamefont {Taylor}\ \emph {et~al.}(2013)\citenamefont {Taylor},
  \citenamefont {Janousek}, \citenamefont {Daria}, \citenamefont {Knittel},
  \citenamefont {Hage}, \citenamefont {Bachor},\ and\ \citenamefont
  {Bowen}}]{Multiquantum_Biological_measurement}%
  \BibitemOpen
  \bibfield  {author} {\bibinfo {author} {\bibfnamefont {Michael~A.}\
  \bibnamefont {Taylor}}, \bibinfo {author} {\bibfnamefont {Jiri}\ \bibnamefont
  {Janousek}}, \bibinfo {author} {\bibfnamefont {Vincent}\ \bibnamefont
  {Daria}}, \bibinfo {author} {\bibfnamefont {Joachim}\ \bibnamefont
  {Knittel}}, \bibinfo {author} {\bibfnamefont {Boris}\ \bibnamefont {Hage}},
  \bibinfo {author} {\bibfnamefont {Hans-A.}\ \bibnamefont {Bachor}}, \ and\
  \bibinfo {author} {\bibfnamefont {Warwick~P.}\ \bibnamefont {Bowen}},\
  }\bibfield  {title} {\enquote {\bibinfo {title} {Biological measurement
  beyond the quantum limit},}\ }\href {\doibase 10.1038/nphoton.2012.346}
  {\bibfield  {journal} {\bibinfo  {journal} {Nature Photonics}\ }\textbf
  {\bibinfo {volume} {7}},\ \bibinfo {pages} {229--233} (\bibinfo {year}
  {2013})}\BibitemShut {NoStop}%
\bibitem [{\citenamefont {Waldecker}\ \emph {et~al.}(2017)\citenamefont
  {Waldecker}, \citenamefont {Bertoni}, \citenamefont {H\"ubener},
  \citenamefont {Brumme}, \citenamefont {Vasileiadis}, \citenamefont {Zahn},
  \citenamefont {Rubio},\ and\ \citenamefont
  {Ernstorfer}}]{Multiquantum_Biological_measurement_1}%
  \BibitemOpen
  \bibfield  {author} {\bibinfo {author} {\bibfnamefont {L.}~\bibnamefont
  {Waldecker}}, \bibinfo {author} {\bibfnamefont {R.}~\bibnamefont {Bertoni}},
  \bibinfo {author} {\bibfnamefont {H.}~\bibnamefont {H\"ubener}}, \bibinfo
  {author} {\bibfnamefont {T.}~\bibnamefont {Brumme}}, \bibinfo {author}
  {\bibfnamefont {T.}~\bibnamefont {Vasileiadis}}, \bibinfo {author}
  {\bibfnamefont {D.}~\bibnamefont {Zahn}}, \bibinfo {author} {\bibfnamefont
  {A.}~\bibnamefont {Rubio}}, \ and\ \bibinfo {author} {\bibfnamefont
  {R.}~\bibnamefont {Ernstorfer}},\ }\bibfield  {title} {\enquote {\bibinfo
  {title} {Momentum-resolved view of electron-phonon coupling in multilayer
  ${\mathrm{wse}}_{2}$},}\ }\href {\doibase 10.1103/PhysRevLett.119.036803}
  {\bibfield  {journal} {\bibinfo  {journal} {Phys. Rev. Lett.}\ }\textbf
  {\bibinfo {volume} {119}},\ \bibinfo {pages} {036803} (\bibinfo {year}
  {2017})}\BibitemShut {NoStop}%
\bibitem [{\citenamefont {Pan}\ \emph {et~al.}(2012)\citenamefont {Pan},
  \citenamefont {Chen}, \citenamefont {Lu}, \citenamefont {Weinfurter},
  \citenamefont {Zeilinger},\ and\ \citenamefont {\ifmmode~\dot{Z}\else
  \.{Z}\fi{}ukowski}}]{Multiquantum_post_selection}%
  \BibitemOpen
  \bibfield  {author} {\bibinfo {author} {\bibfnamefont {Jian-Wei}\
  \bibnamefont {Pan}}, \bibinfo {author} {\bibfnamefont {Zeng-Bing}\
  \bibnamefont {Chen}}, \bibinfo {author} {\bibfnamefont {Chao-Yang}\
  \bibnamefont {Lu}}, \bibinfo {author} {\bibfnamefont {Harald}\ \bibnamefont
  {Weinfurter}}, \bibinfo {author} {\bibfnamefont {Anton}\ \bibnamefont
  {Zeilinger}}, \ and\ \bibinfo {author} {\bibfnamefont {Marek}\ \bibnamefont
  {\ifmmode~\dot{Z}\else \.{Z}\fi{}ukowski}},\ }\bibfield  {title} {\enquote
  {\bibinfo {title} {Multiphoton entanglement and interferometry},}\ }\href
  {\doibase 10.1103/RevModPhys.84.777} {\bibfield  {journal} {\bibinfo
  {journal} {Rev. Mod. Phys.}\ }\textbf {\bibinfo {volume} {84}},\ \bibinfo
  {pages} {777--838} (\bibinfo {year} {2012})}\BibitemShut {NoStop}%
\bibitem [{\citenamefont {Maghrebi}\ \emph {et~al.}(2015)\citenamefont
  {Maghrebi}, \citenamefont {Gullans}, \citenamefont {Bienias}, \citenamefont
  {Choi}, \citenamefont {Martin}, \citenamefont {Firstenberg}, \citenamefont
  {Lukin}, \citenamefont {B\"uchler},\ and\ \citenamefont
  {Gorshkov}}]{Multiquantum_Rydberg_atom}%
  \BibitemOpen
  \bibfield  {author} {\bibinfo {author} {\bibfnamefont {M.~F.}\ \bibnamefont
  {Maghrebi}}, \bibinfo {author} {\bibfnamefont {M.~J.}\ \bibnamefont
  {Gullans}}, \bibinfo {author} {\bibfnamefont {P.}~\bibnamefont {Bienias}},
  \bibinfo {author} {\bibfnamefont {S.}~\bibnamefont {Choi}}, \bibinfo {author}
  {\bibfnamefont {I.}~\bibnamefont {Martin}}, \bibinfo {author} {\bibfnamefont
  {O.}~\bibnamefont {Firstenberg}}, \bibinfo {author} {\bibfnamefont {M.~D.}\
  \bibnamefont {Lukin}}, \bibinfo {author} {\bibfnamefont {H.~P.}\ \bibnamefont
  {B\"uchler}}, \ and\ \bibinfo {author} {\bibfnamefont {A.~V.}\ \bibnamefont
  {Gorshkov}},\ }\bibfield  {title} {\enquote {\bibinfo {title} {Coulomb bound
  states of strongly interacting photons},}\ }\href {\doibase
  10.1103/PhysRevLett.115.123601} {\bibfield  {journal} {\bibinfo  {journal}
  {Phys. Rev. Lett.}\ }\textbf {\bibinfo {volume} {115}},\ \bibinfo {pages}
  {123601} (\bibinfo {year} {2015})}\BibitemShut {NoStop}%
\bibitem [{\citenamefont {Jachymski}\ \emph {et~al.}(2016)\citenamefont
  {Jachymski}, \citenamefont {Bienias},\ and\ \citenamefont
  {B\"uchler}}]{Multiquantum_Rydberg_atom_1}%
  \BibitemOpen
  \bibfield  {author} {\bibinfo {author} {\bibfnamefont {Krzysztof}\
  \bibnamefont {Jachymski}}, \bibinfo {author} {\bibfnamefont {Przemys\l{}aw}\
  \bibnamefont {Bienias}}, \ and\ \bibinfo {author} {\bibfnamefont
  {Hans~Peter}\ \bibnamefont {B\"uchler}},\ }\bibfield  {title} {\enquote
  {\bibinfo {title} {Three-body interaction of rydberg slow-light
  polaritons},}\ }\href {\doibase 10.1103/PhysRevLett.117.053601} {\bibfield
  {journal} {\bibinfo  {journal} {Phys. Rev. Lett.}\ }\textbf {\bibinfo
  {volume} {117}},\ \bibinfo {pages} {053601} (\bibinfo {year}
  {2016})}\BibitemShut {NoStop}%
\bibitem [{\citenamefont {Liang}\ \emph {et~al.}(2018)\citenamefont {Liang},
  \citenamefont {Venkatramani}, \citenamefont {Cantu}, \citenamefont
  {Nicholson}, \citenamefont {Gullans}, \citenamefont {Gorshkov}, \citenamefont
  {Thompson}, \citenamefont {Chin}, \citenamefont {Lukin},\ and\ \citenamefont
  {Vuletić}}]{Multiquantum_Rydberg_atom_2}%
  \BibitemOpen
  \bibfield  {author} {\bibinfo {author} {\bibfnamefont {Qi-Yu}\ \bibnamefont
  {Liang}}, \bibinfo {author} {\bibfnamefont {Aditya~V.}\ \bibnamefont
  {Venkatramani}}, \bibinfo {author} {\bibfnamefont {Sergio~H.}\ \bibnamefont
  {Cantu}}, \bibinfo {author} {\bibfnamefont {Travis~L.}\ \bibnamefont
  {Nicholson}}, \bibinfo {author} {\bibfnamefont {Michael~J.}\ \bibnamefont
  {Gullans}}, \bibinfo {author} {\bibfnamefont {Alexey~V.}\ \bibnamefont
  {Gorshkov}}, \bibinfo {author} {\bibfnamefont {Jeff~D.}\ \bibnamefont
  {Thompson}}, \bibinfo {author} {\bibfnamefont {Cheng}\ \bibnamefont {Chin}},
  \bibinfo {author} {\bibfnamefont {Mikhail~D.}\ \bibnamefont {Lukin}}, \ and\
  \bibinfo {author} {\bibfnamefont {Vladan}\ \bibnamefont {Vuletić}},\
  }\bibfield  {title} {\enquote {\bibinfo {title} {Observation of three-photon
  bound states in a quantum nonlinear medium},}\ }\href {\doibase
  10.1126/science.aao7293} {\bibfield  {journal} {\bibinfo  {journal}
  {Science}\ }\textbf {\bibinfo {volume} {359}},\ \bibinfo {pages} {783--786}
  (\bibinfo {year} {2018})}\BibitemShut {NoStop}%
\bibitem [{\citenamefont {Yu}\ \emph {et~al.}(2025)\citenamefont {Yu},
  \citenamefont {Xu}, \citenamefont {Zhu}, \citenamefont {Bai},\ and\
  \citenamefont {Yang}}]{Rydberg_PB}%
  \BibitemOpen
  \bibfield  {author} {\bibinfo {author} {\bibfnamefont {Z.~Y.}\ \bibnamefont
  {Yu}}, \bibinfo {author} {\bibfnamefont {J.~P.}\ \bibnamefont {Xu}}, \bibinfo
  {author} {\bibfnamefont {C.~J.}\ \bibnamefont {Zhu}}, \bibinfo {author}
  {\bibfnamefont {Z.~Y.}\ \bibnamefont {Bai}}, \ and\ \bibinfo {author}
  {\bibfnamefont {Y.~P.}\ \bibnamefont {Yang}},\ }\bibfield  {title} {\enquote
  {\bibinfo {title} {Strong photon antibunching and photon pair emission in a
  two-atom cavity qed system with van der waals interaction},}\ }\href
  {\doibase 10.1103/3v2q-7g6r} {\bibfield  {journal} {\bibinfo  {journal}
  {Phys. Rev. Res.}\ }\textbf {\bibinfo {volume} {7}},\ \bibinfo {pages}
  {043248} (\bibinfo {year} {2025})}\BibitemShut {NoStop}%
\bibitem [{\citenamefont {Koshino}\ \emph {et~al.}(2013)\citenamefont
  {Koshino}, \citenamefont {Inomata}, \citenamefont {Yamamoto},\ and\
  \citenamefont {Nakamura}}]{Multiquantum_QED_platform}%
  \BibitemOpen
  \bibfield  {author} {\bibinfo {author} {\bibfnamefont {Kazuki}\ \bibnamefont
  {Koshino}}, \bibinfo {author} {\bibfnamefont {Kunihiro}\ \bibnamefont
  {Inomata}}, \bibinfo {author} {\bibfnamefont {Tsuyoshi}\ \bibnamefont
  {Yamamoto}}, \ and\ \bibinfo {author} {\bibfnamefont {Yasunobu}\ \bibnamefont
  {Nakamura}},\ }\bibfield  {title} {\enquote {\bibinfo {title} {Implementation
  of an impedance-matched $\ensuremath{\Lambda}$ system by dressed-state
  engineering},}\ }\href {\doibase 10.1103/PhysRevLett.111.153601} {\bibfield
  {journal} {\bibinfo  {journal} {Phys. Rev. Lett.}\ }\textbf {\bibinfo
  {volume} {111}},\ \bibinfo {pages} {153601} (\bibinfo {year}
  {2013})}\BibitemShut {NoStop}%
\bibitem [{\citenamefont {Douglas}\ \emph {et~al.}(2016)\citenamefont
  {Douglas}, \citenamefont {Caneva},\ and\ \citenamefont
  {Chang}}]{Multiquantum_QED_platform_1}%
  \BibitemOpen
  \bibfield  {author} {\bibinfo {author} {\bibfnamefont {James~S.}\
  \bibnamefont {Douglas}}, \bibinfo {author} {\bibfnamefont {Tommaso}\
  \bibnamefont {Caneva}}, \ and\ \bibinfo {author} {\bibfnamefont {Darrick~E.}\
  \bibnamefont {Chang}},\ }\bibfield  {title} {\enquote {\bibinfo {title}
  {Photon molecules in atomic gases trapped near photonic crystal
  waveguides},}\ }\href {\doibase 10.1103/PhysRevX.6.031017} {\bibfield
  {journal} {\bibinfo  {journal} {Phys. Rev. X}\ }\textbf {\bibinfo {volume}
  {6}},\ \bibinfo {pages} {031017} (\bibinfo {year} {2016})}\BibitemShut
  {NoStop}%
\bibitem [{\citenamefont {Gonz\'alez-Tudela}\ \emph {et~al.}(2017)\citenamefont
  {Gonz\'alez-Tudela}, \citenamefont {Paulisch}, \citenamefont {Kimble},\ and\
  \citenamefont {Cirac}}]{Multiquantum_QED_platform_2}%
  \BibitemOpen
  \bibfield  {author} {\bibinfo {author} {\bibfnamefont {A.}~\bibnamefont
  {Gonz\'alez-Tudela}}, \bibinfo {author} {\bibfnamefont {V.}~\bibnamefont
  {Paulisch}}, \bibinfo {author} {\bibfnamefont {H.~J.}\ \bibnamefont
  {Kimble}}, \ and\ \bibinfo {author} {\bibfnamefont {J.~I.}\ \bibnamefont
  {Cirac}},\ }\bibfield  {title} {\enquote {\bibinfo {title} {Efficient
  multiphoton generation in waveguide quantum electrodynamics},}\ }\href
  {\doibase 10.1103/PhysRevLett.118.213601} {\bibfield  {journal} {\bibinfo
  {journal} {Phys. Rev. Lett.}\ }\textbf {\bibinfo {volume} {118}},\ \bibinfo
  {pages} {213601} (\bibinfo {year} {2017})}\BibitemShut {NoStop}%
\bibitem [{\citenamefont {Chang}\ \emph {et~al.}(2016)\citenamefont {Chang},
  \citenamefont {Gonz\'alez-Tudela}, \citenamefont {S\'anchez Mu\~noz},
  \citenamefont {Navarrete-Benlloch},\ and\ \citenamefont
  {Shi}}]{Multiquantum_QED_platform_3}%
  \BibitemOpen
  \bibfield  {author} {\bibinfo {author} {\bibfnamefont {Yue}\ \bibnamefont
  {Chang}}, \bibinfo {author} {\bibfnamefont {Alejandro}\ \bibnamefont
  {Gonz\'alez-Tudela}}, \bibinfo {author} {\bibfnamefont {Carlos}\ \bibnamefont
  {S\'anchez Mu\~noz}}, \bibinfo {author} {\bibfnamefont {Carlos}\ \bibnamefont
  {Navarrete-Benlloch}}, \ and\ \bibinfo {author} {\bibfnamefont {Tao}\
  \bibnamefont {Shi}},\ }\bibfield  {title} {\enquote {\bibinfo {title}
  {Deterministic down-converter and continuous photon-pair source within the
  bad-cavity limit},}\ }\href {\doibase 10.1103/PhysRevLett.117.203602}
  {\bibfield  {journal} {\bibinfo  {journal} {Phys. Rev. Lett.}\ }\textbf
  {\bibinfo {volume} {117}},\ \bibinfo {pages} {203602} (\bibinfo {year}
  {2016})}\BibitemShut {NoStop}%
\bibitem [{\citenamefont {Hargart}\ \emph {et~al.}(2016)\citenamefont
  {Hargart}, \citenamefont {M\"uller}, \citenamefont {Roy-Choudhury},
  \citenamefont {Portalupi}, \citenamefont {Schneider}, \citenamefont
  {H\"ofling}, \citenamefont {Kamp}, \citenamefont {Hughes},\ and\
  \citenamefont {Michler}}]{Multiquantum_QED_platform_4}%
  \BibitemOpen
  \bibfield  {author} {\bibinfo {author} {\bibfnamefont {F.}~\bibnamefont
  {Hargart}}, \bibinfo {author} {\bibfnamefont {M.}~\bibnamefont {M\"uller}},
  \bibinfo {author} {\bibfnamefont {K.}~\bibnamefont {Roy-Choudhury}}, \bibinfo
  {author} {\bibfnamefont {S.~L.}\ \bibnamefont {Portalupi}}, \bibinfo {author}
  {\bibfnamefont {C.}~\bibnamefont {Schneider}}, \bibinfo {author}
  {\bibfnamefont {S.}~\bibnamefont {H\"ofling}}, \bibinfo {author}
  {\bibfnamefont {M.}~\bibnamefont {Kamp}}, \bibinfo {author} {\bibfnamefont
  {S.}~\bibnamefont {Hughes}}, \ and\ \bibinfo {author} {\bibfnamefont
  {P.}~\bibnamefont {Michler}},\ }\bibfield  {title} {\enquote {\bibinfo
  {title} {Cavity-enhanced simultaneous dressing of quantum dot exciton and
  biexciton states},}\ }\href {\doibase 10.1103/PhysRevB.93.115308} {\bibfield
  {journal} {\bibinfo  {journal} {Phys. Rev. B}\ }\textbf {\bibinfo {volume}
  {93}},\ \bibinfo {pages} {115308} (\bibinfo {year} {2016})}\BibitemShut
  {NoStop}%
\bibitem [{\citenamefont {Satzinger}\ \emph {et~al.}(2018)\citenamefont
  {Satzinger}, \citenamefont {Zhong}, \citenamefont {Chang}, \citenamefont
  {Peairs}, \citenamefont {Bienfait}, \citenamefont {Chou}, \citenamefont
  {Cleland}, \citenamefont {Conner}, \citenamefont {Dumur}, \citenamefont
  {Grebel}, \citenamefont {Gutierrez}, \citenamefont {November}, \citenamefont
  {Povey}, \citenamefont {Whiteley}, \citenamefont {Awschalom}, \citenamefont
  {Schuster},\ and\ \citenamefont {Cleland}}]{Multiquantum_QED_platform_5}%
  \BibitemOpen
  \bibfield  {author} {\bibinfo {author} {\bibfnamefont {K.~J.}\ \bibnamefont
  {Satzinger}}, \bibinfo {author} {\bibfnamefont {Y.~P.}\ \bibnamefont
  {Zhong}}, \bibinfo {author} {\bibfnamefont {H.-S.}\ \bibnamefont {Chang}},
  \bibinfo {author} {\bibfnamefont {G.~A.}\ \bibnamefont {Peairs}}, \bibinfo
  {author} {\bibfnamefont {A.}~\bibnamefont {Bienfait}}, \bibinfo {author}
  {\bibfnamefont {Ming-Han}\ \bibnamefont {Chou}}, \bibinfo {author}
  {\bibfnamefont {A.~Y.}\ \bibnamefont {Cleland}}, \bibinfo {author}
  {\bibfnamefont {C.~R.}\ \bibnamefont {Conner}}, \bibinfo {author}
  {\bibfnamefont {{\'E}.}~\bibnamefont {Dumur}}, \bibinfo {author}
  {\bibfnamefont {J.}~\bibnamefont {Grebel}}, \bibinfo {author} {\bibfnamefont
  {I.}~\bibnamefont {Gutierrez}}, \bibinfo {author} {\bibfnamefont {B.~H.}\
  \bibnamefont {November}}, \bibinfo {author} {\bibfnamefont {R.~G.}\
  \bibnamefont {Povey}}, \bibinfo {author} {\bibfnamefont {S.~J.}\ \bibnamefont
  {Whiteley}}, \bibinfo {author} {\bibfnamefont {D.~D.}\ \bibnamefont
  {Awschalom}}, \bibinfo {author} {\bibfnamefont {D.~I.}\ \bibnamefont
  {Schuster}}, \ and\ \bibinfo {author} {\bibfnamefont {A.~N.}\ \bibnamefont
  {Cleland}},\ }\bibfield  {title} {\enquote {\bibinfo {title} {Quantum control
  of surface acoustic-wave phonons},}\ }\href {\doibase
  10.1038/s41586-018-0719-5} {\bibfield  {journal} {\bibinfo  {journal}
  {Nature}\ }\textbf {\bibinfo {volume} {563}},\ \bibinfo {pages} {661--665}
  (\bibinfo {year} {2018})}\BibitemShut {NoStop}%
\bibitem [{\citenamefont {Bin}\ \emph {et~al.}(2024)\citenamefont {Bin},
  \citenamefont {Jing}, \citenamefont {Wu}, \citenamefont {Nori},\ and\
  \citenamefont {L\"u}}]{Quantum_Entangled_Pairs_Bin-qian}%
  \BibitemOpen
  \bibfield  {author} {\bibinfo {author} {\bibfnamefont {Qian}\ \bibnamefont
  {Bin}}, \bibinfo {author} {\bibfnamefont {Hui}\ \bibnamefont {Jing}},
  \bibinfo {author} {\bibfnamefont {Ying}\ \bibnamefont {Wu}}, \bibinfo
  {author} {\bibfnamefont {Franco}\ \bibnamefont {Nori}}, \ and\ \bibinfo
  {author} {\bibfnamefont {Xin-You}\ \bibnamefont {L\"u}},\ }\bibfield  {title}
  {\enquote {\bibinfo {title} {Nonreciprocal bundle emissions of quantum
  entangled pairs},}\ }\href {\doibase 10.1103/PhysRevLett.133.043601}
  {\bibfield  {journal} {\bibinfo  {journal} {Phys. Rev. Lett.}\ }\textbf
  {\bibinfo {volume} {133}},\ \bibinfo {pages} {043601} (\bibinfo {year}
  {2024})}\BibitemShut {NoStop}%
\bibitem [{\citenamefont {S{\'a}nchez~Mu{\~n}oz}\ \emph
  {et~al.}(2018)\citenamefont {S{\'a}nchez~Mu{\~n}oz}, \citenamefont {Laussy},
  \citenamefont {Valle}, \citenamefont {Tejedor},\ and\ \citenamefont
  {Gonz{\'a}lez-Tudela}}]{Optics_14-26}%
  \BibitemOpen
  \bibfield  {author} {\bibinfo {author} {\bibfnamefont {Carlos}\ \bibnamefont
  {S{\'a}nchez~Mu{\~n}oz}}, \bibinfo {author} {\bibfnamefont {Fabrice~P}\
  \bibnamefont {Laussy}}, \bibinfo {author} {\bibfnamefont {Elena~del}\
  \bibnamefont {Valle}}, \bibinfo {author} {\bibfnamefont {Carlos}\
  \bibnamefont {Tejedor}}, \ and\ \bibinfo {author} {\bibfnamefont {Alejandro}\
  \bibnamefont {Gonz{\'a}lez-Tudela}},\ }\bibfield  {title} {\enquote {\bibinfo
  {title} {Filtering multiphoton emission from state-of-the-art cavity quantum
  electrodynamics},}\ }\href {\doibase 10.1103/PhysRevLett.121.043602}
  {\bibfield  {journal} {\bibinfo  {journal} {Optica}\ }\textbf {\bibinfo
  {volume} {5}},\ \bibinfo {pages} {14--26} (\bibinfo {year}
  {2018})}\BibitemShut {NoStop}%
\bibitem [{\citenamefont {Bin}\ \emph {et~al.}(2021)\citenamefont {Bin},
  \citenamefont {Wu},\ and\ \citenamefont {L\"u}}]{Tau_min_of_Bundle_Emission}%
  \BibitemOpen
  \bibfield  {author} {\bibinfo {author} {\bibfnamefont {Qian}\ \bibnamefont
  {Bin}}, \bibinfo {author} {\bibfnamefont {Ying}\ \bibnamefont {Wu}}, \ and\
  \bibinfo {author} {\bibfnamefont {Xin-You}\ \bibnamefont {L\"u}},\ }\bibfield
   {title} {\enquote {\bibinfo {title} {Parity-symmetry-protected multiphoton
  bundle emission},}\ }\href {\doibase 10.1103/PhysRevLett.127.073602}
  {\bibfield  {journal} {\bibinfo  {journal} {Phys. Rev. Lett.}\ }\textbf
  {\bibinfo {volume} {127}},\ \bibinfo {pages} {073602} (\bibinfo {year}
  {2021})}\BibitemShut {NoStop}%
\bibitem [{\citenamefont {Mu{\~n}oz}\ \emph {et~al.}(2014)\citenamefont
  {Mu{\~n}oz}, \citenamefont {del Valle}, \citenamefont {Tudela}, \citenamefont
  {M{\"u}ller}, \citenamefont {Lichtmannecker}, \citenamefont {Kaniber},
  \citenamefont {Tejedor}, \citenamefont {Finley},\ and\ \citenamefont
  {Laussy}}]{Emitters_of_N-photon_bundles}%
  \BibitemOpen
  \bibfield  {author} {\bibinfo {author} {\bibfnamefont {C.~S{\'a}nchez}\
  \bibnamefont {Mu{\~n}oz}}, \bibinfo {author} {\bibfnamefont {E.}~\bibnamefont
  {del Valle}}, \bibinfo {author} {\bibfnamefont {A.~Gonz{\'a}lez}\
  \bibnamefont {Tudela}}, \bibinfo {author} {\bibfnamefont {K.}~\bibnamefont
  {M{\"u}ller}}, \bibinfo {author} {\bibfnamefont {S.}~\bibnamefont
  {Lichtmannecker}}, \bibinfo {author} {\bibfnamefont {M.}~\bibnamefont
  {Kaniber}}, \bibinfo {author} {\bibfnamefont {C.}~\bibnamefont {Tejedor}},
  \bibinfo {author} {\bibfnamefont {J.~J.}\ \bibnamefont {Finley}}, \ and\
  \bibinfo {author} {\bibfnamefont {F.~P.}\ \bibnamefont {Laussy}},\ }\bibfield
   {title} {\enquote {\bibinfo {title} {Emitters of {$N$}-photon bundles},}\
  }\href {\doibase 10.1038/nphoton.2014.114} {\bibfield  {journal} {\bibinfo
  {journal} {Nature Photonics}\ }\textbf {\bibinfo {volume} {8}},\ \bibinfo
  {pages} {550--555} (\bibinfo {year} {2014})}\BibitemShut {NoStop}%
\bibitem [{\citenamefont {Deng}\ \emph {et~al.}(2021)\citenamefont {Deng},
  \citenamefont {Shi},\ and\ \citenamefont {Yi}}]{deng2021motional}%
  \BibitemOpen
  \bibfield  {author} {\bibinfo {author} {\bibfnamefont {Yuangang}\
  \bibnamefont {Deng}}, \bibinfo {author} {\bibfnamefont {Tao}\ \bibnamefont
  {Shi}}, \ and\ \bibinfo {author} {\bibfnamefont {Su}~\bibnamefont {Yi}},\
  }\bibfield  {title} {\enquote {\bibinfo {title} {Motional n-phonon bundle
  states of a trapped atom with clock transitions},}\ }\href {\doibase
  10.1364/PRJ.427062} {\bibfield  {journal} {\bibinfo  {journal} {Photon.
  Res.}\ }\textbf {\bibinfo {volume} {9}},\ \bibinfo {pages} {1289--1299}
  (\bibinfo {year} {2021})}\BibitemShut {NoStop}%
\bibitem [{\citenamefont {Hoffman}\ \emph {et~al.}(2011)\citenamefont
  {Hoffman}, \citenamefont {Srinivasan}, \citenamefont {Schmidt}, \citenamefont
  {Spietz}, \citenamefont {Aumentado}, \citenamefont {T\"ureci},\ and\
  \citenamefont {Houck}}]{circuit_PB_1}%
  \BibitemOpen
  \bibfield  {author} {\bibinfo {author} {\bibfnamefont {A.~J.}\ \bibnamefont
  {Hoffman}}, \bibinfo {author} {\bibfnamefont {S.~J.}\ \bibnamefont
  {Srinivasan}}, \bibinfo {author} {\bibfnamefont {S.}~\bibnamefont {Schmidt}},
  \bibinfo {author} {\bibfnamefont {L.}~\bibnamefont {Spietz}}, \bibinfo
  {author} {\bibfnamefont {J.}~\bibnamefont {Aumentado}}, \bibinfo {author}
  {\bibfnamefont {H.~E.}\ \bibnamefont {T\"ureci}}, \ and\ \bibinfo {author}
  {\bibfnamefont {A.~A.}\ \bibnamefont {Houck}},\ }\bibfield  {title} {\enquote
  {\bibinfo {title} {Dispersive photon blockade in a superconducting
  circuit},}\ }\href {\doibase 10.1103/PhysRevLett.107.053602} {\bibfield
  {journal} {\bibinfo  {journal} {Phys. Rev. Lett.}\ }\textbf {\bibinfo
  {volume} {107}},\ \bibinfo {pages} {053602} (\bibinfo {year}
  {2011})}\BibitemShut {NoStop}%
\bibitem [{\citenamefont {Lang}\ \emph {et~al.}(2011)\citenamefont {Lang},
  \citenamefont {Bozyigit}, \citenamefont {Eichler}, \citenamefont {Steffen},
  \citenamefont {Fink}, \citenamefont {Abdumalikov}, \citenamefont {Baur},
  \citenamefont {Filipp}, \citenamefont {da~Silva}, \citenamefont {Blais},\
  and\ \citenamefont {Wallraff}}]{circuit_PB_2}%
  \BibitemOpen
  \bibfield  {author} {\bibinfo {author} {\bibfnamefont {C.}~\bibnamefont
  {Lang}}, \bibinfo {author} {\bibfnamefont {D.}~\bibnamefont {Bozyigit}},
  \bibinfo {author} {\bibfnamefont {C.}~\bibnamefont {Eichler}}, \bibinfo
  {author} {\bibfnamefont {L.}~\bibnamefont {Steffen}}, \bibinfo {author}
  {\bibfnamefont {J.~M.}\ \bibnamefont {Fink}}, \bibinfo {author}
  {\bibfnamefont {A.~A.}\ \bibnamefont {Abdumalikov}}, \bibinfo {author}
  {\bibfnamefont {M.}~\bibnamefont {Baur}}, \bibinfo {author} {\bibfnamefont
  {S.}~\bibnamefont {Filipp}}, \bibinfo {author} {\bibfnamefont {M.~P.}\
  \bibnamefont {da~Silva}}, \bibinfo {author} {\bibfnamefont {A.}~\bibnamefont
  {Blais}}, \ and\ \bibinfo {author} {\bibfnamefont {A.}~\bibnamefont
  {Wallraff}},\ }\bibfield  {title} {\enquote {\bibinfo {title} {Observation of
  resonant photon blockade at microwave frequencies using correlation function
  measurements},}\ }\href {\doibase 10.1103/PhysRevLett.106.243601} {\bibfield
  {journal} {\bibinfo  {journal} {Phys. Rev. Lett.}\ }\textbf {\bibinfo
  {volume} {106}},\ \bibinfo {pages} {243601} (\bibinfo {year}
  {2011})}\BibitemShut {NoStop}%
\bibitem [{\citenamefont {Vaneph}\ \emph {et~al.}(2018)\citenamefont {Vaneph},
  \citenamefont {Morvan}, \citenamefont {Aiello}, \citenamefont {F\'echant},
  \citenamefont {Aprili}, \citenamefont {Gabelli},\ and\ \citenamefont
  {Est\`eve}}]{circuit_PB_3}%
  \BibitemOpen
  \bibfield  {author} {\bibinfo {author} {\bibfnamefont {Cyril}\ \bibnamefont
  {Vaneph}}, \bibinfo {author} {\bibfnamefont {Alexis}\ \bibnamefont {Morvan}},
  \bibinfo {author} {\bibfnamefont {Gianluca}\ \bibnamefont {Aiello}}, \bibinfo
  {author} {\bibfnamefont {Mathieu}\ \bibnamefont {F\'echant}}, \bibinfo
  {author} {\bibfnamefont {Marco}\ \bibnamefont {Aprili}}, \bibinfo {author}
  {\bibfnamefont {Julien}\ \bibnamefont {Gabelli}}, \ and\ \bibinfo {author}
  {\bibfnamefont {J\'er\^ome}\ \bibnamefont {Est\`eve}},\ }\bibfield  {title}
  {\enquote {\bibinfo {title} {Observation of the unconventional photon
  blockade in the microwave domain},}\ }\href {\doibase
  10.1103/PhysRevLett.121.043602} {\bibfield  {journal} {\bibinfo  {journal}
  {Phys. Rev. Lett.}\ }\textbf {\bibinfo {volume} {121}},\ \bibinfo {pages}
  {043602} (\bibinfo {year} {2018})}\BibitemShut {NoStop}%
\bibitem [{\citenamefont {Chu}\ \emph {et~al.}(2018)\citenamefont {Chu},
  \citenamefont {Kharel}, \citenamefont {Yoon}, \citenamefont {Frunzio},
  \citenamefont {Rakich},\ and\ \citenamefont
  {Schoelkopf}}]{quantum_metrology}%
  \BibitemOpen
  \bibfield  {author} {\bibinfo {author} {\bibfnamefont {Yiwen}\ \bibnamefont
  {Chu}}, \bibinfo {author} {\bibfnamefont {Prashanta}\ \bibnamefont {Kharel}},
  \bibinfo {author} {\bibfnamefont {Taekwan}\ \bibnamefont {Yoon}}, \bibinfo
  {author} {\bibfnamefont {Luigi}\ \bibnamefont {Frunzio}}, \bibinfo {author}
  {\bibfnamefont {Peter~T.}\ \bibnamefont {Rakich}}, \ and\ \bibinfo {author}
  {\bibfnamefont {Robert~J.}\ \bibnamefont {Schoelkopf}},\ }\bibfield  {title}
  {\enquote {\bibinfo {title} {Creation and control of multi-phonon {Fock}
  states in a bulk acoustic-wave resonator},}\ }\href {\doibase
  10.1038/s41586-018-0717-7} {\bibfield  {journal} {\bibinfo  {journal}
  {Nature}\ }\textbf {\bibinfo {volume} {563}},\ \bibinfo {pages} {666--670}
  (\bibinfo {year} {2018})}\BibitemShut {NoStop}%
\bibitem [{\citenamefont {Ding}\ \emph {et~al.}(2013)\citenamefont {Ding},
  \citenamefont {Zhou}, \citenamefont {Shi},\ and\ \citenamefont
  {Guo}}]{quantum_image}%
  \BibitemOpen
  \bibfield  {author} {\bibinfo {author} {\bibfnamefont {Dong-Sheng}\
  \bibnamefont {Ding}}, \bibinfo {author} {\bibfnamefont {Zhi-Yuan}\
  \bibnamefont {Zhou}}, \bibinfo {author} {\bibfnamefont {Bao-Sen}\
  \bibnamefont {Shi}}, \ and\ \bibinfo {author} {\bibfnamefont {Guang-Can}\
  \bibnamefont {Guo}},\ }\bibfield  {title} {\enquote {\bibinfo {title}
  {Single-photon-level quantum image memory based on cold atomic ensembles},}\
  }\href {\doibase 10.1038/ncomms3527} {\bibfield  {journal} {\bibinfo
  {journal} {Nature Communications}\ }\textbf {\bibinfo {volume} {4}},\
  \bibinfo {pages} {2527} (\bibinfo {year} {2013})}\BibitemShut {NoStop}%
\bibitem [{\citenamefont {Harrington}\ \emph {et~al.}(2022)\citenamefont
  {Harrington}, \citenamefont {Mueller},\ and\ \citenamefont
  {Murch}}]{quantum_information_processing_1}%
  \BibitemOpen
  \bibfield  {author} {\bibinfo {author} {\bibfnamefont {Patrick~M.}\
  \bibnamefont {Harrington}}, \bibinfo {author} {\bibfnamefont {Erich~J.}\
  \bibnamefont {Mueller}}, \ and\ \bibinfo {author} {\bibfnamefont {Kater~W.}\
  \bibnamefont {Murch}},\ }\bibfield  {title} {\enquote {\bibinfo {title}
  {Engineered dissipation for quantum information science},}\ }\href {\doibase
  10.1038/s42254-022-00494-8} {\bibfield  {journal} {\bibinfo  {journal}
  {Nature Reviews Physics}\ }\textbf {\bibinfo {volume} {4}},\ \bibinfo {pages}
  {660--671} (\bibinfo {year} {2022})}\BibitemShut {NoStop}%
\bibitem [{\citenamefont {Souquet}\ \emph {et~al.}(2014)\citenamefont
  {Souquet}, \citenamefont {Woolley}, \citenamefont {Gabelli}, \citenamefont
  {Simon},\ and\ \citenamefont {Clerk}}]{quantum_information_processing_2}%
  \BibitemOpen
  \bibfield  {author} {\bibinfo {author} {\bibfnamefont {J.-R.}\ \bibnamefont
  {Souquet}}, \bibinfo {author} {\bibfnamefont {M.~J.}\ \bibnamefont
  {Woolley}}, \bibinfo {author} {\bibfnamefont {J.}~\bibnamefont {Gabelli}},
  \bibinfo {author} {\bibfnamefont {P.}~\bibnamefont {Simon}}, \ and\ \bibinfo
  {author} {\bibfnamefont {A.~A.}\ \bibnamefont {Clerk}},\ }\bibfield  {title}
  {\enquote {\bibinfo {title} {Photon-assisted tunnelling with nonclassical
  light},}\ }\href {\doibase 10.1038/ncomms6562} {\bibfield  {journal}
  {\bibinfo  {journal} {Nature Communications}\ }\textbf {\bibinfo {volume}
  {5}},\ \bibinfo {pages} {5562} (\bibinfo {year} {2014})}\BibitemShut
  {NoStop}%
\bibitem [{\citenamefont {Galinskiy}\ \emph {et~al.}(2024)\citenamefont
  {Galinskiy}, \citenamefont {Enzian}, \citenamefont {Parniak},\ and\
  \citenamefont {Polzik}}]{Correlations_between_Photons_and_Phonons}%
  \BibitemOpen
  \bibfield  {author} {\bibinfo {author} {\bibfnamefont {Ivan}\ \bibnamefont
  {Galinskiy}}, \bibinfo {author} {\bibfnamefont {Georg}\ \bibnamefont
  {Enzian}}, \bibinfo {author} {\bibfnamefont {Micha\l{}}\ \bibnamefont
  {Parniak}}, \ and\ \bibinfo {author} {\bibfnamefont {Eugene~S.}\ \bibnamefont
  {Polzik}},\ }\bibfield  {title} {\enquote {\bibinfo {title} {Nonclassical
  correlations between photons and phonons of center-of-mass motion of a
  mechanical oscillator},}\ }\href {\doibase 10.1103/PhysRevLett.133.173605}
  {\bibfield  {journal} {\bibinfo  {journal} {Phys. Rev. Lett.}\ }\textbf
  {\bibinfo {volume} {133}},\ \bibinfo {pages} {173605} (\bibinfo {year}
  {2024})}\BibitemShut {NoStop}%
\bibitem [{\citenamefont {Rasmussen}\ \emph {et~al.}(2024)\citenamefont
  {Rasmussen}, \citenamefont {Álvaro Rodríguez~Echarri}, \citenamefont
  {Cox},\ and\ \citenamefont {de~Abajo}}]{photon_pairs_by_free_electrons}%
  \BibitemOpen
  \bibfield  {author} {\bibinfo {author} {\bibfnamefont {Theis~P.}\
  \bibnamefont {Rasmussen}}, \bibinfo {author} {\bibnamefont {Álvaro
  Rodríguez~Echarri}}, \bibinfo {author} {\bibfnamefont {Joel~D.}\
  \bibnamefont {Cox}}, \ and\ \bibinfo {author} {\bibfnamefont
  {F.~Javier~García}\ \bibnamefont {de~Abajo}},\ }\bibfield  {title} {\enquote
  {\bibinfo {title} {Generation of entangled waveguided photon pairs by free
  electrons},}\ }\href {\doibase 10.1126/sciadv.adn6312} {\bibfield  {journal}
  {\bibinfo  {journal} {Science Advances}\ }\textbf {\bibinfo {volume} {10}},\
  \bibinfo {pages} {eadn6312} (\bibinfo {year} {2024})}\BibitemShut {NoStop}%
\bibitem [{\citenamefont {Li}\ \emph {et~al.}(2020{\natexlab{a}})\citenamefont
  {Li}, \citenamefont {Zhou}, \citenamefont {Gao},\ and\ \citenamefont
  {Nori}}]{Hybrid_quantum_resource_1}%
  \BibitemOpen
  \bibfield  {author} {\bibinfo {author} {\bibfnamefont {Peng-Bo}\ \bibnamefont
  {Li}}, \bibinfo {author} {\bibfnamefont {Yuan}\ \bibnamefont {Zhou}},
  \bibinfo {author} {\bibfnamefont {Wei-Bo}\ \bibnamefont {Gao}}, \ and\
  \bibinfo {author} {\bibfnamefont {Franco}\ \bibnamefont {Nori}},\ }\bibfield
  {title} {\enquote {\bibinfo {title} {Enhancing spin-phonon and spin-spin
  interactions using linear resources in a hybrid quantum system},}\ }\href
  {\doibase 10.1103/PhysRevLett.125.153602} {\bibfield  {journal} {\bibinfo
  {journal} {Phys. Rev. Lett.}\ }\textbf {\bibinfo {volume} {125}},\ \bibinfo
  {pages} {153602} (\bibinfo {year} {2020}{\natexlab{a}})}\BibitemShut
  {NoStop}%
\bibitem [{\citenamefont {Fux}\ \emph {et~al.}(2024)\citenamefont {Fux},
  \citenamefont {Tirrito}, \citenamefont {Dalmonte},\ and\ \citenamefont
  {Fazio}}]{Hybrid_quantum_resource_2}%
  \BibitemOpen
  \bibfield  {author} {\bibinfo {author} {\bibfnamefont {Gerald~E.}\
  \bibnamefont {Fux}}, \bibinfo {author} {\bibfnamefont {Emanuele}\
  \bibnamefont {Tirrito}}, \bibinfo {author} {\bibfnamefont {Marcello}\
  \bibnamefont {Dalmonte}}, \ and\ \bibinfo {author} {\bibfnamefont {Rosario}\
  \bibnamefont {Fazio}},\ }\bibfield  {title} {\enquote {\bibinfo {title}
  {Entanglement -- nonstabilizerness separation in hybrid quantum circuits},}\
  }\href {\doibase 10.1103/PhysRevResearch.6.L042030} {\bibfield  {journal}
  {\bibinfo  {journal} {Phys. Rev. Res.}\ }\textbf {\bibinfo {volume} {6}},\
  \bibinfo {pages} {L042030} (\bibinfo {year} {2024})}\BibitemShut {NoStop}%
\bibitem [{\citenamefont {Zhu}\ \emph {et~al.}(2026)\citenamefont {Zhu},
  \citenamefont {Yue}, \citenamefont {Guo}, \citenamefont {Tu},\ and\
  \citenamefont {Li}}]{Hybrid_quantum_resource_4}%
  \BibitemOpen
  \bibfield  {author} {\bibinfo {author} {\bibfnamefont {Xing-Yu}\ \bibnamefont
  {Zhu}}, \bibinfo {author} {\bibfnamefont {Zhu-Cheng}\ \bibnamefont {Yue}},
  \bibinfo {author} {\bibfnamefont {Guang-Can}\ \bibnamefont {Guo}}, \bibinfo
  {author} {\bibfnamefont {Tao}\ \bibnamefont {Tu}}, \ and\ \bibinfo {author}
  {\bibfnamefont {Chuan-Feng}\ \bibnamefont {Li}},\ }\bibfield  {title}
  {\enquote {\bibinfo {title} {Geometric protected quantum bus in a hybrid
  superconducting-spin architecture},}\ }\href {\doibase 10.1103/dftf-g4kc}
  {\bibfield  {journal} {\bibinfo  {journal} {Phys. Rev. Appl.}\ }\textbf
  {\bibinfo {volume} {25}},\ \bibinfo {pages} {044010} (\bibinfo {year}
  {2026})}\BibitemShut {NoStop}%
\bibitem [{\citenamefont {Oppliger}\ \emph {et~al.}(2026)\citenamefont
  {Oppliger}, \citenamefont {Jang}, \citenamefont {Tarascio}, \citenamefont
  {Palma}, \citenamefont {Reichl}, \citenamefont {Wegscheider}, \citenamefont
  {Maisi}, \citenamefont {Zumbühl},\ and\ \citenamefont
  {Scarlino}}]{Hybrid_quantum_resource_5}%
  \BibitemOpen
  \bibfield  {author} {\bibinfo {author} {\bibfnamefont {Fabian}\ \bibnamefont
  {Oppliger}}, \bibinfo {author} {\bibfnamefont {Wonjin}\ \bibnamefont {Jang}},
  \bibinfo {author} {\bibfnamefont {Aldo}\ \bibnamefont {Tarascio}}, \bibinfo
  {author} {\bibfnamefont {Franco~De}\ \bibnamefont {Palma}}, \bibinfo {author}
  {\bibfnamefont {Christian}\ \bibnamefont {Reichl}}, \bibinfo {author}
  {\bibfnamefont {Werner}\ \bibnamefont {Wegscheider}}, \bibinfo {author}
  {\bibfnamefont {Ville~F.}\ \bibnamefont {Maisi}}, \bibinfo {author}
  {\bibfnamefont {Dominik}\ \bibnamefont {Zumbühl}}, \ and\ \bibinfo {author}
  {\bibfnamefont {Pasquale}\ \bibnamefont {Scarlino}},\ }\bibfield  {title}
  {\enquote {\bibinfo {title} {Tunable high-efficiency microwave photon
  detector based on a double quantum dot coupled to a superconducting
  high-impedance cavity},}\ }\href {\doibase 10.1126/sciadv.aeb9784} {\bibfield
   {journal} {\bibinfo  {journal} {Science Advances}\ }\textbf {\bibinfo
  {volume} {12}},\ \bibinfo {pages} {eaeb9784} (\bibinfo {year}
  {2026})}\BibitemShut {NoStop}%
\bibitem [{\citenamefont {Semnani}\ \emph {et~al.}(2026)\citenamefont
  {Semnani}, \citenamefont {Venuturumilli}, \citenamefont {Soltani},
  \citenamefont {Adhikary}, \citenamefont {Pasharavesh}, \citenamefont
  {Videnov}, \citenamefont {Anderson}, \citenamefont {Sarkar}, \citenamefont
  {Muthu},\ and\ \citenamefont {Bajcsy}}]{Hybrid_quantum_resource_6}%
  \BibitemOpen
  \bibfield  {author} {\bibinfo {author} {\bibfnamefont {Behrooz}\ \bibnamefont
  {Semnani}}, \bibinfo {author} {\bibfnamefont {Sai~Sreesh}\ \bibnamefont
  {Venuturumilli}}, \bibinfo {author} {\bibfnamefont {Mohammad}\ \bibnamefont
  {Soltani}}, \bibinfo {author} {\bibfnamefont {Pratik}\ \bibnamefont
  {Adhikary}}, \bibinfo {author} {\bibfnamefont {Abdolreza}\ \bibnamefont
  {Pasharavesh}}, \bibinfo {author} {\bibfnamefont {Nikolay}\ \bibnamefont
  {Videnov}}, \bibinfo {author} {\bibfnamefont {Paul}\ \bibnamefont
  {Anderson}}, \bibinfo {author} {\bibfnamefont {Supratik}\ \bibnamefont
  {Sarkar}}, \bibinfo {author} {\bibfnamefont {Vinodh Raj~Rajagopal}\
  \bibnamefont {Muthu}}, \ and\ \bibinfo {author} {\bibfnamefont {Michal}\
  \bibnamefont {Bajcsy}},\ }\bibfield  {title} {\enquote {\bibinfo {title}
  {Probing individual quantum emitters in bulk semiconductors via photonic
  nanojets},}\ }\href {\doibase 10.1126/sciadv.aea5936} {\bibfield  {journal}
  {\bibinfo  {journal} {Science Advances}\ }\textbf {\bibinfo {volume} {12}},\
  \bibinfo {pages} {eaea5936} (\bibinfo {year} {2026})}\BibitemShut {NoStop}%
\bibitem [{\citenamefont {Laha}\ \emph {et~al.}(2025)\citenamefont {Laha},
  \citenamefont {Yasir},\ and\ \citenamefont {van
  Loock}}]{tripartite_JC_two_quantized_harmonic_oscillators}%
  \BibitemOpen
  \bibfield  {author} {\bibinfo {author} {\bibfnamefont {Pradip}\ \bibnamefont
  {Laha}}, \bibinfo {author} {\bibfnamefont {P.~A.~Ameen}\ \bibnamefont
  {Yasir}}, \ and\ \bibinfo {author} {\bibfnamefont {Peter}\ \bibnamefont {van
  Loock}},\ }\bibfield  {title} {\enquote {\bibinfo {title} {Tripartite
  multiphoton jaynes-cummings model: Analytical solution and wigner
  nonclassicalities},}\ }\href {\doibase 10.1103/PhysRevA.111.013710}
  {\bibfield  {journal} {\bibinfo  {journal} {Phys. Rev. A}\ }\textbf {\bibinfo
  {volume} {111}},\ \bibinfo {pages} {013710} (\bibinfo {year}
  {2025})}\BibitemShut {NoStop}%
\bibitem [{\citenamefont {Lai}\ \emph {et~al.}(2024)\citenamefont {Lai},
  \citenamefont {Miranowicz},\ and\ \citenamefont
  {Nori}}]{tripartite_Topological_Fizeau_Drag_photon}%
  \BibitemOpen
  \bibfield  {author} {\bibinfo {author} {\bibfnamefont {Deng-Gao}\
  \bibnamefont {Lai}}, \bibinfo {author} {\bibfnamefont {Adam}\ \bibnamefont
  {Miranowicz}}, \ and\ \bibinfo {author} {\bibfnamefont {Franco}\ \bibnamefont
  {Nori}},\ }\bibfield  {title} {\enquote {\bibinfo {title} {Nonreciprocal
  topological phonon transfer independent of both device mass and
  exceptional-point encircling direction},}\ }\href {\doibase
  10.1103/PhysRevLett.132.243602} {\bibfield  {journal} {\bibinfo  {journal}
  {Phys. Rev. Lett.}\ }\textbf {\bibinfo {volume} {132}},\ \bibinfo {pages}
  {243602} (\bibinfo {year} {2024})}\BibitemShut {NoStop}%
\bibitem [{\citenamefont {Lau}\ \emph {et~al.}(2025)\citenamefont {Lau},
  \citenamefont {Qiao}, \citenamefont {Clerk},\ and\ \citenamefont
  {Zhong}}]{tripartite_cavity_atom_photon}%
  \BibitemOpen
  \bibfield  {author} {\bibinfo {author} {\bibfnamefont {Hoi-Kwan}\
  \bibnamefont {Lau}}, \bibinfo {author} {\bibfnamefont {Hong}\ \bibnamefont
  {Qiao}}, \bibinfo {author} {\bibfnamefont {Aashish~A.}\ \bibnamefont
  {Clerk}}, \ and\ \bibinfo {author} {\bibfnamefont {Tian}\ \bibnamefont
  {Zhong}},\ }\bibfield  {title} {\enquote {\bibinfo {title} {Efficient in situ
  generation of photon-memory entanglement in a nonlinear cavity},}\ }\href
  {\doibase 10.1103/PhysRevLett.134.053602} {\bibfield  {journal} {\bibinfo
  {journal} {Phys. Rev. Lett.}\ }\textbf {\bibinfo {volume} {134}},\ \bibinfo
  {pages} {053602} (\bibinfo {year} {2025})}\BibitemShut {NoStop}%
\bibitem [{\citenamefont {Wang}\ \emph
  {et~al.}(2025{\natexlab{a}})\citenamefont {Wang}, \citenamefont {Yang},
  \citenamefont {Yang}, \citenamefont {Zhang}, \citenamefont {Luo},
  \citenamefont {Liu}, \citenamefont {Huang}, \citenamefont {Li},\ and\
  \citenamefont {Guo}}]{tripartite_genuinely_entangled_state}%
  \BibitemOpen
  \bibfield  {author} {\bibinfo {author} {\bibfnamefont {Ning-Ning}\
  \bibnamefont {Wang}}, \bibinfo {author} {\bibfnamefont {Xue}\ \bibnamefont
  {Yang}}, \bibinfo {author} {\bibfnamefont {Yan-Han}\ \bibnamefont {Yang}},
  \bibinfo {author} {\bibfnamefont {Chao}\ \bibnamefont {Zhang}}, \bibinfo
  {author} {\bibfnamefont {Ming-Xing}\ \bibnamefont {Luo}}, \bibinfo {author}
  {\bibfnamefont {Bi-Heng}\ \bibnamefont {Liu}}, \bibinfo {author}
  {\bibfnamefont {Yun-Feng}\ \bibnamefont {Huang}}, \bibinfo {author}
  {\bibfnamefont {Chuan-Feng}\ \bibnamefont {Li}}, \ and\ \bibinfo {author}
  {\bibfnamefont {Guang-Can}\ \bibnamefont {Guo}},\ }\bibfield  {title}
  {\enquote {\bibinfo {title} {Simultaneous verification of genuine
  multipartite nonlocality and full network nonlocality},}\ }\href {\doibase
  10.1103/PhysRevLett.134.080202} {\bibfield  {journal} {\bibinfo  {journal}
  {Phys. Rev. Lett.}\ }\textbf {\bibinfo {volume} {134}},\ \bibinfo {pages}
  {080202} (\bibinfo {year} {2025}{\natexlab{a}})}\BibitemShut {NoStop}%
\bibitem [{\citenamefont {Tang}\ and\ \citenamefont
  {Deng}(2024)}]{tripartite_interactions_by_Prof_Deng}%
  \BibitemOpen
  \bibfield  {author} {\bibinfo {author} {\bibfnamefont {Jing}\ \bibnamefont
  {Tang}}\ and\ \bibinfo {author} {\bibfnamefont {Yuangang}\ \bibnamefont
  {Deng}},\ }\bibfield  {title} {\enquote {\bibinfo {title} {Unveiling vacuum
  fluctuations and nonclassical states with cavity-enhanced tripartite
  interactions},}\ }\href {\doibase 10.1063/5.0210936} {\bibfield  {journal}
  {\bibinfo  {journal} {APL Photonics}\ }\textbf {\bibinfo {volume} {9}}
  (\bibinfo {year} {2024}),\ 10.1063/5.0210936}\BibitemShut {NoStop}%
\bibitem [{\citenamefont {Li}\ \emph {et~al.}(2020{\natexlab{b}})\citenamefont
  {Li}, \citenamefont {Zhao}, \citenamefont {Zhang}, \citenamefont {Liu},
  \citenamefont {Yin}, \citenamefont {Zhang}, \citenamefont {Fei},
  \citenamefont {Chen}, \citenamefont {Liu}, \citenamefont {Xu}, \citenamefont
  {Chen}, \citenamefont {Li},\ and\ \citenamefont
  {Pan}}]{tripartite_panjianwei}%
  \BibitemOpen
  \bibfield  {author} {\bibinfo {author} {\bibfnamefont {Zheng-Da}\
  \bibnamefont {Li}}, \bibinfo {author} {\bibfnamefont {Qi}~\bibnamefont
  {Zhao}}, \bibinfo {author} {\bibfnamefont {Rui}\ \bibnamefont {Zhang}},
  \bibinfo {author} {\bibfnamefont {Li-Zheng}\ \bibnamefont {Liu}}, \bibinfo
  {author} {\bibfnamefont {Xu-Fei}\ \bibnamefont {Yin}}, \bibinfo {author}
  {\bibfnamefont {Xingjian}\ \bibnamefont {Zhang}}, \bibinfo {author}
  {\bibfnamefont {Yue-Yang}\ \bibnamefont {Fei}}, \bibinfo {author}
  {\bibfnamefont {Kai}\ \bibnamefont {Chen}}, \bibinfo {author} {\bibfnamefont
  {Nai-Le}\ \bibnamefont {Liu}}, \bibinfo {author} {\bibfnamefont {Feihu}\
  \bibnamefont {Xu}}, \bibinfo {author} {\bibfnamefont {Yu-Ao}\ \bibnamefont
  {Chen}}, \bibinfo {author} {\bibfnamefont {Li}~\bibnamefont {Li}}, \ and\
  \bibinfo {author} {\bibfnamefont {Jian-Wei}\ \bibnamefont {Pan}},\ }\bibfield
   {title} {\enquote {\bibinfo {title} {Measurement-device-independent
  entanglement witness of tripartite entangled states and its applications},}\
  }\href {\doibase 10.1103/PhysRevLett.124.160503} {\bibfield  {journal}
  {\bibinfo  {journal} {Phys. Rev. Lett.}\ }\textbf {\bibinfo {volume} {124}},\
  \bibinfo {pages} {160503} (\bibinfo {year} {2020}{\natexlab{b}})}\BibitemShut
  {NoStop}%
\bibitem [{\citenamefont {Wang}\ \emph
  {et~al.}(2025{\natexlab{b}})\citenamefont {Wang}, \citenamefont {Wu},
  \citenamefont {Jiao}, \citenamefont {Lu}, \citenamefont {Zhang},
  \citenamefont {Jiang}, \citenamefont {Kuang},\ and\ \citenamefont
  {Jing}}]{tripartite_photon-phonon-magnon}%
  \BibitemOpen
  \bibfield  {author} {\bibinfo {author} {\bibfnamefont {Yan}\ \bibnamefont
  {Wang}}, \bibinfo {author} {\bibfnamefont {Jin-Lei}\ \bibnamefont {Wu}},
  \bibinfo {author} {\bibfnamefont {Ya-Feng}\ \bibnamefont {Jiao}}, \bibinfo
  {author} {\bibfnamefont {Tian-Xiang}\ \bibnamefont {Lu}}, \bibinfo {author}
  {\bibfnamefont {Hui-Lai}\ \bibnamefont {Zhang}}, \bibinfo {author}
  {\bibfnamefont {Li-Ying}\ \bibnamefont {Jiang}}, \bibinfo {author}
  {\bibfnamefont {Le-Man}\ \bibnamefont {Kuang}}, \ and\ \bibinfo {author}
  {\bibfnamefont {Hui}\ \bibnamefont {Jing}},\ }\bibfield  {title} {\enquote
  {\bibinfo {title} {Enhancing tripartite photon-phonon-magnon entanglement by
  synergizing parametric amplifications},}\ }\href {\doibase
  10.1103/PhysRevA.111.013709} {\bibfield  {journal} {\bibinfo  {journal}
  {Phys. Rev. A}\ }\textbf {\bibinfo {volume} {111}},\ \bibinfo {pages}
  {013709} (\bibinfo {year} {2025}{\natexlab{b}})}\BibitemShut {NoStop}%
\bibitem [{\citenamefont {Abdi}\ \emph {et~al.}(2015)\citenamefont {Abdi},
  \citenamefont {Pernpeintner}, \citenamefont {Gross}, \citenamefont {Huebl},\
  and\ \citenamefont {Hartmann}}]{tripartite_photons_phonons_qubit_excitation}%
  \BibitemOpen
  \bibfield  {author} {\bibinfo {author} {\bibfnamefont {Mehdi}\ \bibnamefont
  {Abdi}}, \bibinfo {author} {\bibfnamefont {Matthias}\ \bibnamefont
  {Pernpeintner}}, \bibinfo {author} {\bibfnamefont {Rudolf}\ \bibnamefont
  {Gross}}, \bibinfo {author} {\bibfnamefont {Hans}\ \bibnamefont {Huebl}}, \
  and\ \bibinfo {author} {\bibfnamefont {Michael~J.}\ \bibnamefont
  {Hartmann}},\ }\bibfield  {title} {\enquote {\bibinfo {title} {Quantum state
  engineering with circuit electromechanical three-body interactions},}\ }\href
  {\doibase 10.1103/PhysRevLett.114.173602} {\bibfield  {journal} {\bibinfo
  {journal} {Phys. Rev. Lett.}\ }\textbf {\bibinfo {volume} {114}},\ \bibinfo
  {pages} {173602} (\bibinfo {year} {2015})}\BibitemShut {NoStop}%
\bibitem [{\citenamefont {Ren}\ \emph {et~al.}(2026)\citenamefont {Ren},
  \citenamefont {Huang},\ and\ \citenamefont
  {Xiong}}]{complex_quantum_resources_1}%
  \BibitemOpen
  \bibfield  {author} {\bibinfo {author} {\bibfnamefont {Yi}~\bibnamefont
  {Ren}}, \bibinfo {author} {\bibfnamefont {Kai-Wei}\ \bibnamefont {Huang}}, \
  and\ \bibinfo {author} {\bibfnamefont {Hao}\ \bibnamefont {Xiong}},\
  }\bibfield  {title} {\enquote {\bibinfo {title} {Antibunched
  photon-magnon-pair generation via two-photon microwave driving in a
  cavity-magnon system},}\ }\href {\doibase 10.1103/tfkv-1s5p} {\bibfield
  {journal} {\bibinfo  {journal} {Phys. Rev. A}\ }\textbf {\bibinfo {volume}
  {113}},\ \bibinfo {pages} {053723} (\bibinfo {year} {2026})}\BibitemShut
  {NoStop}%
\bibitem [{\citenamefont {Lee}\ \emph {et~al.}(2018)\citenamefont {Lee},
  \citenamefont {Suh},\ and\ \citenamefont
  {Seok}}]{dissipation_nonclassical_1}%
  \BibitemOpen
  \bibfield  {author} {\bibinfo {author} {\bibfnamefont {Jae~Hoon}\
  \bibnamefont {Lee}}, \bibinfo {author} {\bibfnamefont {Junho}\ \bibnamefont
  {Suh}}, \ and\ \bibinfo {author} {\bibfnamefont {Hyojun}\ \bibnamefont
  {Seok}},\ }\bibfield  {title} {\enquote {\bibinfo {title} {Dissipation-driven
  nonclassical-state generation in optomechanics with squeezed light},}\ }\href
  {\doibase 10.1103/PhysRevA.98.043821} {\bibfield  {journal} {\bibinfo
  {journal} {Phys. Rev. A}\ }\textbf {\bibinfo {volume} {98}},\ \bibinfo
  {pages} {043821} (\bibinfo {year} {2018})}\BibitemShut {NoStop}%
\bibitem [{\citenamefont {Liu}\ \emph {et~al.}(2026)\citenamefont {Liu},
  \citenamefont {Wang}, \citenamefont {Tang}, \citenamefont {Chen},
  \citenamefont {Jing}, \citenamefont {Su},\ and\ \citenamefont
  {Nori}}]{dissipation_nonclassical_2}%
  \BibitemOpen
  \bibfield  {author} {\bibinfo {author} {\bibfnamefont {B.-B.}\ \bibnamefont
  {Liu}}, \bibinfo {author} {\bibfnamefont {D.-Y.}\ \bibnamefont {Wang}},
  \bibinfo {author} {\bibfnamefont {J.}~\bibnamefont {Tang}}, \bibinfo {author}
  {\bibfnamefont {G.}~\bibnamefont {Chen}}, \bibinfo {author} {\bibfnamefont
  {H.}~\bibnamefont {Jing}}, \bibinfo {author} {\bibfnamefont {Shi-Lei}\
  \bibnamefont {Su}}, \ and\ \bibinfo {author} {\bibfnamefont {F.}~\bibnamefont
  {Nori}},\ }\bibfield  {title} {\enquote {\bibinfo {title} {Enhancing
  nonreciprocity through squeezing-induced symmetry breaking},}\ }\href
  {\doibase 10.1103/kh36-7z76} {\bibfield  {journal} {\bibinfo  {journal}
  {Phys. Rev. Lett.}\ }\textbf {\bibinfo {volume} {136}},\ \bibinfo {pages}
  {253602} (\bibinfo {year} {2026})}\BibitemShut {NoStop}%
\bibitem [{\citenamefont {Ghasemian}(2023)}]{dissipation_nonclassical_3}%
  \BibitemOpen
  \bibfield  {author} {\bibinfo {author} {\bibfnamefont {E.}~\bibnamefont
  {Ghasemian}},\ }\bibfield  {title} {\enquote {\bibinfo {title} {Dissipative
  dynamics of optomagnonic nonclassical features via anti-{Stokes} optical
  pulses: squeezing, blockade, anti-correlation, and entanglement},}\ }\href
  {\doibase 10.1038/s41598-023-39822-y} {\bibfield  {journal} {\bibinfo
  {journal} {Scientific Reports}\ }\textbf {\bibinfo {volume} {13}},\ \bibinfo
  {pages} {12757} (\bibinfo {year} {2023})}\BibitemShut {NoStop}%
\bibitem [{\citenamefont {Sarkar}\ \emph {et~al.}(2026)\citenamefont {Sarkar},
  \citenamefont {Moitra},\ and\ \citenamefont
  {Sensarma}}]{dissipation_nonclassical_4}%
  \BibitemOpen
  \bibfield  {author} {\bibinfo {author} {\bibfnamefont {Mrinal~Kanti}\
  \bibnamefont {Sarkar}}, \bibinfo {author} {\bibfnamefont {Saranyo}\
  \bibnamefont {Moitra}}, \ and\ \bibinfo {author} {\bibfnamefont {Rajdeep}\
  \bibnamefont {Sensarma}},\ }\bibfield  {title} {\enquote {\bibinfo {title}
  {Entanglement entropy from correlation functions of scalar fields in and out
  of equilibrium},}\ }\href {\doibase 10.1103/jt4g-j6mr} {\bibfield  {journal}
  {\bibinfo  {journal} {Phys. Rev. A}\ }\textbf {\bibinfo {volume} {113}},\
  \bibinfo {pages} {062463} (\bibinfo {year} {2026})}\BibitemShut {NoStop}%
\bibitem [{\citenamefont {Reiter}\ \emph {et~al.}(2017)\citenamefont {Reiter},
  \citenamefont {S{\o}rensen}, \citenamefont {Zoller},\ and\ \citenamefont
  {Muschik}}]{quantum_engineering_1}%
  \BibitemOpen
  \bibfield  {author} {\bibinfo {author} {\bibfnamefont {F.}~\bibnamefont
  {Reiter}}, \bibinfo {author} {\bibfnamefont {A.~S.}\ \bibnamefont
  {S{\o}rensen}}, \bibinfo {author} {\bibfnamefont {P.}~\bibnamefont {Zoller}},
  \ and\ \bibinfo {author} {\bibfnamefont {C.~A.}\ \bibnamefont {Muschik}},\
  }\bibfield  {title} {\enquote {\bibinfo {title} {Dissipative quantum error
  correction and application to quantum sensing with trapped ions},}\ }\href
  {\doibase 10.1038/s41467-017-01895-5} {\bibfield  {journal} {\bibinfo
  {journal} {Nature Communications}\ }\textbf {\bibinfo {volume} {8}},\
  \bibinfo {pages} {1822} (\bibinfo {year} {2017})}\BibitemShut {NoStop}%
\bibitem [{\citenamefont {Di~Candia}\ \emph {et~al.}(2015)\citenamefont
  {Di~Candia}, \citenamefont {Pedernales}, \citenamefont {del Campo},
  \citenamefont {Solano},\ and\ \citenamefont
  {Casanova}}]{quantum_engineering_2}%
  \BibitemOpen
  \bibfield  {author} {\bibinfo {author} {\bibfnamefont {R.}~\bibnamefont
  {Di~Candia}}, \bibinfo {author} {\bibfnamefont {J.~S.}\ \bibnamefont
  {Pedernales}}, \bibinfo {author} {\bibfnamefont {A.}~\bibnamefont {del
  Campo}}, \bibinfo {author} {\bibfnamefont {E.}~\bibnamefont {Solano}}, \ and\
  \bibinfo {author} {\bibfnamefont {J.}~\bibnamefont {Casanova}},\ }\bibfield
  {title} {\enquote {\bibinfo {title} {Quantum simulation of dissipative
  processes without reservoir engineering},}\ }\href {\doibase
  10.1038/srep09981} {\bibfield  {journal} {\bibinfo  {journal} {Scientific
  Reports}\ }\textbf {\bibinfo {volume} {5}},\ \bibinfo {pages} {9981}
  (\bibinfo {year} {2015})}\BibitemShut {NoStop}%
\bibitem [{\citenamefont {Zhao}\ \emph {et~al.}(2025)\citenamefont {Zhao},
  \citenamefont {Tian}, \citenamefont {Ye}, \citenamefont {Wu}, \citenamefont
  {Zhao}, \citenamefont {Chi}, \citenamefont {Tian}, \citenamefont {Yao},
  \citenamefont {Hu}, \citenamefont {Chen},\ and\ \citenamefont
  {Chen}}]{quantum_engineering_3}%
  \BibitemOpen
  \bibfield  {author} {\bibinfo {author} {\bibfnamefont {Yajuan}\ \bibnamefont
  {Zhao}}, \bibinfo {author} {\bibfnamefont {Ye}~\bibnamefont {Tian}}, \bibinfo
  {author} {\bibfnamefont {Jilai}\ \bibnamefont {Ye}}, \bibinfo {author}
  {\bibfnamefont {Yue}\ \bibnamefont {Wu}}, \bibinfo {author} {\bibfnamefont
  {Zihan}\ \bibnamefont {Zhao}}, \bibinfo {author} {\bibfnamefont {Zhihao}\
  \bibnamefont {Chi}}, \bibinfo {author} {\bibfnamefont {Tian}\ \bibnamefont
  {Tian}}, \bibinfo {author} {\bibfnamefont {Hepeng}\ \bibnamefont {Yao}},
  \bibinfo {author} {\bibfnamefont {Jiazhong}\ \bibnamefont {Hu}}, \bibinfo
  {author} {\bibfnamefont {Yu}~\bibnamefont {Chen}}, \ and\ \bibinfo {author}
  {\bibfnamefont {Wenlan}\ \bibnamefont {Chen}},\ }\bibfield  {title} {\enquote
  {\bibinfo {title} {Universal dissipative dynamics in strongly correlated
  quantum gases},}\ }\href {\doibase 10.1038/s41567-025-02800-4} {\bibfield
  {journal} {\bibinfo  {journal} {Nature Physics}\ }\textbf {\bibinfo {volume}
  {21}},\ \bibinfo {pages} {530--535} (\bibinfo {year} {2025})}\BibitemShut
  {NoStop}%
\bibitem [{\citenamefont {T{\'o}th}\ \emph {et~al.}(2017)\citenamefont
  {T{\'o}th}, \citenamefont {Bernier}, \citenamefont {Nunnenkamp},
  \citenamefont {Feofanov},\ and\ \citenamefont
  {Kippenberg}}]{quantum_engineering_4}%
  \BibitemOpen
  \bibfield  {author} {\bibinfo {author} {\bibfnamefont {L.~D.}\ \bibnamefont
  {T{\'o}th}}, \bibinfo {author} {\bibfnamefont {N.~R.}\ \bibnamefont
  {Bernier}}, \bibinfo {author} {\bibfnamefont {A.}~\bibnamefont {Nunnenkamp}},
  \bibinfo {author} {\bibfnamefont {A.~K.}\ \bibnamefont {Feofanov}}, \ and\
  \bibinfo {author} {\bibfnamefont {T.~J.}\ \bibnamefont {Kippenberg}},\
  }\bibfield  {title} {\enquote {\bibinfo {title} {A dissipative quantum
  reservoir for microwave light using a mechanical oscillator},}\ }\href
  {\doibase 10.1038/nphys4121} {\bibfield  {journal} {\bibinfo  {journal}
  {Nature Physics}\ }\textbf {\bibinfo {volume} {13}},\ \bibinfo {pages}
  {787--793} (\bibinfo {year} {2017})}\BibitemShut {NoStop}%
\bibitem [{\citenamefont {Mollow}(1969)}]{Mollow_Triplet_origin}%
  \BibitemOpen
  \bibfield  {author} {\bibinfo {author} {\bibfnamefont {B.~R.}\ \bibnamefont
  {Mollow}},\ }\bibfield  {title} {\enquote {\bibinfo {title} {Power spectrum
  of light scattered by two-level systems},}\ }\href {\doibase
  10.1103/PhysRev.188.1969} {\bibfield  {journal} {\bibinfo  {journal} {Phys.
  Rev.}\ }\textbf {\bibinfo {volume} {188}},\ \bibinfo {pages} {1969--1975}
  (\bibinfo {year} {1969})}\BibitemShut {NoStop}%
\bibitem [{\citenamefont {Bin}\ \emph {et~al.}(2020)\citenamefont {Bin},
  \citenamefont {L\"u}, \citenamefont {Laussy}, \citenamefont {Nori},\ and\
  \citenamefont {Wu}}]{N-Phonon_Bundle_Emission_via_the_Stokes_Process}%
  \BibitemOpen
  \bibfield  {author} {\bibinfo {author} {\bibfnamefont {Qian}\ \bibnamefont
  {Bin}}, \bibinfo {author} {\bibfnamefont {Xin-You}\ \bibnamefont {L\"u}},
  \bibinfo {author} {\bibfnamefont {Fabrice~P.}\ \bibnamefont {Laussy}},
  \bibinfo {author} {\bibfnamefont {Franco}\ \bibnamefont {Nori}}, \ and\
  \bibinfo {author} {\bibfnamefont {Ying}\ \bibnamefont {Wu}},\ }\bibfield
  {title} {\enquote {\bibinfo {title} {N-phonon bundle emission via the stokes
  process},}\ }\href {\doibase 10.1103/PhysRevLett.124.053601} {\bibfield
  {journal} {\bibinfo  {journal} {Phys. Rev. Lett.}\ }\textbf {\bibinfo
  {volume} {124}},\ \bibinfo {pages} {053601} (\bibinfo {year}
  {2020})}\BibitemShut {NoStop}%
\bibitem [{\citenamefont {Xiong}\ \emph {et~al.}(2025)\citenamefont {Xiong},
  \citenamefont {Bin}, \citenamefont {Chao}, \citenamefont {Liu},\ and\
  \citenamefont {L\"u}}]{Two_photon_decay}%
  \BibitemOpen
  \bibfield  {author} {\bibinfo {author} {\bibfnamefont {Biao}\ \bibnamefont
  {Xiong}}, \bibinfo {author} {\bibfnamefont {Qian}\ \bibnamefont {Bin}},
  \bibinfo {author} {\bibfnamefont {Shi-Lei}\ \bibnamefont {Chao}}, \bibinfo
  {author} {\bibfnamefont {Ji-Bing}\ \bibnamefont {Liu}}, \ and\ \bibinfo
  {author} {\bibfnamefont {Xin-You}\ \bibnamefont {L\"u}},\ }\bibfield  {title}
  {\enquote {\bibinfo {title} {Two-photon decay enhanced even photon bundle
  emission},}\ }\href {\doibase 10.1103/PhysRevResearch.7.013238} {\bibfield
  {journal} {\bibinfo  {journal} {Phys. Rev. Res.}\ }\textbf {\bibinfo {volume}
  {7}},\ \bibinfo {pages} {013238} (\bibinfo {year} {2025})}\BibitemShut
  {NoStop}%
\bibitem [{\citenamefont {Yang}\ \emph {et~al.}(2026)\citenamefont {Yang},
  \citenamefont {Chen},\ and\ \citenamefont
  {Zhong}}]{phonon_and_photon_phonon_pairs}%
  \BibitemOpen
  \bibfield  {author} {\bibinfo {author} {\bibfnamefont {Zhe-Qi}\ \bibnamefont
  {Yang}}, \bibinfo {author} {\bibfnamefont {Lei}\ \bibnamefont {Chen}}, \ and\
  \bibinfo {author} {\bibfnamefont {Zhi-Rong}\ \bibnamefont {Zhong}},\
  }\bibfield  {title} {\enquote {\bibinfo {title} {Tunable bundle emissions of
  $n$-phonon and photon-phonon pairs via the barnett effect},}\ }\href
  {\doibase 10.1103/y4v3-yn6v} {\bibfield  {journal} {\bibinfo  {journal}
  {Phys. Rev. A}\ }\textbf {\bibinfo {volume} {113}},\ \bibinfo {pages}
  {063722} (\bibinfo {year} {2026})}\BibitemShut {NoStop}%
\bibitem [{\citenamefont {Peng}\ and\ \citenamefont {Deng}(2022)}]{pengzhong}%
  \BibitemOpen
  \bibfield  {author} {\bibinfo {author} {\bibfnamefont {Zhong}\ \bibnamefont
  {Peng}}\ and\ \bibinfo {author} {\bibfnamefont {Yuangang}\ \bibnamefont
  {Deng}},\ }\bibfield  {title} {\enquote {\bibinfo {title} {Nonclassical
  correlated deterministic single-photon pairs for a trapped atom in bimodal
  cavities},}\ }\href {\doibase 10.1103/PhysRevA.105.043711} {\bibfield
  {journal} {\bibinfo  {journal} {Phys. Rev. A}\ }\textbf {\bibinfo {volume}
  {105}},\ \bibinfo {pages} {043711} (\bibinfo {year} {2022})}\BibitemShut
  {NoStop}%
\bibitem [{\citenamefont {Miranowicz}\ \emph {et~al.}(2010)\citenamefont
  {Miranowicz}, \citenamefont {Bartkowiak}, \citenamefont {Wang}, \citenamefont
  {Liu},\ and\ \citenamefont {Nori}}]{Cauchy_Schwarz_inequality}%
  \BibitemOpen
  \bibfield  {author} {\bibinfo {author} {\bibfnamefont {Adam}\ \bibnamefont
  {Miranowicz}}, \bibinfo {author} {\bibfnamefont {Monika}\ \bibnamefont
  {Bartkowiak}}, \bibinfo {author} {\bibfnamefont {Xiaoguang}\ \bibnamefont
  {Wang}}, \bibinfo {author} {\bibfnamefont {Yu-xi}\ \bibnamefont {Liu}}, \
  and\ \bibinfo {author} {\bibfnamefont {Franco}\ \bibnamefont {Nori}},\
  }\bibfield  {title} {\enquote {\bibinfo {title} {Testing nonclassicality in
  multimode fields: A unified derivation of classical inequalities},}\ }\href
  {\doibase 10.1103/PhysRevA.82.013824} {\bibfield  {journal} {\bibinfo
  {journal} {Phys. Rev. A}\ }\textbf {\bibinfo {volume} {82}},\ \bibinfo
  {pages} {013824} (\bibinfo {year} {2010})}\BibitemShut {NoStop}%
\bibitem [{\citenamefont {Kuzmich}\ \emph {et~al.}(2003)\citenamefont
  {Kuzmich}, \citenamefont {Bowen}, \citenamefont {Boozer}, \citenamefont
  {Boca}, \citenamefont {Chou}, \citenamefont {Duan},\ and\ \citenamefont
  {Kimble}}]{Cauchy_Schwarz_inequality_1}%
  \BibitemOpen
  \bibfield  {author} {\bibinfo {author} {\bibfnamefont {A.}~\bibnamefont
  {Kuzmich}}, \bibinfo {author} {\bibfnamefont {W.~P.}\ \bibnamefont {Bowen}},
  \bibinfo {author} {\bibfnamefont {A.~D.}\ \bibnamefont {Boozer}}, \bibinfo
  {author} {\bibfnamefont {A.}~\bibnamefont {Boca}}, \bibinfo {author}
  {\bibfnamefont {C.~W.}\ \bibnamefont {Chou}}, \bibinfo {author}
  {\bibfnamefont {L.-M.}\ \bibnamefont {Duan}}, \ and\ \bibinfo {author}
  {\bibfnamefont {H.~J.}\ \bibnamefont {Kimble}},\ }\bibfield  {title}
  {\enquote {\bibinfo {title} {Generation of nonclassical photon pairs for
  scalable quantum communication with atomic ensembles},}\ }\href {\doibase
  10.1038/nature01714} {\bibfield  {journal} {\bibinfo  {journal} {Nature}\
  }\textbf {\bibinfo {volume} {423}},\ \bibinfo {pages} {731--734} (\bibinfo
  {year} {2003})}\BibitemShut {NoStop}%
\bibitem [{\citenamefont {Riedinger}\ \emph {et~al.}(2016)\citenamefont
  {Riedinger}, \citenamefont {Hong}, \citenamefont {Norte}, \citenamefont
  {Slater}, \citenamefont {Shang}, \citenamefont {Krause}, \citenamefont
  {Anant}, \citenamefont {Aspelmeyer},\ and\ \citenamefont
  {Gr{\"o}blacher}}]{Cauchy_Schwarz_inequality_2}%
  \BibitemOpen
  \bibfield  {author} {\bibinfo {author} {\bibfnamefont {Ralf}\ \bibnamefont
  {Riedinger}}, \bibinfo {author} {\bibfnamefont {Sungkun}\ \bibnamefont
  {Hong}}, \bibinfo {author} {\bibfnamefont {Richard~A.}\ \bibnamefont
  {Norte}}, \bibinfo {author} {\bibfnamefont {Joshua~A.}\ \bibnamefont
  {Slater}}, \bibinfo {author} {\bibfnamefont {Juying}\ \bibnamefont {Shang}},
  \bibinfo {author} {\bibfnamefont {Alexander~G.}\ \bibnamefont {Krause}},
  \bibinfo {author} {\bibfnamefont {Vikas}\ \bibnamefont {Anant}}, \bibinfo
  {author} {\bibfnamefont {Markus}\ \bibnamefont {Aspelmeyer}}, \ and\ \bibinfo
  {author} {\bibfnamefont {Simon}\ \bibnamefont {Gr{\"o}blacher}},\ }\bibfield
  {title} {\enquote {\bibinfo {title} {Non-classical correlations between
  single photons and phonons from a mechanical oscillator},}\ }\href {\doibase
  10.1038/nature16536} {\bibfield  {journal} {\bibinfo  {journal} {Nature}\
  }\textbf {\bibinfo {volume} {530}},\ \bibinfo {pages} {313--316} (\bibinfo
  {year} {2016})}\BibitemShut {NoStop}%
\bibitem [{\citenamefont {Hoshi}\ \emph {et~al.}(2025)\citenamefont {Hoshi},
  \citenamefont {Nagase}, \citenamefont {Kwon}, \citenamefont {Iyama},
  \citenamefont {Kamiya}, \citenamefont {Fujii}, \citenamefont {Mukai},
  \citenamefont {Ahmed}, \citenamefont {Kockum}, \citenamefont {Watabe},
  \citenamefont {Yoshihara},\ and\ \citenamefont {Tsai}}]{entangled_Wigner}%
  \BibitemOpen
  \bibfield  {author} {\bibinfo {author} {\bibfnamefont {Daisuke}\ \bibnamefont
  {Hoshi}}, \bibinfo {author} {\bibfnamefont {Toshiaki}\ \bibnamefont
  {Nagase}}, \bibinfo {author} {\bibfnamefont {Sangil}\ \bibnamefont {Kwon}},
  \bibinfo {author} {\bibfnamefont {Daisuke}\ \bibnamefont {Iyama}}, \bibinfo
  {author} {\bibfnamefont {Takahiko}\ \bibnamefont {Kamiya}}, \bibinfo {author}
  {\bibfnamefont {Shiori}\ \bibnamefont {Fujii}}, \bibinfo {author}
  {\bibfnamefont {Hiroto}\ \bibnamefont {Mukai}}, \bibinfo {author}
  {\bibfnamefont {Shahnawaz}\ \bibnamefont {Ahmed}}, \bibinfo {author}
  {\bibfnamefont {Anton~Frisk}\ \bibnamefont {Kockum}}, \bibinfo {author}
  {\bibfnamefont {Shohei}\ \bibnamefont {Watabe}}, \bibinfo {author}
  {\bibfnamefont {Fumiki}\ \bibnamefont {Yoshihara}}, \ and\ \bibinfo {author}
  {\bibfnamefont {Jaw-Shen}\ \bibnamefont {Tsai}},\ }\bibfield  {title}
  {\enquote {\bibinfo {title} {Entangling {Schr\"odinger}'s cat states by
  bridging discrete- and continuous-variable encoding},}\ }\href {\doibase
  10.1038/s41467-025-56503-8} {\bibfield  {journal} {\bibinfo  {journal}
  {Nature Communications}\ }\textbf {\bibinfo {volume} {16}},\ \bibinfo {pages}
  {1309} (\bibinfo {year} {2025})}\BibitemShut {NoStop}%
\bibitem [{\citenamefont {Zippilli}\ and\ \citenamefont
  {Vitali}(2021)}]{PhysRevLett.126.020402}%
  \BibitemOpen
  \bibfield  {author} {\bibinfo {author} {\bibfnamefont {Stefano}\ \bibnamefont
  {Zippilli}}\ and\ \bibinfo {author} {\bibfnamefont {David}\ \bibnamefont
  {Vitali}},\ }\bibfield  {title} {\enquote {\bibinfo {title} {Dissipative
  engineering of gaussian entangled states in harmonic lattices with a
  single-site squeezed reservoir},}\ }\href {\doibase
  10.1103/PhysRevLett.126.020402} {\bibfield  {journal} {\bibinfo  {journal}
  {Phys. Rev. Lett.}\ }\textbf {\bibinfo {volume} {126}},\ \bibinfo {pages}
  {020402} (\bibinfo {year} {2021})}\BibitemShut {NoStop}%
\bibitem [{\citenamefont {Mi}\ \emph {et~al.}(2024)\citenamefont {Mi},
  \citenamefont {Michailidis}, \citenamefont {Shabani}, \citenamefont {Miao},
  \citenamefont {Klimov}, \citenamefont {Lloyd}, \citenamefont {Rosenberg},
  \citenamefont {Acharya}, \citenamefont {Aleiner}, \citenamefont {Andersen}
  \emph {et~al.}}]{mi2024stable}%
  \BibitemOpen
  \bibfield  {author} {\bibinfo {author} {\bibfnamefont {Xiao}\ \bibnamefont
  {Mi}}, \bibinfo {author} {\bibfnamefont {AA}~\bibnamefont {Michailidis}},
  \bibinfo {author} {\bibfnamefont {Sara}\ \bibnamefont {Shabani}}, \bibinfo
  {author} {\bibfnamefont {KC}~\bibnamefont {Miao}}, \bibinfo {author}
  {\bibfnamefont {PV}~\bibnamefont {Klimov}}, \bibinfo {author} {\bibfnamefont
  {J}~\bibnamefont {Lloyd}}, \bibinfo {author} {\bibfnamefont {E}~\bibnamefont
  {Rosenberg}}, \bibinfo {author} {\bibfnamefont {R}~\bibnamefont {Acharya}},
  \bibinfo {author} {\bibfnamefont {I}~\bibnamefont {Aleiner}}, \bibinfo
  {author} {\bibfnamefont {TI}~\bibnamefont {Andersen}},  \emph {et~al.},\
  }\bibfield  {title} {\enquote {\bibinfo {title} {Stable quantum-correlated
  many-body states through engineered dissipation},}\ }\href@noop {} {\bibfield
   {journal} {\bibinfo  {journal} {Science}\ }\textbf {\bibinfo {volume}
  {383}},\ \bibinfo {pages} {1332--1337} (\bibinfo {year} {2024})}\BibitemShut
  {NoStop}%
\bibitem [{\citenamefont {Li}\ \emph {et~al.}(2025)\citenamefont {Li},
  \citenamefont {Li}, \citenamefont {Cheng}, \citenamefont {Wang},
  \citenamefont {Zhao}, \citenamefont {Hou}, \citenamefont {Rehan},
  \citenamefont {Zhu}, \citenamefont {Yan}, \citenamefont {Qin} \emph
  {et~al.}}]{li2025programmable}%
  \BibitemOpen
  \bibfield  {author} {\bibinfo {author} {\bibfnamefont {Yue}\ \bibnamefont
  {Li}}, \bibinfo {author} {\bibfnamefont {Yi}~\bibnamefont {Li}}, \bibinfo
  {author} {\bibfnamefont {Xu}~\bibnamefont {Cheng}}, \bibinfo {author}
  {\bibfnamefont {Lingna}\ \bibnamefont {Wang}}, \bibinfo {author}
  {\bibfnamefont {Xingyu}\ \bibnamefont {Zhao}}, \bibinfo {author}
  {\bibfnamefont {Waner}\ \bibnamefont {Hou}}, \bibinfo {author} {\bibfnamefont
  {Kamran}\ \bibnamefont {Rehan}}, \bibinfo {author} {\bibfnamefont {Mingdong}\
  \bibnamefont {Zhu}}, \bibinfo {author} {\bibfnamefont {Lin}\ \bibnamefont
  {Yan}}, \bibinfo {author} {\bibfnamefont {Xi}~\bibnamefont {Qin}},  \emph
  {et~al.},\ }\bibfield  {title} {\enquote {\bibinfo {title} {Programmable
  multi-mode entanglement via dissipative engineering in vibrating trapped
  ions},}\ }\href@noop {} {\bibfield  {journal} {\bibinfo  {journal} {Science
  Advances}\ }\textbf {\bibinfo {volume} {11}},\ \bibinfo {pages} {eadv7838}
  (\bibinfo {year} {2025})}\BibitemShut {NoStop}%
\end{thebibliography}

%

\end{document}